\magnification=\magstep1 \overfullrule=0pt
\parskip=6pt
\baselineskip=15pt
\headline={\ifnum\pageno>1 \hss \number\pageno\ \hss \else\hfill \fi}
\pageno=1
\nopagenumbers
\hbadness=1000000
\vbadness=1000000

\input epsf

\centerline{\bf ON CHARACTERS OF $F_4$ LIE ALGEBRA}
\medskip

\centerline{\bf M. Gungormez} \centerline{Dept. Physics, Fac.
Science, Istanbul Tech. Univ.} \centerline{34469, Maslak, Istanbul,
Turkey } \centerline{e-mail: gungorm@itu.edu.tr}

\medskip

\centerline{\bf H. R. Karadayi} \centerline{Dept. Physics, Fac.
Science, Istanbul Tech. Univ.} \centerline{34469, Maslak, Istanbul,
Turkey } \centerline{e-mail: karadayi@itu.edu.tr } \centerline{and}
\centerline{Dept. Physics, Fac. Science, Istanbul Kultur University}
\centerline{34156, Atakoy, Istanbul, Turkey }

\medskip
\centerline{\bf Abstract}
\vskip5mm

In a previous work, we have given an explicit method to obtain
irreducible characters of finite Lie algebras without referring to
Weyl character formula. Irreducible characters of $G_2$ Lie algebra
has been given as an example. The work is now extended to somewhat
more complicated case of $F_4$ Lie algebra, in the same manner.

\vskip5mm
\vskip5mm
\vskip5mm
\vskip5mm
\vskip5mm
\vskip5mm

\eject

\vskip 3mm
\noindent {\bf{I.\ INTRODUCTION}}
\vskip 3mm

To be complete here, sections I. and II. will be a reminder of our
previous work{\bf [1]}. It is known that the character formula of
Weyl {\bf [2]} gives us a direct way to calculate the character of
irreducible representations of finite Lie algebras. For this, let
$G_r$ be a Lie algebra of rank r, $W(G_r)$ its Weyl group,
$\alpha_i$'s and $\lambda_i$'s be, respectively, its simple roots
and fundamental dominant weights. The notation here and in the
following sections will be as in our previous work {\bf [3]}. For
further reading, we refer to the beautiful book of Humphreys {\bf
[4]}. A dominant weight $\Lambda^+$ is expressed in the form
$$ \Lambda^+ = \sum_{i=1}^r s_i \lambda_i  \eqno(I.1) $$
where $s_i$'s are some positive integers (including zero). An
irreducible representation $V(\Lambda^+)$ can then be attributed to
$\Lambda^+$ . The character $Ch(\Lambda^+)$ of $V(\Lambda^+)$ is
defined by
$$ Ch(\Lambda^+) \equiv \sum_{\lambda^+} \ \sum_{\mu \in W(\lambda^+)}
m_{\Lambda^+}(\mu) \ e^\mu \eqno(I.2) $$ where $m_{\Lambda^+}(\mu)$\
's are multiplicities which count the number of times a weight $\mu$
is repeated for $V(\Lambda^+)$. The first sum here is over
$\Lambda^+$ and all of its sub-dominant weights $\lambda^+$'s while
the second sum is over the elements of their Weyl orbits
$W(\lambda^+)$'s. Formal exponentials are taken just as in the book
of Kac {\bf [5]} and in (I.2) we extend the concept for any weight
$\mu$, in the form $e^\mu$. Note here that multiplicities are
invariant under Weyl group actions and hence it is sufficient to
determine only $m_{\Lambda^+}(\lambda^+)$ for the whole Weyl orbit
$W(\lambda^+)$.

An equivalent form of (I.2) can be given by
$$ Ch(\Lambda^+) = { A(\rho+\Lambda^+) \over A(\rho) }  \eqno(I.3) $$
where $\rho$ is the Weyl vector of $G_r$. (I.3) is the celebrated
{\bf Weyl Character Formula} which gives us the possibility to calculate
characters in the most direct and efficient way. The central objects here
are $A(\rho+\Lambda^+)$'s which include a sum over the whole Weyl group:
$$ A(\rho+\Lambda^+) \equiv \sum_{\sigma \in W(G_r)} \ \epsilon(\sigma) \
e^{\sigma(\rho+\Lambda^+)}   \ \   .  \eqno(I.4) $$
In (I.4), $\sigma$ denotes an element of Weyl group,i.e. a Weyl reflection,
and $\epsilon(\sigma)$ is the corresponding {\bf signature} with values
either +1 or -1.

\eject

The structure of Weyl groups is completely known for finite Lie algebras
in principle. In practice, however, the problem is not so trivial, especially
for Lie algebras of some higher rank. The order of $E_8$ Weyl group is, for
instance, 696729600 and any application of Weyl character formula needs for
$E_8$ an explicit calculation of a sum over 696729600 Weyl reflections.
Our main point here is to overcome this difficulty in an essential manner.

For an actual application of (I.3), an important notice is the {\bf
specialization} of formal exponentials $e^\mu$ as is called in the
book of Kac {\bf [5]}. In its most general form, we consider here
the specialization
$$ e^{\alpha_i} \equiv u_i \  ,  \ i=1,2, \dots, r  .  \eqno(I.5) $$
which allows us to obtain $A(\rho)$ in the form of
$$ A(\rho) = P(u_1,u_2, \dots, u_r)  \eqno(I.6) $$
where $P(u_1,u_2, \dots, u_r)$ is a polynomial in indeterminates
$u_i$'s. We also have
$$ A(\rho+\Lambda^+) = P(u_1,u_2, \dots, u_r;s_1,s_2, \dots ,s_r) \eqno(I.7) $$
where $P(u_1,u_2, \dots, u_r;s_1,s_2, \dots, s_r)$ is another
polynomial of indeterminates $u_i$'s and also parameters $s_i$'s
defined in (I.1). The Weyl formula (I.3) then says that polynomial
(I.7) always factorizes on polynomial (I.6) leaving us with another
polynomial \break $R(u_1,u_2, \dots, u_r;s_1,s_2, \dots, s_r)$ which
is nothing but the character polynomial of $V(\Lambda^+)$.

The specialization (I.5) will always be normalized in such a way
that (I.3) gives us the {\bf Weyl dimension formula} {\bf [6]}, in
the limit $u_i = 1$ for all $i=1,2, \dots, r$. One also expects that
$$ P(u_1,u_2, \dots, u_r;0,0, \dots, 0) \equiv P(u_1,u_2, \dots, u_r) \ .
\eqno(I.8) $$

\vskip 3mm
\noindent {\bf{II.\ RECREATING $A(\rho+\Lambda^+)$ FROM $A(\rho)$ }}
\vskip 3mm

In this section, without any reference to Weyl groups, we give a way
to calculate polynomial (I.7) directly from polynomial (I.6). For this,
we first give the following explicit expression for polynomial (I.6):
$$ A(\rho) = {
\prod_{\alpha \in \Phi^+} \ ( e^\alpha - 1 ) \over
\prod_{i=1}^r \ (e^{\alpha_i})^{k_i}     }   \eqno(II.1)  $$
where
$$ k_i \equiv {1 \over 2} \ (\alpha_i,\alpha_i) \ (\lambda_i,\rho)
\eqno(II.2) $$
and $\Phi^+$ is the positive root system of $G_r$. Exponents $k_i$'s are due
to the fact that the monomial of maximal order is
$$ \prod_{i=1}^r \ (e^{\alpha_i})^{2 k_i}  $$
in the product $\prod_{\alpha \in \Phi^+} \ ( e^\alpha - 1 )$.
All the scalar products like $(\lambda_i,\alpha_j)$ or $(\alpha_i,\alpha_i)$
are the symmetrical ones and they are known to be defined via Cartan matrix of
a Lie algebra. The crucial point, however, is to see that (II.1) is equivalent to
$$ A(\rho) = \prod_{A=1}^{\vert W(G_r) \vert} \ \epsilon_A \ \
(e^{\alpha_i})^{\xi_i^0(A)}    \eqno(II.3)  $$ where $\vert W(G_r)
\vert$ is the order of Weyl group $W(G_r)$ and as is emphasized in
\break section I, $\epsilon_A$'s are signatures with values
$\epsilon_A = \mp 1$. Note here that, by expanding the product
$\prod_{\alpha \in \Phi^+} \ ( e^\alpha - 1 )$, the precise values
of signatures can be determined uniquely.

To get an explicit expression for exponents $\xi_i^0(A)$ in (II.3),
let us define ${\bf \it R^+}$ is composed out of elements of the
form
$$ \beta^+ \equiv \sum_{i=1}^r n_i \ \alpha_i   \eqno(II.4) $$
where $n_i$'s are some positive integers including zero. ${\bf \it R^+}$ is a subset
of the Positive Root Lattice of $G_r$. The main emphasis here is on some special roots
$$ \gamma_i(I_i) \in {\bf \it R^+} $$
which are defined by following conditions
$$ (\lambda_i - \gamma_i(I_i),\lambda_j - \gamma_j(I_j)) = (\lambda_i,\lambda_j) \ \  ,
\ \  i,j = 1,2, \dots, r .   \eqno(II.5)  $$
Note here that $(\lambda_i,\lambda_j)$'s are defined by inverse Cartan matrix.
For the range of indices $ I_j$'s, we assume that they take values from the set
$ \{1,2, \dots, \vert I_j \vert \} $, that is
$$ I_j \in \{1,2, \dots, \vert I_j \vert \} \ \ , \ \ j = 1,2, \dots, r. $$

\noindent We also define the sets

$$ \Gamma(A) \equiv \{ \gamma_1(I_1(A)),\gamma_2(I_2(A)), \dots, \gamma_r(I_r(A)) \} \ \ ,
\ \ A=1,2, \dots, D  \eqno(II.6)  $$

\noindent are chosen by conditions (II.5) where $I_j(A) \in \{1,2,
\dots, \vert I_j \vert \} $ and D is the maximal number of these
sets.

Following two statements are then valid:

\noindent (1) \ \ ${\bf  D = \vert W(G_r) \vert  }$

\noindent (2) \ \ ${\bf \vert I_j \vert  = \vert W(\lambda_j) \vert  }$

\noindent where $ \vert W(\lambda_j) \vert $ is the order, i.e.
the number of elements, of the Weyl orbit $W(\lambda_j)$. As is known,
a Weyl orbit is stable under Weyl reflections and hence all its elements
have the same length. It is interesting to note however that the lengths
of any two elements $\gamma_i(j_1)$ and $\gamma_i(j_2)$ could, in general,
be different while the statement (2) is, still, valid.

The exponents in (II.3) can now be defined by

$$ \xi_i^0(A) \equiv {1 \over 2} \ (\alpha_i,\alpha_i) \
(\lambda_i - \gamma_i(A),\rho)  \ \ . \eqno(II.7) $$

\noindent We then also state that the natural extension of (II.3) is
as in the following:

$$ A(\rho+\Lambda^+) = \prod_{A=1}^{\vert W(G_r) \vert} \ \epsilon_A \ \
(e^{\alpha_i})^{\xi_i(A)}    \eqno(II.8)  $$

\noindent for which the exponents are

$$ \xi_i(A) \equiv {1 \over 2} \ (\alpha_i,\alpha_i) \
(\lambda_i - \gamma_i(A),\rho+\Lambda^+)  \ \ . \eqno(II.9) $$

\noindent as a natural extension of (II.7). Signatures have the same
values in both expressions, (II.3) and (II.8). This reduces the
problem of explicit calculation of character polynomial
$Ch(\Lambda^+)$ to the problem of finding solutions to conditions
(II.5). It is clear that this is more manageable than that of using
Weyl character formula directly.

\vskip 3mm \noindent {\bf{III.\ EXPLICIT CONSTRUCTION OF $F_4$
CHARACTERS }} \vskip 3mm

It is known that $F_4$ is characterized by two different root
lengths
$$ (\alpha_1,\alpha_1)=(\alpha_2,\alpha_2)=4 \ \ , \ \
(\alpha_3,\alpha_3)=(\alpha_4,\alpha_4) = 2 \eqno(III.1)$$ \noindent
and also
$$ (\alpha_1,\alpha_2)=(\alpha_2,\alpha_3)=-2  \ \, \ \ (\alpha_3,\alpha_4)=-1
\eqno(III.2) $$

\eject
in the following normalization of its Dynkin diagram

\epsfxsize=5cm \centerline{\epsfbox{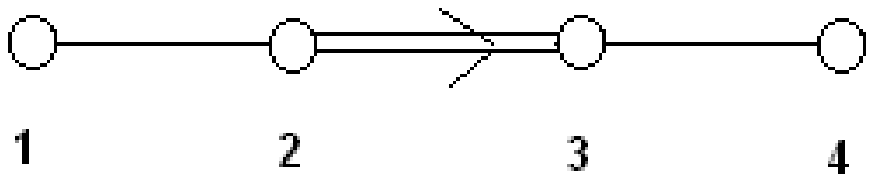}}

\noindent Its positive root system is thus

$$ \eqalign{ \Phi^+ = \{ &\alpha_1 , \ \alpha_2 , \ \alpha_3, \
\alpha_4, \ \alpha_1+\alpha_2, \ \alpha_2+\alpha_3, \
\alpha_3+\alpha_4, \ \alpha_2+2 \ \alpha_3, \
\alpha_1+\alpha_2+\alpha_3, \ \alpha_2+\alpha_3+\alpha_4, \cr
&\alpha_1+\alpha_2+2 \ \alpha_3, \ \alpha_2+2 \ \alpha_3+\alpha_4, \
\alpha_1+\alpha_2+\alpha_3+\alpha_4, \ \alpha_2+2 \ \alpha_3+2 \
\alpha_4, \cr &\alpha_1+2 \ \alpha_2+2 \ \alpha_3, \
\alpha_1+\alpha_2+2 \ \alpha_3+\alpha_4, \ \alpha_1+\alpha_2+2 \
\alpha_3+2 \ \alpha_4, \cr &\alpha_1+2 \ \alpha_2+2 \
\alpha_3+\alpha_4, \ \alpha_1+2 \ \alpha_2+2 \ \alpha_3+2 \
\alpha_4, \ \alpha_1+2 \ \alpha_2+3 \ \alpha_3+\alpha_4, \ \cr
&\alpha_1+2 \ \alpha_2+3 \ \alpha_3+2 \ \alpha_4, \ \alpha_1+2 \
\alpha_2+4 \ \alpha_3+2 \ \alpha_4, \ \alpha_1+3 \ \alpha_2+4 \
\alpha_3+2 \ \alpha_4, \cr &2 \ \alpha_1+3 \ \alpha_2+4 \ \alpha_3+2
\ \alpha_4 \} } \eqno(III.3)  $$

The so-called special roots which are specified as in definitions
(II.5) will be given here in Table-I :

$$ \vbox{\tabskip=0pt \offinterlineskip
\halign to380pt{\strut#& \vrule#\tabskip=0em plus1em& \hfil#&
\vrule#& \hfil#\hfil& \vrule#\tabskip=0pt\cr \noalign{\hrule}

&& $\eqalign{ \hfill \cr \gamma_1( 1)&=0          \cr \gamma_1(
2)&=\alpha_1   \cr \gamma_1( 3)&=\alpha_1 + \alpha_2 \cr \gamma_1(
4)&=\alpha_1 + \alpha_2 + 2 \alpha_3 \cr \gamma_1( 5)&=\alpha_1 +
\alpha_2 + 2 \alpha_3 + 2 \alpha_4 \cr \gamma_1( 6)&=\alpha_1 + 2
\alpha_2 + 2 \alpha_3 \cr \gamma_1( 7)&=\alpha_1 + 2 \alpha_2 + 2
\alpha_3 + 2 \alpha_4 \cr \gamma_1( 8)&=\alpha_1 + 2 \alpha_2 + 4
\alpha_3 + 2 \alpha_4 \cr \gamma_1( 9)&=\alpha_1 + 3 \alpha_2 + 4
\alpha_3 + 2 \alpha_4 \cr \gamma_1( 10)&=2 \alpha_1 + 2 \alpha_2 + 2
\alpha_3 \cr \gamma_1( 11)&=2 \alpha_1 + 2 \alpha_2 + 2 \alpha_3 + 2
\alpha_4 \cr \gamma_1( 12)&=2 \alpha_1 + 2 \alpha_2 + 4 \alpha_3 + 2
\alpha_4 \cr \gamma_1( 13)&=2 \alpha_1 + 4 \alpha_2 + 4 \alpha_3 + 2
\alpha_4 \cr \gamma_1( 14)&=2 \alpha_1 + 4 \alpha_2 + 6 \alpha_3 + 2
\alpha_4 \cr \gamma_1( 15)&=2 \alpha_1 + 4 \alpha_2 + 6 \alpha_3 + 4
\alpha_4 \cr  \hfill \cr } $

&& $\eqalign{ \hfill \cr \gamma_4( 1)&=0  \cr \gamma_4( 2)&=\alpha_4
\cr \gamma_4( 3)&=\alpha_3 + \alpha_4 \cr \gamma_4( 4)&=\alpha_2 +
\alpha_3 + \alpha_4 \cr \gamma_4( 5)&=\alpha_2 + 2 \alpha_3 +
\alpha_4 \cr \gamma_4( 6)&=\alpha_2 + 2 \alpha_3 + 2 \alpha_4 \cr
\gamma_4( 7)&=\alpha_1 + \alpha_2 + \alpha_3 + \alpha_4 \cr
\gamma_4( 8)&=\alpha_1 + \alpha_2 + 2 \alpha_3 + \alpha_4 \cr
\gamma_4( 9)&=\alpha_1 + \alpha_2 + 2 \alpha_3 + 2 \alpha_4 \cr
\gamma_4(10)&=\alpha_1 + 2 \alpha_2 + 2 \alpha_3 + \alpha_4 \cr
\gamma_4(11)&=\alpha_1 + 2 \alpha_2 + 2 \alpha_3 + 2 \alpha_4 \cr
\gamma_4(12)&=\alpha_1 + 2 \alpha_2 + 3 \alpha_3 + \alpha_4 \cr
\gamma_4(13)&=\alpha_1 + 2 \alpha_2 + 3 \alpha_3 + 3 \alpha_4 \cr
\gamma_4(14)&=\alpha_1 + 2 \alpha_2 + 4 \alpha_3 + 2 \alpha_4 \cr
\gamma_4(15)&=\alpha_1 + 2 \alpha_2 + 4 \alpha_3 + 3 \alpha_4 \cr
\hfill \cr } $  & \cr \noalign{\hrule} }} $$

$$ \vbox{\tabskip=0pt \offinterlineskip
\halign to380pt{\strut#& \vrule#\tabskip=0em plus1em& \hfil#&
\vrule#& \hfil#\hfil&
\vrule#\tabskip=0pt\cr \noalign{\hrule}

&& $\eqalign{ \hfill \cr  \gamma_1( 16)&=3 \alpha_1 + 3 \alpha_2 + 4
\alpha_3 + 2 \alpha_4 \cr \gamma_1( 17)&=3 \alpha_1 + 4 \alpha_2 + 4
\alpha_3 + 2 \alpha_4 \cr \gamma_1( 18)&=3 \alpha_1 + 4 \alpha_2 + 6
\alpha_3 + 2 \alpha_4 \cr \gamma_1( 19)&=3 \alpha_1 + 4 \alpha_2 + 6
\alpha_3 + 4 \alpha_4 \cr \gamma_1( 20)&=3 \alpha_1 + 5 \alpha_2 + 6
\alpha_3 + 2 \alpha_4 \cr \gamma_1( 21)&=3 \alpha_1 + 5 \alpha_2 + 6
\alpha_3 + 4 \alpha_4 \cr \gamma_1( 22)&=3 \alpha_1 + 5 \alpha_2 + 8
\alpha_3 + 4 \alpha_4 \cr \gamma_1( 23)&=3 \alpha_1 + 6 \alpha_2 + 8
\alpha_3 + 4 \alpha_4 \cr \gamma_1( 24)&=4 \alpha_1 + 6 \alpha_2 + 8
\alpha_3 + 4 \alpha_4 \cr \hfill \cr } $

&& $\eqalign{ \hfill \cr \gamma_4(16)&=\alpha_1 + 3 \alpha_2 + 4
\alpha_3 + 2 \alpha_4 \cr \gamma_4(17)&=\alpha_1 + 3 \alpha_2 + 4
\alpha_3 + 3 \alpha_4 \cr \gamma_4(18)&=\alpha_1 + 3 \alpha_2 + 5
\alpha_3 + 3 \alpha_4 \cr \gamma_4(19)&=2 \alpha_1 + 3 \alpha_2 + 4
\alpha_3 + 2 \alpha_4 \cr \gamma_4(20)&=2 \alpha_1 + 3 \alpha_2 + 4
\alpha_3 + 3 \alpha_4 \cr \gamma_4(21)&=2 \alpha_1 + 3 \alpha_2 + 5
\alpha_3 + 3 \alpha_4 \cr \gamma_4(22)&=2 \alpha_1 + 4 \alpha_2 + 5
\alpha_3 + 3 \alpha_4 \cr \gamma_4(23)&=2 \alpha_1 + 4 \alpha_2 + 6
\alpha_3 + 3 \alpha_4 \cr \gamma_4(24)&=2 \alpha_1 + 4 \alpha_2 + 6
\alpha_3 + 4 \alpha_4 \cr \hfill \cr } $  & \cr \noalign{\hrule} }}
$$

$$ \vbox{\tabskip=0pt \offinterlineskip
\halign to380pt{\strut#& \vrule#\tabskip=0em plus1em& \hfil#&
\vrule#& \hfil#\hfil& \vrule#\tabskip=0pt\cr \noalign{\hrule} &&
$\eqalign{ \hfill \cr \gamma_2( 1)&=0 \cr \gamma_2( 2)&=\alpha_2 \cr
\gamma_2( 3)&=\alpha_2 + 2 \alpha_3 \cr \gamma_2( 4)&=\alpha_2 + 2
\alpha_3 + 2 \alpha_4 \cr \gamma_2( 5)&=2 \alpha_2 + 2 \alpha_3 \cr
\gamma_2( 6)&=2 \alpha_2 + 2 \alpha_3 + 2 \alpha_4 \cr \gamma_2(
7)&=2 \alpha_2 + 4 \alpha_3 + 2 \alpha_4 \cr \gamma_2( 8)&=3
\alpha_2 + 4 \alpha_3 + 2 \alpha_4 \cr \gamma_2( 9)&=\alpha_1 +
\alpha_2 \cr \gamma_2(10)&=\alpha_1 + \alpha_2 + 2 \alpha_3 \cr
\gamma_2(11)&=\alpha_1 + \alpha_2 + 2 \alpha_3 + 2 \alpha_4 \cr
\gamma_2(12)&=\alpha_1 + 3 \alpha_2 + 2 \alpha_3 \cr
\gamma_2(13)&=\alpha_1 + 3 \alpha_2 + 2 \alpha_3 + 2 \alpha_4 \cr
\gamma_2(14)&=\alpha_1 + 3 \alpha_2 + 4 \alpha_3 \cr
\gamma_2(15)&=\alpha_1 + 3 \alpha_2 + 4 \alpha_3 + 4 \alpha_4 \cr
\gamma_2(16)&=\alpha_1 + 3 \alpha_2 + 6 \alpha_3 + 2 \alpha_4 \cr
\gamma_2(17)&=\alpha_1 + 3 \alpha_2 + 6 \alpha_3 + 4 \alpha_4 \cr
\gamma_2(18)&=\alpha_1 + 5 \alpha_2 + 6 \alpha_3 + 2 \alpha_4 \cr
\gamma_2(19)&=\alpha_1 + 5 \alpha_2 + 6 \alpha_3 + 4 \alpha_4 \cr
\gamma_2(20)&=\alpha_1 + 5 \alpha_2 + 8 \alpha_3 + 4 \alpha_4 \cr
\hfill \cr }$

&& $\eqalign{ \hfill \cr \gamma_3( 1)&=0 \cr \gamma_3( 2)&=\alpha_3
\cr \gamma_3( 3)&=\alpha_3 + \alpha_4 \cr \gamma_3( 4)&=\alpha_2 +
\alpha_3 \cr \gamma_3( 5)&=\alpha_2 + \alpha_3 + \alpha_4 \cr
\gamma_3( 6)&=\alpha_2 + 2 \alpha_3 \cr \gamma_3( 7)&=\alpha_2 + 2
\alpha_3 + 2 \alpha_4   \cr \gamma_3( 8)&=\alpha_2 + 3 \alpha_3 +
\alpha_4 \cr \gamma_3( 9)&=\alpha_2 + 3 \alpha_3 + 2 \alpha_4   \cr
\gamma_3(10)&=2 \alpha_2 + 3 \alpha_3 + \alpha_4   \cr
\gamma_3(11)&=2 \alpha_2 + 3 \alpha_3 + 2 \alpha_4 \cr
\gamma_3(12)&=2 \alpha_2 + 4 \alpha_3 + 2 \alpha_4 \cr
\gamma_3(13)&=\alpha_1 + \alpha_2 + \alpha_3 \cr
\gamma_3(14)&=\alpha_1 + \alpha_2 + \alpha_3 + \alpha_4 \cr
\gamma_3(15)&=\alpha_1 + \alpha_2 + 2 \alpha_3 \cr
\gamma_3(16)&=\alpha_1 + \alpha_2 + 2 \alpha_3 + 2 \alpha_4 \cr
\gamma_3(17)&=\alpha_1 + \alpha_2 + 3 \alpha_3 + \alpha_4 \cr
\gamma_3(18)&=\alpha_1 + \alpha_2 + 3 \alpha_3 + 2 \alpha_4 \cr
\gamma_3(19)&=\alpha_1 + 2 \alpha_2 + 2 \alpha_3 \cr
\gamma_3(20)&=\alpha_1 + 2 \alpha_2 + 2 \alpha_3 + 2 \alpha_4 \cr
\hfill \cr } $  & \cr \noalign{\hrule} }} $$

$$ \vbox{\tabskip=0pt \offinterlineskip
\halign to380pt{\strut#& \vrule#\tabskip=0em plus1em& \hfil#&
\vrule#& \hfil#\hfil& \vrule#\tabskip=0pt\cr \noalign{\hrule} &&
$\eqalign{ \hfill \cr \gamma_2(21)&=2 \alpha_1 + 2 \alpha_2 + 2
\alpha_3 \cr \gamma_2(22)&=2 \alpha_1 + 2 \alpha_2 + 2 \alpha_3 + 2
\alpha_4 \cr \gamma_2(23)&=2 \alpha_1 + 2 \alpha_2 + 4 \alpha_3 + 2
\alpha_4 \cr \gamma_2(24)&=2 \alpha_1 + 3 \alpha_2 + 2 \alpha_3 \cr
\gamma_2(25)&=2 \alpha_1 + 3 \alpha_2 + 2 \alpha_3 + 2 \alpha_4 \cr
\gamma_2(26)&=2 \alpha_1 + 3 \alpha_2 + 4 \alpha_3 \cr
\gamma_2(27)&=2 \alpha_1 + 3 \alpha_2 + 4 \alpha_3 + 4 \alpha_4 \cr
\gamma_2(28)&=2 \alpha_1 + 3 \alpha_2 + 6 \alpha_3 + 2 \alpha_4 \cr
\gamma_2(29)&=2 \alpha_1 + 3 \alpha_2 + 6 \alpha_3 + 4 \alpha_4 \cr
\gamma_2(30)&=2 \alpha_1 + 4 \alpha_2 + 4 \alpha_3 \cr
\gamma_2(31)&=2 \alpha_1 + 4 \alpha_2 + 4 \alpha_3 + 4 \alpha_4 \cr
\gamma_2(32)&=2 \alpha_1 + 4 \alpha_2 + 8 \alpha_3 + 4 \alpha_4 \cr
\gamma_2(33)&=2 \alpha_1 + 5 \alpha_2 + 4 \alpha_3 + 2 \alpha_4 \cr
\gamma_2(34)&=2 \alpha_1 + 5 \alpha_2 + 8 \alpha_3 + 2 \alpha_4 \cr
\gamma_2(35)&=2 \alpha_1 + 5 \alpha_2 + 8 \alpha_3 + 6 \alpha_4 \cr
\gamma_2(36)&=2 \alpha_1 + 6 \alpha_2 + 6 \alpha_3 + 2 \alpha_4 \cr
\gamma_2(37)&=2 \alpha_1 + 6 \alpha_2 + 6 \alpha_3 + 4 \alpha_4 \cr
\gamma_2(38)&=2 \alpha_1 + 6 \alpha_2 + 8 \alpha_3 + 2 \alpha_4 \cr
\gamma_2(39)&=2 \alpha_1 + 6 \alpha_2 + 8 \alpha_3 + 6 \alpha_4 \cr
\gamma_2(40)&=2 \alpha_1 + 6 \alpha_2 + 10 \alpha_3 + 4 \alpha_4 \cr
\gamma_2(41)&=2 \alpha_1 + 6 \alpha_2 + 10 \alpha_3 + 6 \alpha_4 \cr
\gamma_2(42)&=2 \alpha_1 + 7 \alpha_2 + 8 \alpha_3 + 4 \alpha_4 \cr
\gamma_2(43)&=2 \alpha_1 + 7 \alpha_2 + 10 \alpha_3 + 4 \alpha_4 \cr
\gamma_2(44)&=2 \alpha_1 + 7 \alpha_2 + 10 \alpha_3 + 6 \alpha_4 \cr
\gamma_2(45)&=3 \alpha_1 + 3 \alpha_2 + 4 \alpha_3 + 2 \alpha_4 \cr
\gamma_2(46)&=3 \alpha_1 + 5 \alpha_2 + 4 \alpha_3 + 2 \alpha_4 \cr
\gamma_2(47)&=3 \alpha_1 + 5 \alpha_2 + 8 \alpha_3 + 2 \alpha_4 \cr
\gamma_2(48)&=3 \alpha_1 + 5 \alpha_2 + 8 \alpha_3 + 6 \alpha_4 \cr
\gamma_2(49)&=3 \alpha_1 + 7 \alpha_2 + 8 \alpha_3 + 2 \alpha_4 \cr
\gamma_2(50)&=3 \alpha_1 + 7 \alpha_2 + 8 \alpha_3 + 6 \alpha_4 \cr
\gamma_2(51)&=3 \alpha_1 + 7 \alpha_2 + 12 \alpha_3 + 6 \alpha_4 \cr
\gamma_2(52)&=3 \alpha_1 + 9 \alpha_2 + 12 \alpha_3 + 6 \alpha_4 \cr
\gamma_2(53)&=4 \alpha_1 + 5 \alpha_2 + 6 \alpha_3 + 2 \alpha_4 \cr
\gamma_2(54)&=4 \alpha_1 + 5 \alpha_2 + 6 \alpha_3 + 4 \alpha_4 \cr
\gamma_2(55)&=4 \alpha_1 + 5 \alpha_2 + 8 \alpha_3 + 4 \alpha_4 \cr
\gamma_2(56)&=4 \alpha_1 + 6 \alpha_2 + 6 \alpha_3 + 2 \alpha_4 \cr
\hfill \cr } $

&& $\eqalign{ \hfill \cr \gamma_3(21)&=\alpha_1 + 2 \alpha_2 + 3
\alpha_3  \cr \gamma_3(22)&=\alpha_1 + 2 \alpha_2 + 3 \alpha_3 + 3
\alpha_4 \cr \gamma_3(23)&=\alpha_1 + 2 \alpha_2 + 5 \alpha_3 + 2
\alpha_4 \cr \gamma_3(24)&=\alpha_1 + 2 \alpha_2 + 5 \alpha_3 + 3
\alpha_4 \cr \gamma_3(25)&=\alpha_1 + 3 \alpha_2 + 3 \alpha_3 +
\alpha_4 \cr \gamma_3(26)&=\alpha_1 + 3 \alpha_2 + 3 \alpha_3 + 2
\alpha_4 \cr \gamma_3(27)&=\alpha_1 + 3 \alpha_2 + 5 \alpha_3 +
\alpha_4 \cr \gamma_3(28)&=\alpha_1 + 3 \alpha_2 + 5 \alpha_3 + 4
\alpha_4 \cr \gamma_3(29)&=\alpha_1 + 3 \alpha_2 + 6 \alpha_3 + 2
\alpha_4 \cr \gamma_3(30)&=\alpha_1 + 3 \alpha_2 + 6 \alpha_3 + 4
\alpha_4 \cr \gamma_3(31)&=\alpha_1 + 4 \alpha_2 + 5 \alpha_3 + 2
\alpha_4 \cr \gamma_3(32)&=\alpha_1 + 4 \alpha_2 + 5 \alpha_3 + 3
\alpha_4 \cr \gamma_3(33)&=\alpha_1 + 4 \alpha_2 + 6 \alpha_3 + 2
\alpha_4 \cr \gamma_3(34)&=\alpha_1 + 4 \alpha_2 + 6 \alpha_3 + 4
\alpha_4 \cr \gamma_3(35)&=\alpha_1 + 4 \alpha_2 + 7 \alpha_3 + 3
\alpha_4 \cr \gamma_3(36)&=\alpha_1 + 4 \alpha_2 + 7 \alpha_3 + 4
\alpha_4 \cr \gamma_3(37)&=2 \alpha_1 + 2 \alpha_2 + 3 \alpha_3 +
\alpha_4 \cr \gamma_3(38)&=2 \alpha_1 + 2 \alpha_2 + 3 \alpha_3 + 2
\alpha_4 \cr \gamma_3(39)&=2 \alpha_1 + 2 \alpha_2 + 4 \alpha_3 + 2
\alpha_4 \cr \gamma_3(40)&=2 \alpha_1 + 3 \alpha_2 + 3 \alpha_3 +
\alpha_4 \cr \gamma_3(41)&=2 \alpha_1 + 3 \alpha_2 + 3 \alpha_3 + 2
\alpha_4 \cr \gamma_3(42)&=2 \alpha_1 + 3 \alpha_2 + 5 \alpha_3 +
\alpha_4 \cr \gamma_3(43)&=2 \alpha_1 + 3 \alpha_2 + 5 \alpha_3 + 4
\alpha_4 \cr \gamma_3(44)&=2 \alpha_1 + 3 \alpha_2 + 6 \alpha_3 + 2
\alpha_4 \cr \gamma_3(45)&=2 \alpha_1 + 3 \alpha_2 + 6 \alpha_3 + 4
\alpha_4 \cr \gamma_3(46)&=2 \alpha_1 + 4 \alpha_2 + 4 \alpha_3 + 2
\alpha_4 \cr \gamma_3(47)&=2 \alpha_1 + 4 \alpha_2 + 5 \alpha_3 +
\alpha_4 \cr \gamma_3(48)&=2 \alpha_1 + 4 \alpha_2 + 5 \alpha_3 + 4
\alpha_4 \cr \gamma_3(49)&=2 \alpha_1 + 4 \alpha_2 + 7 \alpha_3 + 2
\alpha_4 \cr \gamma_3(50)&=2 \alpha_1 + 4 \alpha_2 + 7 \alpha_3 + 5
\alpha_4 \cr \gamma_3(51)&=2 \alpha_1 + 4 \alpha_2 + 8 \alpha_3 + 4
\alpha_4 \cr \gamma_3(52)&=2 \alpha_1 + 5 \alpha_2 + 6 \alpha_3 + 2
\alpha_4 \cr \gamma_3(53)&=2 \alpha_1 + 5 \alpha_2 + 6 \alpha_3 + 4
\alpha_4 \cr \gamma_3(54)&=2 \alpha_1 + 5 \alpha_2 + 7 \alpha_3 + 2
\alpha_4 \cr \gamma_3(55)&=2 \alpha_1 + 5 \alpha_2 + 7 \alpha_3 + 5
\alpha_4 \cr \gamma_3(56)&=2 \alpha_1 + 5 \alpha_2 + 9 \alpha_3 + 4
\alpha_4 \cr \hfill \cr } $  & \cr \noalign{\hrule} }} $$

$$ \vbox{\tabskip=0pt \offinterlineskip
\halign to380pt{\strut#& \vrule#\tabskip=0em plus1em& \hfil#&
\vrule#& \hfil#\hfil& \vrule#\tabskip=0pt\cr \noalign{\hrule} &&
$\eqalign{ \hfill \cr \gamma_2(57)&=4 \alpha_1 + 6 \alpha_2 + 6
\alpha_3 + 4 \alpha_4 \cr \gamma_2(58)&=4 \alpha_1 + 6 \alpha_2 + 8
\alpha_3 + 2 \alpha_4 \cr \gamma_2(59)&=4 \alpha_1 + 6 \alpha_2 + 8
\alpha_3 + 6 \alpha_4 \cr \gamma_2(60)&=4 \alpha_1 + 6 \alpha_2 + 10
\alpha_3 + 4 \alpha_4 \cr \gamma_2(61)&=4 \alpha_1 + 6 \alpha_2 + 10
\alpha_3 + 6 \alpha_4 \cr \gamma_2(62)&=4 \alpha_1 + 7 \alpha_2 + 8
\alpha_3 + 2 \alpha_4 \cr \gamma_2(63)&=4 \alpha_1 + 7 \alpha_2 + 8
\alpha_3 + 6 \alpha_4 \cr \gamma_2(64)&=4 \alpha_1 + 7 \alpha_2 + 12
\alpha_3 + 6 \alpha_4 \cr \gamma_2(65)&=4 \alpha_1 + 8 \alpha_2 + 8
\alpha_3 + 4 \alpha_4 \cr \gamma_2(66)&=4 \alpha_1 + 8 \alpha_2 + 12
\alpha_3 + 4 \alpha_4 \cr \gamma_2(67)&=4 \alpha_1 + 8 \alpha_2 + 12
\alpha_3 + 8 \alpha_4 \cr \gamma_2(68)&=4 \alpha_1 + 9 \alpha_2 + 10
\alpha_3 + 4 \alpha_4 \cr \gamma_2(69)&=4 \alpha_1 + 9 \alpha_2 + 10
\alpha_3 + 6 \alpha_4 \cr \gamma_2(70)&=4 \alpha_1 + 9 \alpha_2 + 12
\alpha_3 + 4 \alpha_4 \cr \gamma_2(71)&=4 \alpha_1 + 9 \alpha_2 + 12
\alpha_3 + 8 \alpha_4 \cr \gamma_2(72)&=4 \alpha_1 + 9 \alpha_2 + 14
\alpha_3 + 6 \alpha_4 \cr \gamma_2(73)&=4 \alpha_1 + 9 \alpha_2 + 14
\alpha_3 + 8 \alpha_4 \cr \gamma_2(74)&=4 \alpha_1 + 10 \alpha_2 +
12 \alpha_3 + 6 \alpha_4 \cr \gamma_2(75)&=4 \alpha_1 + 10 \alpha_2
+ 14 \alpha_3 + 6 \alpha_4 \cr \gamma_2(76)&=4 \alpha_1 + 10
\alpha_2 + 14 \alpha_3 + 8 \alpha_4 \cr \gamma_2(77)&=5 \alpha_1 + 7
\alpha_2 + 8 \alpha_3 + 4 \alpha_4 \cr \gamma_2(78)&=5 \alpha_1 + 7
\alpha_2 + 10 \alpha_3 + 4 \alpha_4 \cr \gamma_2(79)&=5 \alpha_1 + 7
\alpha_2 + 10 \alpha_3 + 6 \alpha_4 \cr \gamma_2(80)&=5 \alpha_1 + 9
\alpha_2 + 10 \alpha_3 + 4 \alpha_4 \cr \gamma_2(81)&=5 \alpha_1 + 9
\alpha_2 + 10 \alpha_3 + 6 \alpha_4 \cr \gamma_2(82)&=5 \alpha_1 + 9
\alpha_2 + 12 \alpha_3 + 4 \alpha_4 \cr \gamma_2(83)&=5 \alpha_1 + 9
\alpha_2 + 12 \alpha_3 + 8 \alpha_4 \cr \gamma_2(84)&=5 \alpha_1 + 9
\alpha_2 + 14 \alpha_3 + 6 \alpha_4 \cr \gamma_2(85)&=5 \alpha_1 + 9
\alpha_2 + 14 \alpha_3 + 8 \alpha_4 \cr \gamma_2(86)&=5 \alpha_1 +
11 \alpha_2 + 14 \alpha_3 + 6 \alpha_4 \cr \gamma_2(87)&=5 \alpha_1
+ 11 \alpha_2 + 14 \alpha_3 + 8 \alpha_4 \cr \gamma_2(88)&=5
\alpha_1 + 11 \alpha_2 + 16 \alpha_3 + 8 \alpha_4  \cr
\gamma_2(89)&=6 \alpha_1 + 9 \alpha_2 + 12 \alpha_3 + 6 \alpha_4 \cr
\gamma_2(90)&=6 \alpha_1 + 10 \alpha_2 + 12 \alpha_3 + 6 \alpha_4
\cr  \hfill \cr } $

&& $\eqalign{ \hfill \cr \gamma_3(57)&=2 \alpha_1 + 5 \alpha_2 + 9
\alpha_3 + 5 \alpha_4 \cr \gamma_3(58)&=2 \alpha_1 + 6 \alpha_2 + 8
\alpha_3 + 4 \alpha_4 \cr \gamma_3(59)&=2 \alpha_1 + 6 \alpha_2 + 9
\alpha_3 + 4 \alpha_4 \cr \gamma_3(60)&=2 \alpha_1 + 6 \alpha_2 + 9
\alpha_3 + 5 \alpha_4 \cr \gamma_3(61)&=3 \alpha_1 + 4 \alpha_2 + 5
\alpha_3 + 2 \alpha_4 \cr \gamma_3(62)&=3 \alpha_1 + 4 \alpha_2 + 5
\alpha_3 + 3 \alpha_4 \cr \gamma_3(63)&=3 \alpha_1 + 4 \alpha_2 + 6
\alpha_3 + 2 \alpha_4 \cr \gamma_3(64)&=3 \alpha_1 + 4 \alpha_2 + 6
\alpha_3 + 4 \alpha_4 \cr \gamma_3(65)&=3 \alpha_1 + 4 \alpha_2 + 7
\alpha_3 + 3 \alpha_4 \cr \gamma_3(66)&=3 \alpha_1 + 4 \alpha_2 + 7
\alpha_3 + 4 \alpha_4 \cr \gamma_3(67)&=3 \alpha_1 + 5 \alpha_2 + 6
\alpha_3 + 2 \alpha_4 \cr \gamma_3(68)&=3 \alpha_1 + 5 \alpha_2 + 6
\alpha_3 + 4 \alpha_4 \cr \gamma_3(69)&=3 \alpha_1 + 5 \alpha_2 + 7
\alpha_3 + 2 \alpha_4 \cr \gamma_3(70)&=3 \alpha_1 + 5 \alpha_2 + 7
\alpha_3 + 5 \alpha_4 \cr \gamma_3(71)&=3 \alpha_1 + 5 \alpha_2 + 9
\alpha_3 + 4 \alpha_4 \cr \gamma_3(72)&=3 \alpha_1 + 5 \alpha_2 + 9
\alpha_3 + 5 \alpha_4 \cr \gamma_3(73)&=3 \alpha_1 + 6 \alpha_2 + 7
\alpha_3 + 3 \alpha_4 \cr \gamma_3(74)&=3 \alpha_1 + 6 \alpha_2 + 7
\alpha_3 + 4 \alpha_4 \cr \gamma_3(75)&=3 \alpha_1 + 6 \alpha_2 + 9
\alpha_3 + 3 \alpha_4 \cr \gamma_3(76)&=3 \alpha_1 + 6 \alpha_2 + 9
\alpha_3 + 6 \alpha_4 \cr \gamma_3(77)&=3 \alpha_1 + 6 \alpha_2 + 10
\alpha_3 + 4 \alpha_4 \cr \gamma_3(78)&=3 \alpha_1 + 6 \alpha_2 + 10
\alpha_3 + 6 \alpha_4 \cr \gamma_3(79)&=3 \alpha_1 + 7 \alpha_2 + 9
\alpha_3 + 4 \alpha_4 \cr \gamma_3(80)&=3 \alpha_1 + 7 \alpha_2 + 9
\alpha_3 + 5 \alpha_4 \cr \gamma_3(81)&=3 \alpha_1 + 7 \alpha_2 + 10
\alpha_3 + 4 \alpha_4 \cr \gamma_3(82)&=3 \alpha_1 + 7 \alpha_2 + 10
\alpha_3 + 6 \alpha_4 \cr \gamma_3(83)&=3 \alpha_1 + 7 \alpha_2 + 11
\alpha_3 + 5 \alpha_4 \cr \gamma_3(84)&=3 \alpha_1 + 7 \alpha_2 + 11
\alpha_3 + 6 \alpha_4 \cr \gamma_3(85)&=4 \alpha_1 + 6 \alpha_2 + 8
\alpha_3 + 4 \alpha_4 \cr \gamma_3(86)&=4 \alpha_1 + 6 \alpha_2 + 9
\alpha_3 + 4 \alpha_4 \cr \gamma_3(87)&=4 \alpha_1 + 6 \alpha_2 + 9
\alpha_3 + 5 \alpha_4 \cr \gamma_3(88)&=4 \alpha_1 + 7 \alpha_2 + 9
\alpha_3 + 4 \alpha_4 \cr \gamma_3(89)&=4 \alpha_1 + 7 \alpha_2 + 9
\alpha_3 + 5 \alpha_4 \cr \gamma_3(90)&=4 \alpha_1 + 7 \alpha_2 + 10
\alpha_3 + 4 \alpha_4 \cr \hfill \cr } $  & \cr \noalign{\hrule} }}
$$

$$ \vbox{\tabskip=0pt \offinterlineskip
\halign to380pt{\strut#& \vrule#\tabskip=0em plus1em& \hfil#&
\vrule#& \hfil#\hfil& \vrule#\tabskip=0pt\cr \noalign{\hrule} &&
$\eqalign{ \hfill \cr \gamma_2(91)&=6 \alpha_1 + 10 \alpha_2 + 14
\alpha_3 + 6 \alpha_4 \cr \gamma_2(92)&=6 \alpha_1 + 10 \alpha_2 +
14 \alpha_3 + 8 \alpha_4 \cr \gamma_2(93)&=6 \alpha_1 + 11 \alpha_2
+ 14 \alpha_3 + 6 \alpha_4  \cr \gamma_2(94)&=6 \alpha_1 + 11
\alpha_2 + 14 \alpha_3 + 8 \alpha_4  \cr \gamma_2(95)&=6 \alpha_1 +
11 \alpha_2 + 16 \alpha_3 + 8 \alpha_4  \cr \gamma_2(96)&=6 \alpha_1
+ 12 \alpha_2 + 16 \alpha_3 + 8 \alpha_4 \cr \hfill \cr } $

&& $\eqalign{ \hfill \cr \gamma_3(91)&=4 \alpha_1 + 7 \alpha_2 + 10
\alpha_3 + 6 \alpha_4 \cr \gamma_3(92)&=4 \alpha_1 + 7 \alpha_2 + 11
\alpha_3 + 5 \alpha_4 \cr \gamma_3(93)&=4 \alpha_1 + 7 \alpha_2 + 11
\alpha_3 + 6 \alpha_4 \cr \gamma_3(94)&=4 \alpha_1 + 8 \alpha_2 + 11
\alpha_3 + 5 \alpha_4 \cr \gamma_3(95)&=4 \alpha_1 + 8 \alpha_2 + 11
\alpha_3 + 6 \alpha_4 \cr \gamma_3(96)&=4 \alpha_1 + 8 \alpha_2 + 12
\alpha_3 + 6 \alpha_4 \cr\hfill \cr } $  & \cr \noalign{\hrule} }}
$$
\centerline{Table-I}
\vskip 5mm

\noindent By the aid of Table-I, the sets $ \Gamma_{A} $ of (II.6)
are given, for $ A=1,2, \dots, 1152$, in the following Table-II :

$$ \vbox{\tabskip=0pt \offinterlineskip
\halign to380pt{\strut#& \vrule#\tabskip=0em plus1em& \hfil#&
\vrule#& \hfil#\hfil& \vrule#& \hfil#\hfil& \vrule#& \hfil#\hfil&
\vrule#\tabskip=0pt\cr \noalign{\hrule}

&& $\eqalign{ \hfill \cr \Gamma_{1}&= \{1,1,1,1 \}  \cr \Gamma_{2}&=
\{1,1,1,2 \}      \cr \Gamma_{3}&= \{1,1,2,1 \} \cr \Gamma_{4}&=
\{1,1,2,3 \}      \cr \Gamma_{5}&= \{1,1,3,2 \} \cr \Gamma_{6}&=
\{1,1,3,3 \}      \cr \Gamma_{7}&= \{1,2,1,1 \} \cr \Gamma_{8}&=
\{1,2,1,2 \}      \cr \Gamma_{9}&= \{1,2,4,1 \} \cr \Gamma_{10}&=
\{1,2,4,4 \}      \cr \Gamma_{11}&= \{1,2,5,2 \} \cr \Gamma_{12}&=
\{1,2,5,4 \}      \cr \Gamma_{13}&= \{1,3,2,1 \} \cr \Gamma_{14}&=
\{1,3,2,3 \}      \cr \Gamma_{15}&= \{1,3,6,1 \} \cr \Gamma_{16}&=
\{1,3,6,5 \}      \cr \Gamma_{17}&= \{1,3,8,3 \} \cr \Gamma_{18}&=
\{1,3,8,5 \}      \cr  \hfill \cr } $

&& $\eqalign{ \hfill \cr \Gamma_{19}&= \{1,4,3,2 \} \cr
\Gamma_{20}&= \{1,4,3,3 \}      \cr \Gamma_{21}&= \{1,4,7,2 \} \cr
\Gamma_{22}&= \{1,4,7,6 \}      \cr \Gamma_{23}&= \{1,4,9,3 \} \cr
\Gamma_{24}&= \{1,4,9,6 \}      \cr \Gamma_{25}&= \{1,5,4,1 \} \cr
\Gamma_{26}&= \{1,5,4,4 \}      \cr \Gamma_{27}&= \{1,5,6,1 \} \cr
\Gamma_{28}&= \{1,5,6,5 \}      \cr \Gamma_{29}&= \{1,5,10,4 \} \cr
\Gamma_{30}&= \{1,5,10,5 \}      \cr \Gamma_{31}&= \{1,6,5,2 \} \cr
\Gamma_{32}&= \{1,6,5,4 \}      \cr \Gamma_{33}&= \{1,6,7,2 \} \cr
\Gamma_{34}&= \{1,6,7,6 \}      \cr \Gamma_{35}&= \{1,6,11,4 \} \cr
\Gamma_{36}&= \{1,6,11,6 \}      \cr \hfill \cr } $

&& $\eqalign{ \hfill \cr  \Gamma_{37}&= \{1,7,8,3 \} \cr
\Gamma_{38}&= \{1,7,8,5 \} \cr \Gamma_{39}&= \{1,7,9,3 \} \cr
\Gamma_{40}&= \{1,7,9,6 \} \cr \Gamma_{41}&= \{1,7,12,5 \} \cr
\Gamma_{42}&= \{1,7,12,6 \} \cr \Gamma_{43}&= \{1,8,10,4 \} \cr
\Gamma_{44}&= \{1,8,10,5 \} \cr \Gamma_{45}&= \{1,8,11,4 \} \cr
\Gamma_{46}&= \{1,8,11,6 \} \cr \Gamma_{47}&= \{1,8,12,5 \} \cr
\Gamma_{48}&= \{1,8,12,6 \} \cr \Gamma_{49}&= \{2,1,1,1 \} \cr
\Gamma_{50}&= \{2,1,1,2 \} \cr \Gamma_{51}&= \{2,1,2,1 \} \cr
\Gamma_{52}&= \{2,1,2,3 \} \cr \Gamma_{53}&= \{2,1,3,2 \} \cr
\Gamma_{54}&= \{2,1,3,3 \} \cr \hfill \cr } $

&& $\eqalign{ \hfill \cr \Gamma_{55}&= \{2,9,1,1 \} \cr
\Gamma_{56}&= \{2,9,1,2 \} \cr \Gamma_{57}&= \{2,9,13,1 \} \cr
\Gamma_{58}&= \{2,9,13,7 \} \cr \Gamma_{59}&= \{2,9,14,2 \} \cr
\Gamma_{60}&= \{2,9,14,7 \} \cr \Gamma_{61}&= \{2,10,2,1 \} \cr
\Gamma_{62}&= \{2,10,2,3 \} \cr \Gamma_{63}&= \{2,10,15,1 \} \cr
\Gamma_{64}&= \{2,10,15,8 \} \cr \Gamma_{65}&= \{2,10,17,3 \} \cr
\Gamma_{66}&= \{2,10,17,8 \} \cr \Gamma_{67}&= \{2,11,3,2 \} \cr
\Gamma_{68}&= \{2,11,3,3 \} \cr \Gamma_{69}&= \{2,11,16,2 \} \cr
\Gamma_{70}&= \{2,11,16,9 \} \cr \Gamma_{71}&= \{2,11,18,3 \} \cr
\Gamma_{72}&= \{2,11,18,9 \} \cr \hfill \cr } $ & \cr
\noalign{\hrule} }}
$$

$$ \vbox{\tabskip=0pt \offinterlineskip
\halign to380pt{\strut#& \vrule#\tabskip=0em plus1em& \hfil#&
\vrule#& \hfil#\hfil& \vrule#& \hfil#\hfil& \vrule#& \hfil#\hfil&
\vrule#\tabskip=0pt\cr \noalign{\hrule}

&& $\eqalign{ \hfill \cr \Gamma_{73}&= \{2,21,13,1 \} \cr
\Gamma_{74}&= \{2,21,13,7 \} \cr \Gamma_{75}&= \{2,21,15,1 \} \cr
\Gamma_{76}&= \{2,21,15,8 \} \cr \Gamma_{77}&= \{2,21,37,7 \} \cr
\Gamma_{78}&= \{2,21,37,8 \} \cr \Gamma_{79}&= \{2,22,14,2 \} \cr
\Gamma_{80}&= \{2,22,14,7 \} \cr \Gamma_{81}&= \{2,22,16,2 \} \cr
\Gamma_{82}&= \{2,22,16,9 \} \cr \Gamma_{83}&= \{2,22,38,7 \} \cr
\Gamma_{84}&= \{2,22,38,9 \} \cr \Gamma_{85}&= \{2,23,17,3 \} \cr
\Gamma_{86}&= \{2,23,17,8 \} \cr \Gamma_{87}&= \{2,23,18,3 \} \cr
\Gamma_{88}&= \{2,23,18,9 \} \cr \Gamma_{89}&= \{2,23,39,8 \} \cr
\Gamma_{90}&= \{2,23,39,9 \} \cr \Gamma_{91}&= \{2,45,37,7 \} \cr
\Gamma_{92}&= \{2,45,37,8 \} \cr \Gamma_{93}&= \{2,45,38,7 \} \cr
\Gamma_{94}&= \{2,45,38,9 \} \cr \Gamma_{95}&= \{2,45,39,8 \} \cr
\Gamma_{96}&= \{2,45,39,9 \} \cr \Gamma_{97}&= \{3,2,1,1 \} \cr
\Gamma_{98}&= \{3,2,1,2 \} \cr \Gamma_{99}&= \{3,2,4,1 \} \cr
\Gamma_{100}&= \{3,2,4,4 \} \cr \Gamma_{101}&= \{3,2,5,2 \} \cr
\Gamma_{102}&= \{3,2,5,4 \} \cr \Gamma_{103}&= \{3,9,1,1 \} \cr
\Gamma_{104}&= \{3,9,1,2 \} \cr \Gamma_{105}&= \{3,9,13,1 \} \cr
\Gamma_{106}&= \{3,9,13,7 \} \cr \Gamma_{107}&= \{3,9,14,2 \} \cr
\Gamma_{108}&= \{3,9,14,7 \} \cr \Gamma_{109}&= \{3,12,4,1 \} \cr
\hfill \cr } $

&& $\eqalign{ \hfill \cr \Gamma_{110}&= \{3,12,4,4 \} \cr
\Gamma_{111}&= \{3,12,19,1 \} \cr \Gamma_{112}&= \{3,12,19,10 \} \cr
\Gamma_{113}&= \{3,12,25,4 \}      \cr \Gamma_{114}&= \{3,12,25,10
\}      \cr \Gamma_{115}&= \{3,13,5,2 \} \cr \Gamma_{116}&=
\{3,13,5,4 \} \cr \Gamma_{117}&= \{3,13,20,2 \} \cr \Gamma_{118}&=
\{3,13,20,11 \} \cr \Gamma_{119}&= \{3,13,26,4 \} \cr \Gamma_{120}&=
\{3,13,26,11 \}      \cr \Gamma_{121}&= \{3,24,13,1 \}      \cr
\Gamma_{122}&= \{3,24,13,7 \} \cr \Gamma_{123}&= \{3,24,19,1 \} \cr
\Gamma_{124}&= \{3,24,19,10 \} \cr \Gamma_{125}&= \{3,24,40,7 \} \cr
\Gamma_{126}&= \{3,24,40,10 \} \cr \Gamma_{127}&= \{3,25,14,2 \} \cr
\Gamma_{128}&= \{3,25,14,7 \} \cr \Gamma_{129}&= \{3,25,20,2 \} \cr
\Gamma_{130}&= \{3,25,20,11 \} \cr \Gamma_{131}&= \{3,25,41,7 \} \cr
\Gamma_{132}&= \{3,25,41,11 \} \cr \Gamma_{133}&= \{3,33,25,4 \} \cr
\Gamma_{134}&= \{3,33,25,10 \}      \cr \Gamma_{135}&= \{3,33,26,4
\}      \cr \Gamma_{136}&= \{3,33,26,11 \}      \cr \Gamma_{137}&=
\{3,33,46,10 \}      \cr \Gamma_{138}&= \{3,33,46,11 \}      \cr
\Gamma_{139}&= \{3,46,40,7 \}      \cr \Gamma_{140}&= \{3,46,40,10
\}      \cr \Gamma_{141}&= \{3,46,41,7 \}      \cr \Gamma_{142}&=
\{3,46,41,11 \}      \cr \Gamma_{143}&= \{3,46,46,10 \}      \cr
\Gamma_{144}&= \{3,46,46,11 \}      \cr \Gamma_{145}&= \{4,3,2,1 \}
\cr \Gamma_{146}&= \{4,3,2,3 \} \cr
\hfill \cr } $

&& $\eqalign{ \hfill \cr \Gamma_{147}&= \{4,3,6,1 \} \cr
\Gamma_{148}&= \{4,3,6,5 \} \cr \Gamma_{149}&= \{4,3,8,3 \}      \cr
\Gamma_{150}&= \{4,3,8,5 \} \cr \Gamma_{151}&= \{4,10,2,1 \} \cr
\Gamma_{152}&= \{4,10,2,3 \} \cr \Gamma_{153}&= \{4,10,15,1 \} \cr
\Gamma_{154}&= \{4,10,15,8 \} \cr \Gamma_{155}&= \{4,10,17,3 \} \cr
\Gamma_{156}&= \{4,10,17,8 \}      \cr \Gamma_{157}&= \{4,14,6,1 \}
\cr \Gamma_{158}&= \{4,14,6,5 \} \cr \Gamma_{159}&= \{4,14,21,1 \}
\cr \Gamma_{160}&= \{4,14,21,12 \}      \cr \Gamma_{161}&=
\{4,14,27,5 \}      \cr \Gamma_{162}&= \{4,14,27,12 \}      \cr
\Gamma_{163}&= \{4,16,8,3 \} \cr \Gamma_{164}&= \{4,16,8,5 \} \cr
\Gamma_{165}&= \{4,16,23,3 \}      \cr \Gamma_{166}&= \{4,16,23,14
\}      \cr \Gamma_{167}&= \{4,16,29,5 \}      \cr \Gamma_{168}&=
\{4,16,29,14 \}      \cr \Gamma_{169}&= \{4,26,15,1 \}      \cr
\Gamma_{170}&= \{4,26,15,8 \} \cr \Gamma_{171}&= \{4,26,21,1 \} \cr
\Gamma_{172}&= \{4,26,21,12 \}      \cr \Gamma_{173}&= \{4,26,42,8
\}      \cr \Gamma_{174}&= \{4,26,42,12 \}      \cr \Gamma_{175}&=
\{4,28,17,3 \}      \cr \Gamma_{176}&= \{4,28,17,8 \} \cr
\Gamma_{177}&= \{4,28,23,3 \}      \cr \Gamma_{178}&= \{4,28,23,14
\}      \cr \Gamma_{179}&= \{4,28,44,8 \}      \cr \Gamma_{180}&=
\{4,28,44,14 \}      \cr \Gamma_{181}&= \{4,34,27,5 \}      \cr
\Gamma_{182}&= \{4,34,27,12 \}      \cr \Gamma_{183}&= \{4,34,29,5
\} \cr \hfill \cr } $

&& $\eqalign{ \hfill \cr  \Gamma_{184}&= \{4,34,29,14 \}      \cr
\Gamma_{185}&= \{4,34,49,12 \}  \cr \Gamma_{186}&= \{4,34,49,14 \}
\cr \Gamma_{187}&= \{4,47,42,8 \}      \cr \Gamma_{188}&=
\{4,47,42,12 \}      \cr \Gamma_{189}&= \{4,47,44,8 \}      \cr
\Gamma_{190}&= \{4,47,44,14 \}      \cr \Gamma_{191}&= \{4,47,49,12
\}      \cr \Gamma_{192}&= \{4,47,49,14 \}      \cr \Gamma_{193}&=
\{5,4,3,2 \} \cr \Gamma_{194}&= \{5,4,3,3 \}      \cr \Gamma_{195}&=
\{5,4,7,2 \} \cr \Gamma_{196}&= \{5,4,7,6 \}      \cr \Gamma_{197}&=
\{5,4,9,3 \} \cr \Gamma_{198}&= \{5,4,9,6 \}      \cr \Gamma_{199}&=
\{5,11,3,2 \} \cr \Gamma_{200}&= \{5,11,3,3 \}      \cr
\Gamma_{201}&= \{5,11,16,2 \}      \cr \Gamma_{202}&= \{5,11,16,9 \}
\cr \Gamma_{203}&= \{5,11,18,3 \} \cr \Gamma_{204}&= \{5,11,18,9 \}
\cr \Gamma_{205}&= \{5,15,7,2 \} \cr \Gamma_{206}&= \{5,15,7,6 \}
\cr \Gamma_{207}&= \{5,15,22,2 \}      \cr \Gamma_{208}&=
\{5,15,22,13 \}      \cr \Gamma_{209}&= \{5,15,28,6 \}      \cr
\Gamma_{210}&= \{5,15,28,13 \}      \cr \Gamma_{211}&= \{5,17,9,3 \}
\cr \Gamma_{212}&= \{5,17,9,6 \} \cr \Gamma_{213}&= \{5,17,24,3 \}
\cr \Gamma_{214}&= \{5,17,24,15 \}      \cr \Gamma_{215}&=
\{5,17,30,6 \}      \cr \Gamma_{216}&= \{5,17,30,15 \}      \cr
\Gamma_{217}&= \{5,27,16,2 \}      \cr \Gamma_{218}&= \{5,27,16,9 \}
\cr \Gamma_{219}&= \{5,27,22,2 \} \cr \Gamma_{220}&= \{5,27,22,13 \}
\cr \hfill \cr } $ & \cr \noalign{\hrule} }}
$$

$$ \vbox{\tabskip=0pt \offinterlineskip
\halign to405pt{\strut#& \vrule#\tabskip=0em plus1em& \hfil#&
\vrule#& \hfil#\hfil& \vrule#& \hfil#\hfil& \vrule#& \hfil#\hfil&
\vrule#\tabskip=0pt\cr \noalign{\hrule}

&& $\eqalign{ \hfill \cr  \Gamma_{221}&= \{5,27,43,9 \}      \cr
\Gamma_{222}&= \{5,27,43,13 \} \cr \Gamma_{223}&= \{5,29,18,3 \} \cr
\Gamma_{224}&= \{5,29,18,9 \} \cr \Gamma_{225}&= \{5,29,24,3 \} \cr
\Gamma_{226}&= \{5,29,24,15 \} \cr \Gamma_{227}&= \{5,29,45,9 \} \cr
\Gamma_{228}&= \{5,29,45,15 \}      \cr \Gamma_{229}&= \{5,35,28,6
\}      \cr \Gamma_{230}&= \{5,35,28,13 \}      \cr \Gamma_{231}&=
\{5,35,30,6 \}      \cr \Gamma_{232}&= \{5,35,30,15 \}      \cr
\Gamma_{233}&= \{5,35,50,13 \}      \cr \Gamma_{234}&= \{5,35,50,15
\}      \cr \Gamma_{235}&= \{5,48,43,9 \}      \cr \Gamma_{236}&=
\{5,48,43,13 \}      \cr \Gamma_{237}&= \{5,48,45,9 \}      \cr
\Gamma_{238}&= \{5,48,45,15 \}      \cr \Gamma_{239}&= \{5,48,50,13
\}      \cr \Gamma_{240}&= \{5,48,50,15 \}      \cr \Gamma_{241}&=
\{6,5,4,1 \} \cr \Gamma_{242}&= \{6,5,4,4 \}      \cr \Gamma_{243}&=
\{6,5,6,1 \} \cr \Gamma_{244}&= \{6,5,6,5 \}      \cr \Gamma_{245}&=
\{6,5,10,4 \} \cr \Gamma_{246}&= \{6,5,10,5 \} \cr \Gamma_{247}&=
\{6,12,4,1 \} \cr \Gamma_{248}&= \{6,12,4,4 \} \cr \Gamma_{249}&=
\{6,12,19,1 \}      \cr \Gamma_{250}&= \{6,12,19,10 \}      \cr
\Gamma_{251}&= \{6,12,25,4 \}      \cr \Gamma_{252}&= \{6,12,25,10
\}      \cr \Gamma_{253}&= \{6,14,6,1 \} \cr \Gamma_{254}&=
\{6,14,6,5 \} \cr \Gamma_{255}&= \{6,14,21,1 \} \cr \Gamma_{256}&=
\{6,14,21,12 \} \cr \Gamma_{257}&= \{6,14,27,5 \} \cr \hfill \cr } $

&& $\eqalign{ \hfill \cr \Gamma_{258}&= \{6,14,27,12 \}      \cr
\Gamma_{259}&= \{6,18,10,4 \} \cr \Gamma_{260}&= \{6,18,10,5 \} \cr
\Gamma_{261}&= \{6,18,31,4 \} \cr \Gamma_{262}&= \{6,18,31,16 \} \cr
\Gamma_{263}&= \{6,18,33,5 \}      \cr \Gamma_{264}&= \{6,18,33,16
\}      \cr \Gamma_{265}&= \{6,30,19,1 \}      \cr \Gamma_{266}&=
\{6,30,19,10 \}      \cr \Gamma_{267}&= \{6,30,21,1 \}      \cr
\Gamma_{268}&= \{6,30,21,12 \}      \cr \Gamma_{269}&= \{6,30,47,10
\}      \cr \Gamma_{270}&= \{6,30,47,12 \}      \cr \Gamma_{271}&=
\{6,36,25,4 \}      \cr \Gamma_{272}&= \{6,36,25,10 \}      \cr
\Gamma_{273}&= \{6,36,31,4 \}      \cr \Gamma_{274}&= \{6,36,31,16
\}      \cr \Gamma_{275}&= \{6,36,52,10 \}      \cr \Gamma_{276}&=
\{6,36,52,16 \}      \cr \Gamma_{277}&= \{6,38,27,5 \}      \cr
\Gamma_{278}&= \{6,38,27,12 \}      \cr \Gamma_{279}&= \{6,38,33,5
\}      \cr \Gamma_{280}&= \{6,38,33,16 \}      \cr \Gamma_{281}&=
\{6,38,54,12 \}      \cr \Gamma_{282}&= \{6,38,54,16 \}      \cr
\Gamma_{283}&= \{6,49,47,10 \}      \cr \Gamma_{284}&= \{6,49,47,12
\}      \cr \Gamma_{285}&= \{6,49,52,10 \}      \cr \Gamma_{286}&=
\{6,49,52,16 \}      \cr \Gamma_{287}&= \{6,49,54,12 \}      \cr
\Gamma_{288}&= \{6,49,54,16 \}      \cr \Gamma_{289}&= \{7,6,5,2 \}
\cr \Gamma_{290}&= \{7,6,5,4 \} \cr \Gamma_{291}&= \{7,6,7,2 \}
\cr \Gamma_{292}&= \{7,6,7,6 \} \cr \Gamma_{293}&= \{7,6,11,4 \} \cr
\Gamma_{294}&= \{7,6,11,6 \} \cr \hfill \cr } $

&& $\eqalign{ \hfill \cr \Gamma_{295}&= \{7,13,5,2 \} \cr
\Gamma_{296}&= \{7,13,5,4 \} \cr \Gamma_{297}&= \{7,13,20,2 \} \cr
\Gamma_{298}&= \{7,13,20,11 \}      \cr \Gamma_{299}&= \{7,13,26,4
\}      \cr \Gamma_{300}&= \{7,13,26,11 \}      \cr \Gamma_{301}&=
\{7,15,7,2 \}      \cr \Gamma_{302}&= \{7,15,7,6 \} \cr
\Gamma_{303}&= \{7,15,22,2 \} \cr \Gamma_{304}&= \{7,15,22,13 \} \cr
\Gamma_{305}&= \{7,15,28,6 \}      \cr \Gamma_{306}&= \{7,15,28,13
\}      \cr \Gamma_{307}&= \{7,19,11,4 \}      \cr \Gamma_{308}&=
\{7,19,11,6 \} \cr \Gamma_{309}&= \{7,19,32,4 \} \cr \Gamma_{310}&=
\{7,19,32,17 \} \cr \Gamma_{311}&= \{7,19,34,6 \} \cr \Gamma_{312}&=
\{7,19,34,17 \}      \cr \Gamma_{313}&= \{7,31,20,2 \}      \cr
\Gamma_{314}&= \{7,31,20,11 \}      \cr \Gamma_{315}&= \{7,31,22,2
\}      \cr \Gamma_{316}&= \{7,31,22,13 \}      \cr \Gamma_{317}&=
\{7,31,48,11 \}      \cr \Gamma_{318}&= \{7,31,48,13 \}      \cr
\Gamma_{319}&= \{7,37,26,4 \}      \cr \Gamma_{320}&= \{7,37,26,11
\}      \cr \Gamma_{321}&= \{7,37,32,4 \}      \cr \Gamma_{322}&=
\{7,37,32,17 \}      \cr \Gamma_{323}&= \{7,37,53,11 \}      \cr
\Gamma_{324}&= \{7,37,53,17 \}      \cr \Gamma_{325}&= \{7,39,28,6
\}      \cr \Gamma_{326}&= \{7,39,28,13 \}      \cr \Gamma_{327}&=
\{7,39,34,6 \}      \cr \Gamma_{328}&= \{7,39,34,17 \}      \cr
\Gamma_{329}&= \{7,39,55,13 \}      \cr \Gamma_{330}&= \{7,39,55,17
\}      \cr \Gamma_{331}&= \{7,50,48,11 \} \cr \hfill \cr } $

&& $\eqalign{ \hfill \cr \Gamma_{332}&= \{7,50,48,13 \}      \cr
\Gamma_{333}&= \{7,50,53,11 \} \cr \Gamma_{334}&= \{7,50,53,17 \}
\cr \Gamma_{335}&= \{7,50,55,13 \}      \cr \Gamma_{336}&=
\{7,50,55,17 \}      \cr \Gamma_{337}&= \{8,7,8,3 \} \cr
\Gamma_{338}&= \{8,7,8,5 \}      \cr \Gamma_{339}&= \{8,7,9,3 \} \cr
\Gamma_{340}&= \{8,7,9,6 \}      \cr \Gamma_{341}&= \{8,7,12,5 \}
\cr \Gamma_{342}&= \{8,7,12,6 \}      \cr \Gamma_{343}&= \{8,16,8,3
\}      \cr \Gamma_{344}&= \{8,16,8,5 \}      \cr \Gamma_{345}&=
\{8,16,23,3 \} \cr \Gamma_{346}&= \{8,16,23,14 \}      \cr
\Gamma_{347}&= \{8,16,29,5 \}      \cr \Gamma_{348}&= \{8,16,29,14
\}      \cr \Gamma_{349}&= \{8,17,9,3 \}      \cr \Gamma_{350}&=
\{8,17,9,6 \} \cr \Gamma_{351}&= \{8,17,24,3 \} \cr \Gamma_{352}&=
\{8,17,24,15 \} \cr \Gamma_{353}&= \{8,17,30,6 \}      \cr
\Gamma_{354}&= \{8,17,30,15 \}      \cr \Gamma_{355}&= \{8,20,12,5
\}      \cr \Gamma_{356}&= \{8,20,12,6 \} \cr \Gamma_{357}&=
\{8,20,35,5 \} \cr \Gamma_{358}&= \{8,20,35,18 \}      \cr
\Gamma_{359}&= \{8,20,36,6 \}      \cr \Gamma_{360}&= \{8,20,36,18
\}      \cr \Gamma_{361}&= \{8,32,23,3 \}      \cr \Gamma_{362}&=
\{8,32,23,14 \}      \cr \Gamma_{363}&= \{8,32,24,3 \}      \cr
\Gamma_{364}&= \{8,32,24,15 \}      \cr \Gamma_{365}&= \{8,32,51,14
\}      \cr \Gamma_{366}&= \{8,32,51,15 \}      \cr \Gamma_{367}&=
\{8,40,29,5 \}      \cr \Gamma_{368}&= \{8,40,29,14 \}      \cr
\hfill \cr } $  & \cr \noalign{\hrule} }}
$$

$$ \vbox{\tabskip=0pt \offinterlineskip
\halign to405pt{\strut#& \vrule#\tabskip=0em plus1em& \hfil#&
\vrule#& \hfil#\hfil& \vrule#& \hfil#\hfil& \vrule#& \hfil#\hfil&
\vrule#\tabskip=0pt\cr \noalign{\hrule}

&& $\eqalign{ \hfill \cr \Gamma_{369}&= \{8,40,35,5 \}      \cr
\Gamma_{370}&= \{8,40,35,18 \} \cr \Gamma_{371}&= \{8,40,56,14 \}
\cr \Gamma_{372}&= \{8,40,56,18 \}      \cr \Gamma_{373}&=
\{8,41,30,6 \}      \cr \Gamma_{374}&= \{8,41,30,15 \}      \cr
\Gamma_{375}&= \{8,41,36,6 \}      \cr \Gamma_{376}&= \{8,41,36,18
\}      \cr \Gamma_{377}&= \{8,41,57,15 \}      \cr \Gamma_{378}&=
\{8,41,57,18 \}      \cr \Gamma_{379}&= \{8,51,51,14 \}      \cr
\Gamma_{380}&= \{8,51,51,15 \}      \cr \Gamma_{381}&= \{8,51,56,14
\}      \cr \Gamma_{382}&= \{8,51,56,18 \}      \cr \Gamma_{383}&=
\{8,51,57,15 \}      \cr \Gamma_{384}&= \{8,51,57,18 \}      \cr
\Gamma_{385}&= \{9,8,10,4 \} \cr \Gamma_{386}&= \{9,8,10,5 \} \cr
\Gamma_{387}&= \{9,8,11,4 \}      \cr \Gamma_{388}&= \{9,8,11,6 \}
\cr \Gamma_{389}&= \{9,8,12,5 \}      \cr \Gamma_{390}&= \{9,8,12,6
\}      \cr \Gamma_{391}&= \{9,18,10,4 \}      \cr \Gamma_{392}&=
\{9,18,10,5 \} \cr \Gamma_{393}&= \{9,18,31,4 \} \cr \Gamma_{394}&=
\{9,18,31,16 \}      \cr \Gamma_{395}&= \{9,18,33,5 \}      \cr
\Gamma_{396}&= \{9,18,33,16 \}      \cr \Gamma_{397}&= \{9,19,11,4
\}      \cr \Gamma_{398}&= \{9,19,11,6 \} \cr \Gamma_{399}&=
\{9,19,32,4 \}      \cr \Gamma_{400}&= \{9,19,32,17 \}      \cr
\Gamma_{401}&= \{9,19,34,6 \}      \cr \Gamma_{402}&= \{9,19,34,17
\}      \cr \Gamma_{403}&= \{9,20,12,5 \}      \cr \Gamma_{404}&=
\{9,20,12,6 \}      \cr \Gamma_{405}&= \{9,20,35,5 \} \cr \hfill \cr
} $

&& $\eqalign{ \hfill \cr \Gamma_{406}&= \{9,20,35,18 \}      \cr
\Gamma_{407}&= \{9,20,36,6 \} \cr \Gamma_{408}&= \{9,20,36,18 \} \cr
\Gamma_{409}&= \{9,42,31,4 \}      \cr \Gamma_{410}&= \{9,42,31,16
\}      \cr \Gamma_{411}&= \{9,42,32,4 \}      \cr \Gamma_{412}&=
\{9,42,32,17 \}      \cr \Gamma_{413}&= \{9,42,58,16 \}      \cr
\Gamma_{414}&= \{9,42,58,17 \}      \cr \Gamma_{415}&= \{9,43,33,5
\}      \cr \Gamma_{416}&= \{9,43,33,16 \}      \cr \Gamma_{417}&=
\{9,43,35,5 \}      \cr \Gamma_{418}&= \{9,43,35,18 \}      \cr
\Gamma_{419}&= \{9,43,59,16 \}      \cr \Gamma_{420}&= \{9,43,59,18
\}      \cr \Gamma_{421}&= \{9,44,34,6 \}      \cr \Gamma_{422}&=
\{9,44,34,17 \}      \cr \Gamma_{423}&= \{9,44,36,6 \}      \cr
\Gamma_{424}&= \{9,44,36,18 \}      \cr \Gamma_{425}&= \{9,44,60,17
\}      \cr \Gamma_{426}&= \{9,44,60,18 \}      \cr \Gamma_{427}&=
\{9,52,58,16 \}      \cr \Gamma_{428}&= \{9,52,58,17 \}      \cr
\Gamma_{429}&= \{9,52,59,16 \}      \cr \Gamma_{430}&= \{9,52,59,18
\}      \cr \Gamma_{431}&= \{9,52,60,17 \}      \cr \Gamma_{432}&=
\{9,52,60,18 \}      \cr \Gamma_{433}&= \{10,21,13,1 \}      \cr
\Gamma_{434}&= \{10,21,13,7 \}      \cr \Gamma_{435}&= \{10,21,15,1
\}      \cr \Gamma_{436}&= \{10,21,15,8 \}      \cr \Gamma_{437}&=
\{10,21,37,7 \}      \cr \Gamma_{438}&= \{10,21,37,8 \}      \cr
\Gamma_{439}&= \{10,24,13,1 \}      \cr \Gamma_{440}&= \{10,24,13,7
\}      \cr \Gamma_{441}&= \{10,24,19,1 \}      \cr \Gamma_{442}&=
\{10,24,19,10 \}  \cr \hfill \cr } $

&& $\eqalign{ \hfill \cr  \Gamma_{443}&= \{10,24,40,7 \}      \cr
\Gamma_{444}&= \{10,24,40,10 \} \cr \Gamma_{445}&= \{10,26,15,1 \}
\cr \Gamma_{446}&= \{10,26,15,8 \}      \cr \Gamma_{447}&=
\{10,26,21,1 \}      \cr \Gamma_{448}&= \{10,26,21,12 \}      \cr
\Gamma_{449}&= \{10,26,42,8 \}      \cr \Gamma_{450}&= \{10,26,42,12
\}      \cr \Gamma_{451}&= \{10,30,19,1 \}      \cr \Gamma_{452}&=
\{10,30,19,10 \}      \cr \Gamma_{453}&= \{10,30,21,1 \}      \cr
\Gamma_{454}&= \{10,30,21,12 \}      \cr \Gamma_{455}&=
\{10,30,47,10 \}      \cr \Gamma_{456}&= \{10,30,47,12 \}      \cr
\Gamma_{457}&= \{10,53,37,7 \}      \cr \Gamma_{458}&= \{10,53,37,8
\}      \cr \Gamma_{459}&= \{10,53,61,7 \}      \cr \Gamma_{460}&=
\{10,53,61,19 \}      \cr \Gamma_{461}&= \{10,53,63,8 \}      \cr
\Gamma_{462}&= \{10,53,63,19 \}      \cr \Gamma_{463}&= \{10,56,40,7
\}      \cr \Gamma_{464}&= \{10,56,40,10 \}      \cr \Gamma_{465}&=
\{10,56,61,7 \}      \cr \Gamma_{466}&= \{10,56,61,19 \}      \cr
\Gamma_{467}&= \{10,56,67,10 \}      \cr \Gamma_{468}&=
\{10,56,67,19 \}      \cr \Gamma_{469}&= \{10,58,42,8 \}      \cr
\Gamma_{470}&= \{10,58,42,12 \}      \cr \Gamma_{471}&= \{10,58,63,8
\}      \cr \Gamma_{472}&= \{10,58,63,19 \}      \cr \Gamma_{473}&=
\{10,58,69,12 \}      \cr \Gamma_{474}&= \{10,58,69,19 \}      \cr
\Gamma_{475}&= \{10,62,47,10 \}      \cr \Gamma_{476}&=
\{10,62,47,12 \}      \cr \Gamma_{477}&= \{10,62,67,10 \}      \cr
\Gamma_{478}&= \{10,62,67,19 \}      \cr \Gamma_{479}&=
\{10,62,69,12 \}      \cr \hfill \cr } $

&& $\eqalign{ \hfill \cr  \Gamma_{480}&= \{10,62,69,19 \}      \cr
\Gamma_{481}&= \{11,22,14,2 \} \cr \Gamma_{482}&= \{11,22,14,7 \}
\cr \Gamma_{483}&= \{11,22,16,2 \}      \cr \Gamma_{484}&=
\{11,22,16,9 \}      \cr \Gamma_{485}&= \{11,22,38,7 \}      \cr
\Gamma_{486}&= \{11,22,38,9 \}      \cr \Gamma_{487}&= \{11,25,14,2
\}      \cr \Gamma_{488}&= \{11,25,14,7 \}      \cr \Gamma_{489}&=
\{11,25,20,2 \}      \cr \Gamma_{490}&= \{11,25,20,11 \}      \cr
\Gamma_{491}&= \{11,25,41,7 \}      \cr \Gamma_{492}&= \{11,25,41,11
\}      \cr \Gamma_{493}&= \{11,27,16,2 \}      \cr \Gamma_{494}&=
\{11,27,16,9 \}      \cr \Gamma_{495}&= \{11,27,22,2 \}      \cr
\Gamma_{496}&= \{11,27,22,13 \}      \cr \Gamma_{497}&= \{11,27,43,9
\}      \cr \Gamma_{498}&= \{11,27,43,13 \}      \cr \Gamma_{499}&=
\{11,31,20,2 \}      \cr \Gamma_{500}&= \{11,31,20,11 \}      \cr
\Gamma_{501}&= \{11,31,22,2 \}      \cr \Gamma_{502}&= \{11,31,22,13
\}      \cr \Gamma_{503}&= \{11,31,48,11 \}      \cr \Gamma_{504}&=
\{11,31,48,13 \}      \cr \Gamma_{505}&= \{11,54,38,7 \}      \cr
\Gamma_{506}&= \{11,54,38,9 \}      \cr \Gamma_{507}&= \{11,54,62,7
\}      \cr \Gamma_{508}&= \{11,54,62,20 \}      \cr \Gamma_{509}&=
\{11,54,64,9 \}      \cr \Gamma_{510}&= \{11,54,64,20 \}      \cr
\Gamma_{511}&= \{11,57,41,7 \}      \cr \Gamma_{512}&= \{11,57,41,11
\}      \cr \Gamma_{513}&= \{11,57,62,7 \}      \cr \Gamma_{514}&=
\{11,57,62,20 \}      \cr \Gamma_{515}&= \{11,57,68,11 \}      \cr
\Gamma_{516}&= \{11,57,68,20 \}      \cr \hfill \cr } $  & \cr
\noalign{\hrule} }}
$$

$$ \vbox{\tabskip=0pt \offinterlineskip
\halign to410pt{\strut#& \vrule#\tabskip=0em plus1em& \hfil#&
\vrule#& \hfil#\hfil& \vrule#& \hfil#\hfil& \vrule#& \hfil#\hfil&
\vrule#\tabskip=0pt\cr \noalign{\hrule}

&& $\eqalign{ \hfill \cr \Gamma_{517}&= \{11,59,43,9 \}      \cr
\Gamma_{518}&= \{11,59,43,13 \} \cr \Gamma_{519}&= \{11,59,64,9 \}
\cr \Gamma_{520}&= \{11,59,64,20 \}      \cr \Gamma_{521}&=
\{11,59,70,13 \}      \cr \Gamma_{522}&= \{11,59,70,20 \}      \cr
\Gamma_{523}&= \{11,63,48,11 \}      \cr \Gamma_{524}&=
\{11,63,48,13 \}      \cr \Gamma_{525}&= \{11,63,68,11 \}      \cr
\Gamma_{526}&= \{11,63,68,20 \}      \cr \Gamma_{527}&=
\{11,63,70,13 \}      \cr \Gamma_{528}&= \{11,63,70,20 \}      \cr
\Gamma_{529}&= \{12,23,17,3 \}      \cr \Gamma_{530}&= \{12,23,17,8
\}      \cr \Gamma_{531}&= \{12,23,18,3 \}      \cr \Gamma_{532}&=
\{12,23,18,9 \}      \cr \Gamma_{533}&= \{12,23,39,8 \}      \cr
\Gamma_{534}&= \{12,23,39,9 \}      \cr \Gamma_{535}&= \{12,28,17,3
\}      \cr \Gamma_{536}&= \{12,28,17,8 \}      \cr \Gamma_{537}&=
\{12,28,23,3 \}      \cr \Gamma_{538}&= \{12,28,23,14 \}      \cr
\Gamma_{539}&= \{12,28,44,8 \}      \cr \Gamma_{540}&= \{12,28,44,14
\}      \cr \Gamma_{541}&= \{12,29,18,3 \}      \cr \Gamma_{542}&=
\{12,29,18,9 \}      \cr \Gamma_{543}&= \{12,29,24,3 \}      \cr
\Gamma_{544}&= \{12,29,24,15 \}      \cr \Gamma_{545}&= \{12,29,45,9
\}      \cr \Gamma_{546}&= \{12,29,45,15 \}      \cr \Gamma_{547}&=
\{12,32,23,3 \}      \cr \Gamma_{548}&= \{12,32,23,14 \}      \cr
\Gamma_{549}&= \{12,32,24,3 \}      \cr \Gamma_{550}&= \{12,32,24,15
\}      \cr \Gamma_{551}&= \{12,32,51,14 \}      \cr \Gamma_{552}&=
\{12,32,51,15 \}      \cr \Gamma_{553}&= \{12,55,39,8 \} \cr \hfill
\cr } $

&& $\eqalign{ \hfill \cr  \Gamma_{554}&= \{12,55,39,9 \}      \cr
\Gamma_{555}&= \{12,55,65,8 \}      \cr  \Gamma_{556}&=
\{12,55,65,21 \}      \cr \Gamma_{557}&= \{12,55,66,9 \}      \cr
\Gamma_{558}&= \{12,55,66,21 \}      \cr \Gamma_{559}&= \{12,60,44,8
\}      \cr \Gamma_{560}&= \{12,60,44,14 \}      \cr \Gamma_{561}&=
\{12,60,65,8 \}      \cr \Gamma_{562}&= \{12,60,65,21 \}      \cr
\Gamma_{563}&= \{12,60,71,14 \}      \cr \Gamma_{564}&=
\{12,60,71,21 \}      \cr \Gamma_{565}&= \{12,61,45,9 \}      \cr
\Gamma_{566}&= \{12,61,45,15 \}      \cr \Gamma_{567}&= \{12,61,66,9
\}      \cr \Gamma_{568}&= \{12,61,66,21 \}      \cr \Gamma_{569}&=
\{12,61,72,15 \}      \cr \Gamma_{570}&= \{12,61,72,21 \}      \cr
\Gamma_{571}&= \{12,64,51,14 \}      \cr \Gamma_{572}&=
\{12,64,51,15 \}      \cr \Gamma_{573}&= \{12,64,71,14 \}      \cr
\Gamma_{574}&= \{12,64,71,21 \}      \cr \Gamma_{575}&=
\{12,64,72,15 \}      \cr \Gamma_{576}&= \{12,64,72,21 \}      \cr
\Gamma_{577}&= \{13,33,25,4 \}      \cr \Gamma_{578}&= \{13,33,25,10
\}      \cr \Gamma_{579}&= \{13,33,26,4 \}      \cr \Gamma_{580}&=
\{13,33,26,11 \}      \cr \Gamma_{581}&= \{13,33,46,10 \}      \cr
\Gamma_{582}&= \{13,33,46,11 \}      \cr \Gamma_{583}&= \{13,36,25,4
\}      \cr \Gamma_{584}&= \{13,36,25,10 \}      \cr \Gamma_{585}&=
\{13,36,31,4 \}      \cr \Gamma_{586}&= \{13,36,31,16 \}      \cr
\Gamma_{587}&= \{13,36,52,10 \}      \cr \Gamma_{588}&=
\{13,36,52,16 \}      \cr \Gamma_{589}&= \{13,37,26,4 \}      \cr
\Gamma_{590}&= \{13,37,26,11 \}   \cr \hfill \cr } $

&& $\eqalign{ \hfill \cr \Gamma_{591}&= \{13,37,32,4 \}      \cr
\Gamma_{592}&= \{13,37,32,17 \} \cr \Gamma_{593}&= \{13,37,53,11 \}
\cr \Gamma_{594}&= \{13,37,53,17 \}      \cr \Gamma_{595}&=
\{13,42,31,4 \}      \cr \Gamma_{596}&= \{13,42,31,16 \}      \cr
\Gamma_{597}&= \{13,42,32,4 \}      \cr \Gamma_{598}&= \{13,42,32,17
\}      \cr \Gamma_{599}&= \{13,42,58,16 \}      \cr \Gamma_{600}&=
\{13,42,58,17 \}      \cr \Gamma_{601}&= \{13,65,46,10 \}      \cr
\Gamma_{602}&= \{13,65,46,11 \}      \cr \Gamma_{603}&=
\{13,65,73,10 \}      \cr \Gamma_{604}&= \{13,65,73,22 \}      \cr
\Gamma_{605}&= \{13,65,74,11 \}      \cr \Gamma_{606}&=
\{13,65,74,22 \}      \cr \Gamma_{607}&= \{13,68,52,10 \}      \cr
\Gamma_{608}&= \{13,68,52,16 \}      \cr \Gamma_{609}&=
\{13,68,73,10 \}      \cr \Gamma_{610}&= \{13,68,73,22 \}      \cr
\Gamma_{611}&= \{13,68,79,16 \}      \cr \Gamma_{612}&=
\{13,68,79,22 \}      \cr \Gamma_{613}&= \{13,69,53,11 \}      \cr
\Gamma_{614}&= \{13,69,53,17 \}      \cr \Gamma_{615}&=
\{13,69,74,11 \}      \cr \Gamma_{616}&= \{13,69,74,22 \}      \cr
\Gamma_{617}&= \{13,69,80,17 \}      \cr \Gamma_{618}&=
\{13,69,80,22 \}      \cr \Gamma_{619}&= \{13,74,58,16 \}      \cr
\Gamma_{620}&= \{13,74,58,17 \}      \cr \Gamma_{621}&=
\{13,74,79,16 \}      \cr \Gamma_{622}&= \{13,74,79,22 \}      \cr
\Gamma_{623}&= \{13,74,80,17 \}      \cr \Gamma_{624}&=
\{13,74,80,22 \}      \cr \Gamma_{625}&= \{14,34,27,5 \}      \cr
\Gamma_{626}&= \{14,34,27,12 \}      \cr \Gamma_{627}&= \{14,34,29,5
\}      \cr \hfill \cr } $

&& $\eqalign{ \hfill \cr \Gamma_{628}&= \{14,34,29,14 \}      \cr
\Gamma_{629}&= \{14,34,49,12 \} \cr \Gamma_{630}&= \{14,34,49,14 \}
\cr \Gamma_{631}&= \{14,38,27,5 \}      \cr \Gamma_{632}&=
\{14,38,27,12 \}      \cr \Gamma_{633}&= \{14,38,33,5 \}      \cr
\Gamma_{634}&= \{14,38,33,16 \}      \cr \Gamma_{635}&=
\{14,38,54,12 \}      \cr \Gamma_{636}&= \{14,38,54,16 \}      \cr
\Gamma_{637}&= \{14,40,29,5 \}      \cr \Gamma_{638}&= \{14,40,29,14
\}      \cr \Gamma_{639}&= \{14,40,35,5 \}      \cr \Gamma_{640}&=
\{14,40,35,18 \}      \cr \Gamma_{641}&= \{14,40,56,14 \}      \cr
\Gamma_{642}&= \{14,40,56,18 \}      \cr \Gamma_{643}&= \{14,43,33,5
\}      \cr \Gamma_{644}&= \{14,43,33,16 \}      \cr \Gamma_{645}&=
\{14,43,35,5 \}      \cr \Gamma_{646}&= \{14,43,35,18 \}      \cr
\Gamma_{647}&= \{14,43,59,16 \}      \cr \Gamma_{648}&=
\{14,43,59,18 \}      \cr \Gamma_{649}&= \{14,66,49,12 \}      \cr
\Gamma_{650}&= \{14,66,49,14 \}      \cr \Gamma_{651}&=
\{14,66,75,12 \}      \cr \Gamma_{652}&= \{14,66,75,23 \}      \cr
\Gamma_{653}&= \{14,66,77,14 \}      \cr \Gamma_{654}&=
\{14,66,77,23 \}      \cr \Gamma_{655}&= \{14,70,54,12 \}      \cr
\Gamma_{656}&= \{14,70,54,16 \}      \cr \Gamma_{657}&=
\{14,70,75,12 \}      \cr \Gamma_{658}&= \{14,70,75,23 \}      \cr
\Gamma_{659}&= \{14,70,81,16 \}      \cr \Gamma_{660}&=
\{14,70,81,23 \}      \cr \Gamma_{661}&= \{14,72,56,14 \}      \cr
\Gamma_{662}&= \{14,72,56,18 \}      \cr \Gamma_{663}&=
\{14,72,77,14 \}      \cr \Gamma_{664}&= \{14,72,77,23 \}      \cr
 \hfill \cr } $ & \cr \noalign{\hrule} }}
$$

$$ \vbox{\tabskip=0pt \offinterlineskip
\halign to410pt{\strut#& \vrule#\tabskip=0em plus1em& \hfil#&
\vrule#& \hfil#\hfil& \vrule#& \hfil#\hfil& \vrule#& \hfil#\hfil&
\vrule#\tabskip=0pt\cr \noalign{\hrule}

&& $\eqalign{ \hfill \cr \Gamma_{665}&= \{14,72,83,18 \}      \cr
\Gamma_{666}&= \{14,72,83,23 \} \cr \Gamma_{667}&= \{14,75,59,16 \}
\cr \Gamma_{668}&= \{14,75,59,18 \}      \cr \Gamma_{669}&=
\{14,75,81,16 \}      \cr \Gamma_{670}&= \{14,75,81,23 \}      \cr
\Gamma_{671}&= \{14,75,83,18 \}      \cr \Gamma_{672}&=
\{14,75,83,23 \}      \cr \Gamma_{673}&= \{15,35,28,6 \}      \cr
\Gamma_{674}&= \{15,35,28,13 \}      \cr \Gamma_{675}&= \{15,35,30,6
\}      \cr \Gamma_{676}&= \{15,35,30,15 \}      \cr \Gamma_{677}&=
\{15,35,50,13 \}      \cr \Gamma_{678}&= \{15,35,50,15 \}      \cr
\Gamma_{679}&= \{15,39,28,6 \}      \cr \Gamma_{680}&= \{15,39,28,13
\}      \cr \Gamma_{681}&= \{15,39,34,6 \}      \cr \Gamma_{682}&=
\{15,39,34,17 \}      \cr \Gamma_{683}&= \{15,39,55,13 \}      \cr
\Gamma_{684}&= \{15,39,55,17 \}      \cr \Gamma_{685}&= \{15,41,30,6
\}      \cr \Gamma_{686}&= \{15,41,30,15 \}      \cr \Gamma_{687}&=
\{15,41,36,6 \}      \cr \Gamma_{688}&= \{15,41,36,18 \}      \cr
\Gamma_{689}&= \{15,41,57,15 \}      \cr \Gamma_{690}&=
\{15,41,57,18 \}      \cr \Gamma_{691}&= \{15,44,34,6 \}      \cr
\Gamma_{692}&= \{15,44,34,17 \}      \cr \Gamma_{693}&= \{15,44,36,6
\}      \cr \Gamma_{694}&= \{15,44,36,18 \}      \cr \Gamma_{695}&=
\{15,44,60,17 \}      \cr \Gamma_{696}&= \{15,44,60,18 \}      \cr
\Gamma_{697}&= \{15,67,50,13 \}      \cr \Gamma_{698}&=
\{15,67,50,15 \}      \cr \Gamma_{699}&= \{15,67,76,13 \}      \cr
\Gamma_{700}&= \{15,67,76,24 \}      \cr \Gamma_{701}&=
\{15,67,78,15 \} \cr \hfill \cr } $

&& $\eqalign{ \hfill \cr  \Gamma_{702}&= \{15,67,78,24 \}      \cr
\Gamma_{703}&= \{15,71,55,13 \}  \cr \Gamma_{704}&= \{15,71,55,17 \}
\cr \Gamma_{705}&= \{15,71,76,13 \}      \cr \Gamma_{706}&=
\{15,71,76,24 \}      \cr \Gamma_{707}&= \{15,71,82,17 \}      \cr
\Gamma_{708}&= \{15,71,82,24 \}      \cr \Gamma_{709}&=
\{15,73,57,15 \}      \cr \Gamma_{710}&= \{15,73,57,18 \}      \cr
\Gamma_{711}&= \{15,73,78,15 \}      \cr \Gamma_{712}&=
\{15,73,78,24 \}      \cr \Gamma_{713}&= \{15,73,84,18 \}      \cr
\Gamma_{714}&= \{15,73,84,24 \}      \cr \Gamma_{715}&=
\{15,76,60,17 \}      \cr \Gamma_{716}&= \{15,76,60,18 \}      \cr
\Gamma_{717}&= \{15,76,82,17 \}      \cr \Gamma_{718}&=
\{15,76,82,24 \}      \cr \Gamma_{719}&= \{15,76,84,18 \}      \cr
\Gamma_{720}&= \{15,76,84,24 \}      \cr \Gamma_{721}&= \{16,45,37,7
\}      \cr \Gamma_{722}&= \{16,45,37,8 \}      \cr \Gamma_{723}&=
\{16,45,38,7 \}      \cr \Gamma_{724}&= \{16,45,38,9 \}      \cr
\Gamma_{725}&= \{16,45,39,8 \}      \cr \Gamma_{726}&= \{16,45,39,9
\}      \cr \Gamma_{727}&= \{16,53,37,7 \}      \cr \Gamma_{728}&=
\{16,53,37,8 \}      \cr \Gamma_{729}&= \{16,53,61,7 \}      \cr
\Gamma_{730}&= \{16,53,61,19 \}      \cr \Gamma_{731}&= \{16,53,63,8
\}      \cr \Gamma_{732}&= \{16,53,63,19 \}      \cr \Gamma_{733}&=
\{16,54,38,7 \}      \cr \Gamma_{734}&= \{16,54,38,9 \}      \cr
\Gamma_{735}&= \{16,54,62,7 \}      \cr \Gamma_{736}&= \{16,54,62,20
\}      \cr \Gamma_{737}&= \{16,54,64,9 \}      \cr \Gamma_{738}&=
\{16,54,64,20 \}      \cr \hfill \cr } $

&& $\eqalign{ \hfill \cr  \Gamma_{739}&= \{16,55,39,8 \}      \cr
\Gamma_{740}&= \{16,55,39,9 \}  \cr  \Gamma_{741}&= \{16,55,65,8 \}
\cr \Gamma_{742}&= \{16,55,65,21 \}      \cr \Gamma_{743}&=
\{16,55,66,9 \}      \cr \Gamma_{744}&= \{16,55,66,21 \}      \cr
\Gamma_{745}&= \{16,77,61,7 \}      \cr \Gamma_{746}&= \{16,77,61,19
\}      \cr \Gamma_{747}&= \{16,77,62,7 \}      \cr \Gamma_{748}&=
\{16,77,62,20 \}      \cr \Gamma_{749}&= \{16,77,85,19 \}      \cr
\Gamma_{750}&= \{16,77,85,20 \}      \cr \Gamma_{751}&= \{16,78,63,8
\}      \cr \Gamma_{752}&= \{16,78,63,19 \}      \cr \Gamma_{753}&=
\{16,78,65,8 \}      \cr \Gamma_{754}&= \{16,78,65,21 \}      \cr
\Gamma_{755}&= \{16,78,86,19 \}      \cr \Gamma_{756}&=
\{16,78,86,21 \}      \cr \Gamma_{757}&= \{16,79,64,9 \}      \cr
\Gamma_{758}&= \{16,79,64,20 \}      \cr \Gamma_{759}&= \{16,79,66,9
\}      \cr \Gamma_{760}&= \{16,79,66,21 \}      \cr \Gamma_{761}&=
\{16,79,87,20 \}      \cr \Gamma_{762}&= \{16,79,87,21 \}      \cr
\Gamma_{763}&= \{16,89,85,19 \}      \cr \Gamma_{764}&=
\{16,89,85,20 \}      \cr \Gamma_{765}&= \{16,89,86,19 \}      \cr
\Gamma_{766}&= \{16,89,86,21 \}      \cr \Gamma_{767}&=
\{16,89,87,20 \}      \cr \Gamma_{768}&= \{16,89,87,21 \}      \cr
\Gamma_{769}&= \{17,46,40,7 \}      \cr \Gamma_{770}&= \{17,46,40,10
\}      \cr \Gamma_{771}&= \{17,46,41,7 \}      \cr \Gamma_{772}&=
\{17,46,41,11 \}      \cr \Gamma_{773}&= \{17,46,46,10 \}      \cr
\Gamma_{774}&= \{17,46,46,11 \}      \cr \Gamma_{775}&= \{17,56,40,7
\}      \cr \hfill \cr } $

&& $\eqalign{ \hfill \cr  \Gamma_{776}&= \{17,56,40,10 \}      \cr
\Gamma_{777}&= \{17,56,61,7 \}  \cr \Gamma_{778}&= \{17,56,61,19 \}
\cr \Gamma_{779}&= \{17,56,67,10 \}      \cr \Gamma_{780}&=
\{17,56,67,19 \}      \cr \Gamma_{781}&= \{17,57,41,7 \}      \cr
\Gamma_{782}&= \{17,57,41,11 \}      \cr \Gamma_{783}&= \{17,57,62,7
\}      \cr \Gamma_{784}&= \{17,57,62,20 \}      \cr \Gamma_{785}&=
\{17,57,68,11 \}      \cr \Gamma_{786}&= \{17,57,68,20 \}      \cr
\Gamma_{787}&= \{17,65,46,10 \}      \cr \Gamma_{788}&=
\{17,65,46,11 \}      \cr \Gamma_{789}&= \{17,65,73,10 \}      \cr
\Gamma_{790}&= \{17,65,73,22 \}      \cr \Gamma_{791}&=
\{17,65,74,11 \}      \cr \Gamma_{792}&= \{17,65,74,22 \}      \cr
\Gamma_{793}&= \{17,77,61,7 \}      \cr \Gamma_{794}&= \{17,77,61,19
\}      \cr \Gamma_{795}&= \{17,77,62,7 \}      \cr \Gamma_{796}&=
\{17,77,62,20 \}      \cr \Gamma_{797}&= \{17,77,85,19 \}      \cr
\Gamma_{798}&= \{17,77,85,20 \}      \cr \Gamma_{799}&=
\{17,80,67,10 \}      \cr \Gamma_{800}&= \{17,80,67,19 \}      \cr
\Gamma_{801}&= \{17,80,73,10 \}      \cr \Gamma_{802}&=
\{17,80,73,22 \}      \cr \Gamma_{803}&= \{17,80,88,19 \}      \cr
\Gamma_{804}&= \{17,80,88,22 \}      \cr \Gamma_{805}&=
\{17,81,68,11 \}      \cr \Gamma_{806}&= \{17,81,68,20 \}      \cr
\Gamma_{807}&= \{17,81,74,11 \}      \cr \Gamma_{808}&=
\{17,81,74,22 \}      \cr \Gamma_{809}&= \{17,81,89,20 \}      \cr
\Gamma_{810}&= \{17,81,89,22 \}      \cr \Gamma_{811}&=
\{17,90,85,19 \}      \cr \Gamma_{812}&= \{17,90,85,20 \}      \cr
 \hfill \cr } $ & \cr
\noalign{\hrule} }}
$$

$$ \vbox{\tabskip=0pt \offinterlineskip
\halign to410pt{\strut#& \vrule#\tabskip=0em plus1em& \hfil#&
\vrule#& \hfil#\hfil& \vrule#& \hfil#\hfil& \vrule#& \hfil#\hfil&
\vrule#\tabskip=0pt\cr \noalign{\hrule}

&& $\eqalign{ \hfill \cr \Gamma_{813}&= \{17,90,88,19 \}      \cr
\Gamma_{814}&= \{17,90,88,22 \}  \cr \Gamma_{815}&= \{17,90,89,20 \}
\cr \Gamma_{816}&= \{17,90,89,22 \}      \cr \Gamma_{817}&=
\{18,47,42,8 \}      \cr \Gamma_{818}&= \{18,47,42,12 \}      \cr
\Gamma_{819}&= \{18,47,44,8 \}      \cr \Gamma_{820}&= \{18,47,44,14
\}      \cr \Gamma_{821}&= \{18,47,49,12 \}      \cr \Gamma_{822}&=
\{18,47,49,14 \}      \cr \Gamma_{823}&= \{18,58,42,8 \}      \cr
\Gamma_{824}&= \{18,58,42,12 \}      \cr \Gamma_{825}&= \{18,58,63,8
\}      \cr \Gamma_{826}&= \{18,58,63,19 \}      \cr \Gamma_{827}&=
\{18,58,69,12 \}      \cr \Gamma_{828}&= \{18,58,69,19 \}      \cr
\Gamma_{829}&= \{18,60,44,8 \}      \cr \Gamma_{830}&= \{18,60,44,14
\}      \cr \Gamma_{831}&= \{18,60,65,8 \}      \cr \Gamma_{832}&=
\{18,60,65,21 \}      \cr \Gamma_{833}&= \{18,60,71,14 \}      \cr
\Gamma_{834}&= \{18,60,71,21 \}      \cr \Gamma_{835}&=
\{18,66,49,12 \}      \cr \Gamma_{836}&= \{18,66,49,14 \}      \cr
\Gamma_{837}&= \{18,66,75,12 \}      \cr \Gamma_{838}&=
\{18,66,75,23 \}      \cr \Gamma_{839}&= \{18,66,77,14 \}      \cr
\Gamma_{840}&= \{18,66,77,23 \}      \cr \Gamma_{841}&= \{18,78,63,8
\}      \cr \Gamma_{842}&= \{18,78,63,19 \}      \cr \Gamma_{843}&=
\{18,78,65,8 \}      \cr \Gamma_{844}&= \{18,78,65,21 \}      \cr
\Gamma_{845}&= \{18,78,86,19 \}      \cr \Gamma_{846}&=
\{18,78,86,21 \}      \cr \Gamma_{847}&= \{18,82,69,12 \}      \cr
\Gamma_{848}&= \{18,82,69,19 \}      \cr \Gamma_{849}&=
\{18,82,75,12 \}      \cr \hfill \cr } $

&& $\eqalign{ \hfill \cr  \Gamma_{850}&= \{18,82,75,23 \}      \cr
\Gamma_{851}&= \{18,82,90,19 \}      \cr \Gamma_{852}&=
\{18,82,90,23 \}      \cr \Gamma_{853}&= \{18,84,71,14 \}      \cr
\Gamma_{854}&= \{18,84,71,21 \}      \cr \Gamma_{855}&=
\{18,84,77,14 \}      \cr \Gamma_{856}&= \{18,84,77,23 \}      \cr
\Gamma_{857}&= \{18,84,92,21 \}      \cr \Gamma_{858}&=
\{18,84,92,23 \}      \cr \Gamma_{859}&= \{18,91,86,19 \}      \cr
\Gamma_{860}&= \{18,91,86,21 \}      \cr \Gamma_{861}&=
\{18,91,90,19 \}      \cr \Gamma_{862}&= \{18,91,90,23 \}      \cr
\Gamma_{863}&= \{18,91,92,21 \}      \cr \Gamma_{864}&=
\{18,91,92,23 \}      \cr \Gamma_{865}&= \{19,48,43,9 \}      \cr
\Gamma_{866}&= \{19,48,43,13 \}      \cr \Gamma_{867}&= \{19,48,45,9
\}      \cr \Gamma_{868}&= \{19,48,45,15 \}      \cr \Gamma_{869}&=
\{19,48,50,13 \}      \cr \Gamma_{870}&= \{19,48,50,15 \}      \cr
\Gamma_{871}&= \{19,59,43,9 \}      \cr \Gamma_{872}&= \{19,59,43,13
\}      \cr \Gamma_{873}&= \{19,59,64,9 \}      \cr \Gamma_{874}&=
\{19,59,64,20 \}      \cr \Gamma_{875}&= \{19,59,70,13 \}      \cr
\Gamma_{876}&= \{19,59,70,20 \}      \cr \Gamma_{877}&= \{19,61,45,9
\}      \cr \Gamma_{878}&= \{19,61,45,15 \}      \cr \Gamma_{879}&=
\{19,61,66,9 \}      \cr \Gamma_{880}&= \{19,61,66,21 \}      \cr
\Gamma_{881}&= \{19,61,72,15 \}      \cr \Gamma_{882}&=
\{19,61,72,21 \}      \cr \Gamma_{883}&= \{19,67,50,13 \}      \cr
\Gamma_{884}&= \{19,67,50,15 \}      \cr \Gamma_{885}&=
\{19,67,76,13 \}      \cr \Gamma_{886}&= \{19,67,76,24 \}      \cr
\hfill \cr } $

&& $\eqalign{ \hfill \cr  \Gamma_{887}&= \{19,67,78,15 \}      \cr
\Gamma_{888}&= \{19,67,78,24 \} \cr \Gamma_{889}&= \{19,79,64,9 \}
\cr \Gamma_{890}&= \{19,79,64,20 \}      \cr \Gamma_{891}&=
\{19,79,66,9 \}      \cr \Gamma_{892}&= \{19,79,66,21 \}      \cr
\Gamma_{893}&= \{19,79,87,20 \}      \cr \Gamma_{894}&=
\{19,79,87,21 \}      \cr \Gamma_{895}&= \{19,83,70,13 \}      \cr
\Gamma_{896}&= \{19,83,70,20 \}      \cr \Gamma_{897}&=
\{19,83,76,13 \}      \cr \Gamma_{898}&= \{19,83,76,24 \}      \cr
\Gamma_{899}&= \{19,83,91,20 \}      \cr \Gamma_{900}&=
\{19,83,91,24 \}      \cr \Gamma_{901}&= \{19,85,72,15 \}      \cr
\Gamma_{902}&= \{19,85,72,21 \}      \cr \Gamma_{903}&=
\{19,85,78,15 \}      \cr \Gamma_{904}&= \{19,85,78,24 \}      \cr
\Gamma_{905}&= \{19,85,93,21 \}      \cr \Gamma_{906}&=
\{19,85,93,24 \}      \cr \Gamma_{907}&= \{19,92,87,20 \}      \cr
\Gamma_{908}&= \{19,92,87,21 \}      \cr \Gamma_{909}&=
\{19,92,91,20 \}      \cr \Gamma_{910}&= \{19,92,91,24 \}      \cr
\Gamma_{911}&= \{19,92,93,21 \}      \cr \Gamma_{912}&=
\{19,92,93,24 \}      \cr \Gamma_{913}&= \{20,49,47,10 \}      \cr
\Gamma_{914}&= \{20,49,47,12 \}      \cr \Gamma_{915}&=
\{20,49,52,10 \}      \cr \Gamma_{916}&= \{20,49,52,16 \}      \cr
\Gamma_{917}&= \{20,49,54,12 \}      \cr \Gamma_{918}&=
\{20,49,54,16 \}      \cr \Gamma_{919}&= \{20,62,47,10 \}      \cr
\Gamma_{920}&= \{20,62,47,12 \}      \cr \Gamma_{921}&=
\{20,62,67,10 \}      \cr \Gamma_{922}&= \{20,62,67,19 \}      \cr
\Gamma_{923}&= \{20,62,69,12 \}      \cr \hfill \cr } $

&& $\eqalign{ \hfill \cr \Gamma_{924}&= \{20,62,69,19 \}      \cr
\Gamma_{925}&= \{20,68,52,10 \} \cr \Gamma_{926}&= \{20,68,52,16 \}
\cr \Gamma_{927}&= \{20,68,73,10 \}      \cr \Gamma_{928}&=
\{20,68,73,22 \}      \cr \Gamma_{929}&= \{20,68,79,16 \}      \cr
\Gamma_{930}&= \{20,68,79,22 \}      \cr \Gamma_{931}&=
\{20,70,54,12 \}      \cr \Gamma_{932}&= \{20,70,54,16 \}      \cr
\Gamma_{933}&= \{20,70,75,12 \}      \cr \Gamma_{934}&=
\{20,70,75,23 \}      \cr \Gamma_{935}&= \{20,70,81,16 \}      \cr
\Gamma_{936}&= \{20,70,81,23 \}      \cr \Gamma_{937}&=
\{20,80,67,10 \}      \cr \Gamma_{938}&= \{20,80,67,19 \}      \cr
\Gamma_{939}&= \{20,80,73,10 \}      \cr \Gamma_{940}&=
\{20,80,73,22 \}      \cr \Gamma_{941}&= \{20,80,88,19 \}      \cr
\Gamma_{942}&= \{20,80,88,22 \}      \cr \Gamma_{943}&=
\{20,82,69,12 \}      \cr \Gamma_{944}&= \{20,82,69,19 \}      \cr
\Gamma_{945}&= \{20,82,75,12 \}      \cr \Gamma_{946}&=
\{20,82,75,23 \}      \cr \Gamma_{947}&= \{20,82,90,19 \}      \cr
\Gamma_{948}&= \{20,82,90,23 \}      \cr \Gamma_{949}&=
\{20,86,79,16 \}      \cr \Gamma_{950}&= \{20,86,79,22 \}      \cr
\Gamma_{951}&= \{20,86,81,16 \}      \cr \Gamma_{952}&=
\{20,86,81,23 \}      \cr \Gamma_{953}&= \{20,86,94,22 \}      \cr
\Gamma_{954}&= \{20,86,94,23 \}      \cr \Gamma_{955}&=
\{20,93,88,19 \}      \cr \Gamma_{956}&= \{20,93,88,22 \}      \cr
\Gamma_{957}&= \{20,93,90,19 \}      \cr \Gamma_{958}&=
\{20,93,90,23 \}      \cr \Gamma_{959}&= \{20,93,94,22 \}      \cr
\Gamma_{960}&= \{20,93,94,23 \}      \cr \hfill \cr } $ & \cr
\noalign{\hrule} }}
$$

$$ \vbox{\tabskip=0pt \offinterlineskip
\halign to426pt{\strut#& \vrule#\tabskip=0em plus1em& \hfil#&
\vrule#& \hfil#\hfil& \vrule#& \hfil#\hfil& \vrule#& \hfil#\hfil&
\vrule#\tabskip=0pt\cr \noalign{\hrule}

&& $\eqalign{ \hfill \cr  \Gamma_{961}&= \{21,50,48,11 \}      \cr
\Gamma_{962}&= \{21,50,48,13 \} \cr \Gamma_{963}&= \{21,50,53,11 \}
\cr \Gamma_{964}&= \{21,50,53,17 \}      \cr \Gamma_{965}&=
\{21,50,55,13 \}      \cr \Gamma_{966}&= \{21,50,55,17 \}      \cr
\Gamma_{967}&= \{21,63,48,11 \}      \cr \Gamma_{968}&=
\{21,63,48,13 \}      \cr \Gamma_{969}&= \{21,63,68,11 \}      \cr
\Gamma_{970}&= \{21,63,68,20 \}      \cr \Gamma_{971}&=
\{21,63,70,13 \}      \cr \Gamma_{972}&= \{21,63,70,20 \}      \cr
\Gamma_{973}&= \{21,69,53,11 \}      \cr \Gamma_{974}&=
\{21,69,53,17 \}      \cr \Gamma_{975}&= \{21,69,74,11 \}      \cr
\Gamma_{976}&= \{21,69,74,22 \}      \cr \Gamma_{977}&=
\{21,69,80,17 \}      \cr \Gamma_{978}&= \{21,69,80,22 \}      \cr
\Gamma_{979}&= \{21,71,55,13 \}      \cr \Gamma_{980}&=
\{21,71,55,17 \}      \cr \Gamma_{981}&= \{21,71,76,13 \}      \cr
\Gamma_{982}&= \{21,71,76,24 \}      \cr \Gamma_{983}&=
\{21,71,82,17 \}      \cr \Gamma_{984}&= \{21,71,82,24 \}      \cr
\Gamma_{985}&= \{21,81,68,11 \}      \cr \Gamma_{986}&=
\{21,81,68,20 \}      \cr \Gamma_{987}&= \{21,81,74,11 \}      \cr
\Gamma_{988}&= \{21,81,74,22 \}      \cr \Gamma_{989}&=
\{21,81,89,20 \}      \cr \Gamma_{990}&= \{21,81,89,22 \}      \cr
\Gamma_{991}&= \{21,83,70,13 \}      \cr \Gamma_{992}&=
\{21,83,70,20 \}      \cr \Gamma_{993}&= \{21,83,76,13 \}      \cr
\Gamma_{994}&= \{21,83,76,24 \}      \cr \Gamma_{995}&=
\{21,83,91,20 \}      \cr \Gamma_{996}&= \{21,83,91,24 \}      \cr
\Gamma_{997}&= \{21,87,80,17 \}      \cr \hfill \cr } $

&& $\eqalign{ \hfill \cr \Gamma_{998}&= \{21,87,80,22 \}      \cr
\Gamma_{999}&= \{21,87,82,17 \}      \cr \Gamma_{1000}&=
\{21,87,82,24 \}      \cr \Gamma_{1001}&= \{21,87,95,22 \}      \cr
\Gamma_{1002}&= \{21,87,95,24 \}      \cr \Gamma_{1003}&=
\{21,94,89,20 \}      \cr \Gamma_{1004}&= \{21,94,89,22 \}      \cr
\Gamma_{1005}&= \{21,94,91,20 \}      \cr \Gamma_{1006}&=
\{21,94,91,24 \}      \cr \Gamma_{1007}&= \{21,94,95,22 \}      \cr
\Gamma_{1008}&= \{21,94,95,24 \}      \cr \Gamma_{1009}&=
\{22,51,51,14 \}      \cr \Gamma_{1010}&= \{22,51,51,15 \}      \cr
\Gamma_{1011}&= \{22,51,56,14 \}      \cr \Gamma_{1012}&=
\{22,51,56,18 \}      \cr \Gamma_{1013}&= \{22,51,57,15 \}      \cr
\Gamma_{1014}&= \{22,51,57,18 \}      \cr \Gamma_{1015}&=
\{22,64,51,14 \}      \cr \Gamma_{1016}&= \{22,64,51,15 \}      \cr
\Gamma_{1017}&= \{22,64,71,14 \}      \cr \Gamma_{1018}&=
\{22,64,71,21 \}      \cr \Gamma_{1019}&= \{22,64,72,15 \}      \cr
\Gamma_{1020}&= \{22,64,72,21 \}      \cr \Gamma_{1021}&=
\{22,72,56,14 \}      \cr \Gamma_{1022}&= \{22,72,56,18 \}      \cr
\Gamma_{1023}&= \{22,72,77,14 \}      \cr \Gamma_{1024}&=
\{22,72,77,23 \}      \cr \Gamma_{1025}&= \{22,72,83,18 \}      \cr
\Gamma_{1026}&= \{22,72,83,23 \}      \cr \Gamma_{1027}&=
\{22,73,57,15 \}      \cr \Gamma_{1028}&= \{22,73,57,18 \}      \cr
\Gamma_{1029}&= \{22,73,78,15 \}      \cr \Gamma_{1030}&=
\{22,73,78,24 \}      \cr \Gamma_{1031}&= \{22,73,84,18 \}      \cr
\Gamma_{1032}&= \{22,73,84,24 \}      \cr \Gamma_{1033}&=
\{22,84,71,14 \}      \cr \Gamma_{1034}&= \{22,84,71,21 \}      \cr
\hfill \cr } $

&& $\eqalign{ \hfill \cr  \Gamma_{1035}&= \{22,84,77,14 \}      \cr
\Gamma_{1036}&= \{22,84,77,23 \} \cr \Gamma_{1037}&= \{22,84,92,21
\}      \cr \Gamma_{1038}&= \{22,84,92,23 \}      \cr
\Gamma_{1039}&= \{22,85,72,15 \}      \cr \Gamma_{1040}&=
\{22,85,72,21 \}      \cr \Gamma_{1041}&= \{22,85,78,15 \}      \cr
\Gamma_{1042}&= \{22,85,78,24 \}      \cr \Gamma_{1043}&=
\{22,85,93,21 \}      \cr \Gamma_{1044}&= \{22,85,93,24 \}      \cr
\Gamma_{1045}&= \{22,88,83,18 \}      \cr \Gamma_{1046}&=
\{22,88,83,23 \}      \cr \Gamma_{1047}&= \{22,88,84,18 \}      \cr
\Gamma_{1048}&= \{22,88,84,24 \}      \cr \Gamma_{1049}&=
\{22,88,96,23 \}      \cr \Gamma_{1050}&= \{22,88,96,24 \}      \cr
\Gamma_{1051}&= \{22,95,92,21 \}      \cr \Gamma_{1052}&=
\{22,95,92,23 \}      \cr \Gamma_{1053}&= \{22,95,93,21 \}      \cr
\Gamma_{1054}&= \{22,95,93,24 \}      \cr \Gamma_{1055}&=
\{22,95,96,23 \}      \cr \Gamma_{1056}&= \{22,95,96,24 \}      \cr
\Gamma_{1057}&= \{23,52,58,16 \}      \cr \Gamma_{1058}&=
\{23,52,58,17 \}      \cr \Gamma_{1059}&= \{23,52,59,16 \}      \cr
\Gamma_{1060}&= \{23,52,59,18 \}      \cr \Gamma_{1061}&=
\{23,52,60,17 \}      \cr \Gamma_{1062}&= \{23,52,60,18 \}      \cr
\Gamma_{1063}&= \{23,74,58,16 \}      \cr \Gamma_{1064}&=
\{23,74,58,17 \}      \cr \Gamma_{1065}&= \{23,74,79,16 \}  \cr
\Gamma_{1066}&= \{23,74,79,22 \}      \cr \Gamma_{1067}&=
\{23,74,80,17 \}      \cr \Gamma_{1068}&= \{23,74,80,22 \}      \cr
\Gamma_{1069}&= \{23,75,59,16 \}      \cr \Gamma_{1070}&=
\{23,75,59,18 \}      \cr \Gamma_{1071}&= \{23,75,81,16 \}      \cr
\hfill \cr } $

&& $\eqalign{ \hfill \cr \Gamma_{1072}&= \{23,75,81,23 \}      \cr
\Gamma_{1073}&= \{23,75,83,18 \}      \cr \Gamma_{1074}&=
\{23,75,83,23 \}      \cr \Gamma_{1075}&= \{23,76,60,17 \}      \cr
\Gamma_{1076}&= \{23,76,60,18 \}      \cr \Gamma_{1077}&=
\{23,76,82,17 \}      \cr \Gamma_{1078}&= \{23,76,82,24 \}      \cr
\Gamma_{1079}&= \{23,76,84,18 \}      \cr \Gamma_{1080}&=
\{23,76,84,24 \}      \cr \Gamma_{1081}&= \{23,86,79,16 \}      \cr
\Gamma_{1082}&= \{23,86,79,22 \}      \cr \Gamma_{1083}&=
\{23,86,81,16 \}      \cr \Gamma_{1084}&= \{23,86,81,23 \}      \cr
\Gamma_{1085}&= \{23,86,94,22 \}      \cr \Gamma_{1086}&=
\{23,86,94,23 \}      \cr \Gamma_{1087}&= \{23,87,80,17 \}      \cr
\Gamma_{1088}&= \{23,87,80,22 \}      \cr \Gamma_{1089}&=
\{23,87,82,17 \}      \cr \Gamma_{1090}&= \{23,87,82,24 \}      \cr
\Gamma_{1091}&= \{23,87,95,22 \}      \cr \Gamma_{1092}&=
\{23,87,95,24 \} \cr \Gamma_{1093}&= \{23,88,83,18 \}  \cr
\Gamma_{1094}&= \{23,88,83,23 \}  \cr \Gamma_{1095}&= \{23,88,84,18
\}      \cr \Gamma_{1096}&= \{23,88,84,24 \}      \cr
\Gamma_{1097}&= \{23,88,96,23 \}      \cr \Gamma_{1098}&=
\{23,88,96,24 \}      \cr \Gamma_{1099}&= \{23,96,94,22 \}      \cr
\Gamma_{1100}&= \{23,96,94,23 \}      \cr \Gamma_{1101}&=
\{23,96,95,22 \}      \cr \Gamma_{1102}&= \{23,96,95,24 \}      \cr
\Gamma_{1103}&= \{23,96,96,23 \}      \cr \Gamma_{1104}&=
\{23,96,96,24 \}      \cr \Gamma_{1105}&= \{24,89,85,19 \}      \cr
\Gamma_{1106}&= \{24,89,85,20 \}      \cr \Gamma_{1107}&=
\{24,89,86,19 \}      \cr \Gamma_{1108}&= \{24,89,86,21 \}      \cr
\hfill \cr } $ & \cr \noalign{\hrule} }}
$$

$$ \vbox{\tabskip=0pt \offinterlineskip
\halign to426pt{\strut#& \vrule#\tabskip=0em plus1em& \hfil#&
\vrule#& \hfil#\hfil& \vrule#& \hfil#\hfil& \vrule#& \hfil#\hfil&
\vrule#\tabskip=0pt\cr \noalign{\hrule}

&& $\eqalign{ \hfill \cr \Gamma_{1109}&= \{24,89,87,20 \}      \cr
\Gamma_{1110}&= \{24,89,87,21 \}      \cr \Gamma_{1111}&=
\{24,90,85,19 \}      \cr \Gamma_{1112}&= \{24,90,85,20 \}      \cr
\Gamma_{1113}&= \{24,90,88,19 \}      \cr \Gamma_{1114}&=
\{24,90,88,22 \}      \cr \Gamma_{1115}&= \{24,90,89,20 \}      \cr
\Gamma_{1116}&= \{24,90,89,22 \}      \cr \Gamma_{1117}&=
\{24,91,86,19 \}      \cr \Gamma_{1118}&= \{24,91,86,21 \}      \cr
\Gamma_{1119}&= \{24,91,90,19 \}      \cr \hfill \cr } $

&& $\eqalign{ \hfill \cr \Gamma_{1120}&= \{24,91,90,23 \}      \cr
\Gamma_{1121}&= \{24,91,92,21 \}      \cr \Gamma_{1122}&=
\{24,91,92,23 \}      \cr \Gamma_{1123}&= \{24,92,87,20 \}  \cr
\Gamma_{1124}&= \{24,92,87,21 \}      \cr \Gamma_{1125}&=
\{24,92,91,20 \}      \cr \Gamma_{1126}&= \{24,92,91,24 \}      \cr
\Gamma_{1127}&= \{24,92,93,21 \}      \cr \Gamma_{1128}&=
\{24,92,93,24 \}      \cr \Gamma_{1129}&= \{24,93,88,19 \}      \cr
\Gamma_{1130}&= \{24,93,88,22 \}      \cr \hfill \cr } $

&& $\eqalign{ \hfill \cr \Gamma_{1131}&= \{24,93,90,19 \}      \cr
\Gamma_{1132}&= \{24,93,90,23 \}      \cr \Gamma_{1133}&=
\{24,93,94,22 \}      \cr \Gamma_{1134}&= \{24,93,94,23 \}      \cr
\Gamma_{1135}&= \{24,94,89,20 \}      \cr \Gamma_{1136}&=
\{24,94,89,22 \}      \cr \Gamma_{1137}&= \{24,94,91,20 \}      \cr
\Gamma_{1138}&= \{24,94,91,24 \}      \cr \Gamma_{1139}&=
\{24,94,95,22 \}      \cr \Gamma_{1140}&= \{24,94,95,24 \}      \cr
\Gamma_{1141}&= \{24,95,92,21 \}      \cr \hfill \cr } $

&& $\eqalign{ \hfill \cr \Gamma_{1142}&= \{24,95,92,23 \}      \cr
\Gamma_{1143}&= \{24,95,93,21 \}      \cr \Gamma_{1144}&=
\{24,95,93,24 \}      \cr \Gamma_{1145}&= \{24,95,96,23 \}      \cr
\Gamma_{1146}&= \{24,95,96,24 \}      \cr \Gamma_{1147}&=
\{24,96,94,22 \}      \cr \Gamma_{1148}&= \{24,96,94,23 \}      \cr
\Gamma_{1149}&= \{24,96,95,22 \}      \cr \Gamma_{1150}&=
\{24,96,95,24 \}      \cr \Gamma_{1151}&= \{24,96,96,23 \}      \cr
\Gamma_{1152}&= \{24,96,96,24 \}      \cr \hfill \cr } $ & \cr
\noalign{\hrule} }}
$$
\centerline{Table-II}
\vskip 3mm

\noindent In view of (II.6), the notation in Table-II is briefly
given by
$$ \Gamma_A=\{I_1,I_2,I_3,I_4 \} \equiv \Gamma(A) = \{ \gamma_1(I_1(A)),\gamma_2(I_2(A)),
\gamma_3(I_3(A)),\gamma_4(I_4(A)) \} $$

\noindent Note here that $\vert W(F_4) \vert = 1152$, $\vert
W(\lambda_1) \vert = \vert W(\lambda_4) \vert=24 $ and $\vert
W(\lambda_2) \vert = \vert W(\lambda_3) \vert =96$ show, for $F_4$,
the validity of our two statements mentioned above.

Corresponding signatures are also given by the following Table-III :

$$ \vbox{\tabskip=0pt \offinterlineskip
\halign to420pt{\strut#& \vrule#\tabskip=0em plus1em& \hfil#&
\vrule#& \hfil#\hfil& \vrule#& \hfil#\hfil& \vrule#& \hfil#\hfil&
\vrule#& \hfil#\hfil& \vrule#& \hfil#\hfil& \vrule#& \hfil#\hfil&
\vrule#& \hfil#\hfil& \vrule#\tabskip=0pt\cr \noalign{\hrule}

&& $\eqalign{ \hfill \cr \epsilon_{1}&=+1    \cr \epsilon_{2}&=-1
\cr \epsilon_{3}&=-1   \cr \epsilon_{4}&=+1    \cr \epsilon_{5}&=+1
\cr \epsilon_{6}&=-1   \cr \epsilon_{7}&=-1   \cr \epsilon_{8}&=+1
\cr \epsilon_{9}&=+1    \cr \epsilon_{10}&=-1  \cr \hfill \cr } $

&& $\eqalign{ \hfill \cr \epsilon_{11}&=-1 \cr \epsilon_{12}&=+1
\cr \epsilon_{13}&=+1   \cr \epsilon_{14}&=-1 \cr \epsilon_{15}&=-1
\cr \epsilon_{16}&=+1   \cr \epsilon_{17}&=+1 \cr \epsilon_{18}&=-1
\cr \epsilon_{19}&=-1  \cr \epsilon_{20}&=+1 \cr  \hfill \cr } $

&& $\eqalign{ \hfill \cr \epsilon_{21}&=+1   \cr \epsilon_{22}&=-1
\cr \epsilon_{23}&=-1 \cr \epsilon_{24}&=+1   \cr \epsilon_{25}&=-1
\cr \epsilon_{26}&=+1 \cr \epsilon_{27}&=+1   \cr \epsilon_{28}&=-1
\cr \epsilon_{29}&=-1 \cr \epsilon_{30}&=+1   \cr \hfill \cr } $

&& $\eqalign{ \hfill \cr  \epsilon_{31}&=+1 \cr \epsilon_{32}&=-1
\cr \epsilon_{33}&=-1  \cr \epsilon_{34}&=+1 \cr \epsilon_{35}&=+1
\cr \epsilon_{36}&=-1  \cr \epsilon_{37}&=-1 \cr \epsilon_{38}&=+1
\cr \epsilon_{39}&=+1 \cr \epsilon_{40}&=-1  \cr \hfill \cr } $

&& $\eqalign{ \hfill \cr \epsilon_{41}&=-1  \cr \epsilon_{42}&=+1
\cr \epsilon_{43}&=+1   \cr \epsilon_{44}&=-1  \cr \epsilon_{45}&=-1
\cr \epsilon_{46}&=+1   \cr \epsilon_{47}&=+1   \cr
\epsilon_{48}&=-1 \cr \epsilon_{49}&=-1  \cr \epsilon_{50}&=+1   \cr
\hfill \cr } $

&& $\eqalign{ \hfill \cr \epsilon_{51}&=+1 \cr \epsilon_{52}&=-1
\cr \epsilon_{53}&=-1  \cr \epsilon_{54}&=+1 \cr \epsilon_{55}&=+1
\cr \epsilon_{56}&=-1  \cr \epsilon_{57}&=-1 \cr \epsilon_{58}&=+1
\cr \epsilon_{59}&=+1 \cr \epsilon_{60}&=-1 \cr
 \hfill \cr } $

&& $\eqalign{ \hfill \cr \epsilon_{61}&=-1  \cr \epsilon_{62}&=+1
\cr \epsilon_{63}&=+1 \cr \epsilon_{64}&=-1  \cr \epsilon_{65}&=-1
\cr \epsilon_{66}&=+1 \cr \epsilon_{67}&=+1   \cr \epsilon_{68}&=-1
\cr \epsilon_{69}&=-1 \cr \epsilon_{70}&=+1   \cr \hfill \cr } $

&& $\eqalign{ \hfill \cr \epsilon_{71}&=+1 \cr \epsilon_{72}&=-1 \cr
\epsilon_{73}&=+1   \cr \epsilon_{74}&=-1 \cr \epsilon_{75}&=-1  \cr
\epsilon_{76}&=+1 \cr \epsilon_{77}&=+1   \cr \epsilon_{78}&=-1  \cr
\epsilon_{79}&=-1 \cr \epsilon_{80}&=+1   \cr \hfill \cr } $ & \cr
\noalign{\hrule} }}
$$

$$ \vbox{\tabskip=0pt \offinterlineskip
\halign to420pt{\strut#& \vrule#\tabskip=0em plus1em& \hfil#&
\vrule#& \hfil#\hfil& \vrule#& \hfil#\hfil& \vrule#& \hfil#\hfil&
\vrule#& \hfil#\hfil& \vrule#& \hfil#\hfil& \vrule#& \hfil#\hfil&
\vrule#& \hfil#\hfil& \vrule#\tabskip=0pt\cr \noalign{\hrule}

&& $\eqalign{ \hfill \cr \epsilon_{81}&=+1   \cr \epsilon_{82}&=-1
\cr \epsilon_{83}&=-1  \cr \epsilon_{84}&=+1   \cr \epsilon_{85}&=+1
\cr \epsilon_{86}&=-1  \cr \epsilon_{87}&=-1  \cr \epsilon_{88}&=+1
\cr \epsilon_{89}&=+1   \cr \epsilon_{90}&=-1  \cr \epsilon_{91}&=-1
\cr \epsilon_{92}&=+1   \cr \epsilon_{93}&=+1 \cr \epsilon_{94}&=-1
\cr \epsilon_{95}&=-1  \cr \epsilon_{96}&=+1 \cr \epsilon_{97}&=+1
\cr \epsilon_{98}&=-1  \cr \epsilon_{99}&=-1 \cr \epsilon_{100}&=+1
\cr \epsilon_{101}&=+1  \cr \epsilon_{102}&=-1 \cr
\epsilon_{103}&=-1 \cr \epsilon_{104}&=+1 \cr \epsilon_{105}&=+1 \cr
\epsilon_{106}&=-1 \cr \epsilon_{107}&=-1 \cr \epsilon_{108}&=+1 \cr
\epsilon_{109}&=+1 \cr \epsilon_{110}&=-1 \cr \epsilon_{111}&=-1 \cr
\epsilon_{112}&=+1  \cr \epsilon_{113}&=+1 \cr \epsilon_{114}&=-1
\cr \epsilon_{115}&=-1 \cr \epsilon_{116}&=+1 \cr \epsilon_{117}&=+1
\cr \epsilon_{118}&=-1 \cr \hfill \cr } $

&& $\eqalign{ \hfill \cr \epsilon_{119}&=-1 \cr \epsilon_{120}&=+1
\cr \epsilon_{121}&=-1 \cr \epsilon_{122}&=+1 \cr \epsilon_{123}&=+1
\cr \epsilon_{124}&=-1 \cr \epsilon_{125}&=-1 \cr \epsilon_{126}&=+1
\cr \epsilon_{127}&=+1  \cr \epsilon_{128}&=-1 \cr
\epsilon_{129}&=-1 \cr \epsilon_{130}&=+1  \cr \epsilon_{131}&=+1
\cr \epsilon_{132}&=-1 \cr \epsilon_{133}&=-1 \cr \epsilon_{134}&=+1
\cr \epsilon_{135}&=+1  \cr \epsilon_{136}&=-1 \cr
\epsilon_{137}&=-1 \cr \epsilon_{138}&=+1  \cr \epsilon_{139}&=+1
\cr \epsilon_{140}&=-1 \cr \epsilon_{141}&=-1 \cr \epsilon_{142}&=+1
\cr \epsilon_{143}&=+1 \cr \epsilon_{144}&=-1 \cr \epsilon_{145}&=-1
\cr \epsilon_{146}&=+1 \cr \epsilon_{147}&=+1  \cr
\epsilon_{148}&=-1 \cr \epsilon_{149}&=-1 \cr \epsilon_{150}&=+1 \cr
\epsilon_{151}&=+1  \cr \epsilon_{152}&=-1 \cr \epsilon_{153}&=-1
\cr \epsilon_{154}&=+1  \cr \epsilon_{155}&=+1 \cr
\epsilon_{156}&=-1 \cr \hfill \cr } $

&& $\eqalign{ \hfill \cr \epsilon_{157}&=-1 \cr \epsilon_{158}&=+1
\cr \epsilon_{159}&=+1 \cr \epsilon_{160}&=-1 \cr \epsilon_{161}&=-1
\cr \epsilon_{162}&=+1 \cr \epsilon_{163}&=+1 \cr \epsilon_{164}&=-1
\cr \epsilon_{165}&=-1 \cr \epsilon_{166}&=+1 \cr \epsilon_{167}&=+1
\cr \epsilon_{168}&=-1 \cr \epsilon_{169}&=+1 \cr \epsilon_{170}&=-1
\cr \epsilon_{171}&=-1 \cr \epsilon_{172}&=+1 \cr \epsilon_{173}&=+1
\cr \epsilon_{174}&=-1 \cr \epsilon_{175}&=-1 \cr \epsilon_{176}&=+1
\cr \epsilon_{177}&=+1 \cr \epsilon_{178}&=-1 \cr \epsilon_{179}&=-1
\cr \epsilon_{180}&=+1 \cr \epsilon_{181}&=+1  \cr
\epsilon_{182}&=-1 \cr \epsilon_{183}&=-1 \cr \epsilon_{184}&=+1 \cr
\epsilon_{185}&=+1 \cr \epsilon_{186}&=-1 \cr \epsilon_{187}&=-1 \cr
\epsilon_{188}&=+1  \cr \epsilon_{189}&=+1  \cr \epsilon_{190}&=-1
\cr \epsilon_{191}&=-1 \cr \epsilon_{192}&=+1 \cr \epsilon_{193}&=+1
\cr \epsilon_{194}&=-1 \cr \hfill \cr } $

&& $\eqalign{ \hfill \cr \epsilon_{195}&=-1 \cr \epsilon_{196}&=+1
\cr \epsilon_{197}&=+1  \cr \epsilon_{198}&=-1 \cr
\epsilon_{199}&=-1 \cr \epsilon_{200}&=+1 \cr \epsilon_{201}&=+1 \cr
\epsilon_{202}&=-1 \cr \epsilon_{203}&=-1 \cr \epsilon_{204}&=+1 \cr
\epsilon_{205}&=+1 \cr \epsilon_{206}&=-1 \cr \epsilon_{207}&=-1 \cr
\epsilon_{208}&=+1 \cr \epsilon_{209}&=+1  \cr \epsilon_{210}&=-1
\cr \epsilon_{211}&=-1 \cr \epsilon_{212}&=+1 \cr \epsilon_{213}&=+1
\cr \epsilon_{214}&=-1 \cr \epsilon_{215}&=-1 \cr \epsilon_{216}&=+1
\cr \epsilon_{217}&=-1 \cr \epsilon_{218}&=+1 \cr \epsilon_{219}&=+1
\cr \epsilon_{220}&=-1 \cr \epsilon_{221}&=-1 \cr \epsilon_{222}&=+1
\cr  \epsilon_{223}&=+1  \cr \epsilon_{224}&=-1 \cr
\epsilon_{225}&=-1 \cr \epsilon_{226}&=+1 \cr \epsilon_{227}&=+1 \cr
\epsilon_{228}&=-1 \cr \epsilon_{229}&=-1 \cr \epsilon_{230}&=+1 \cr
\epsilon_{231}&=+1  \cr \epsilon_{232}&=-1 \cr \hfill \cr } $

&& $\eqalign{ \hfill \cr \epsilon_{233}&=-1 \cr \epsilon_{234}&=+1
\cr \epsilon_{235}&=+1 \cr \epsilon_{236}&=-1 \cr \epsilon_{237}&=-1
\cr \epsilon_{238}&=+1 \cr \epsilon_{239}&=+1 \cr \epsilon_{240}&=-1
\cr \epsilon_{241}&=+1 \cr \epsilon_{242}&=-1 \cr \epsilon_{243}&=-1
\cr \epsilon_{244}&=+1 \cr \epsilon_{245}&=+1 \cr \epsilon_{246}&=-1
\cr \epsilon_{247}&=-1 \cr \epsilon_{248}&=+1 \cr \epsilon_{249}&=+1
\cr \epsilon_{250}&=-1 \cr \epsilon_{251}&=-1 \cr \epsilon_{252}&=+1
\cr \epsilon_{253}&=+1  \cr \epsilon_{254}&=-1 \cr
\epsilon_{255}&=-1 \cr \epsilon_{256}&=+1 \cr \epsilon_{257}&=+1 \cr
\epsilon_{258}&=-1 \cr \epsilon_{259}&=-1 \cr  \epsilon_{260}&=+1
\cr \epsilon_{261}&=+1 \cr \epsilon_{262}&=-1 \cr \epsilon_{263}&=-1
\cr \epsilon_{264}&=+1 \cr \epsilon_{265}&=-1 \cr \epsilon_{266}&=+1
\cr \epsilon_{267}&=+1 \cr \epsilon_{268}&=-1 \cr \epsilon_{269}&=-1
\cr \epsilon_{270}&=+1 \cr \hfill \cr } $

&& $\eqalign{ \hfill \cr \epsilon_{271}&=+1  \cr \epsilon_{272}&=-1
\cr \epsilon_{273}&=-1 \cr \epsilon_{274}&=+1 \cr \epsilon_{275}&=+1
\cr \epsilon_{276}&=-1 \cr \epsilon_{277}&=-1 \cr \epsilon_{278}&=+1
\cr \epsilon_{279}&=+1 \cr \epsilon_{280}&=-1 \cr \epsilon_{281}&=-1
\cr \epsilon_{282}&=+1 \cr \epsilon_{283}&=+1 \cr \epsilon_{284}&=-1
\cr \epsilon_{285}&=-1 \cr \epsilon_{286}&=+1 \cr \epsilon_{287}&=+1
\cr \epsilon_{288}&=-1 \cr \epsilon_{289}&=-1 \cr \epsilon_{290}&=+1
\cr \epsilon_{291}&=+1 \cr \epsilon_{292}&=-1 \cr \epsilon_{293}&=-1
\cr \epsilon_{294}&=+1 \cr \epsilon_{295}&=+1 \cr \epsilon_{296}&=-1
\cr \epsilon_{297}&=-1 \cr \epsilon_{298}&=+1 \cr \epsilon_{299}&=+1
\cr \epsilon_{300}&=-1 \cr \epsilon_{301}&=-1 \cr \epsilon_{302}&=+1
\cr \epsilon_{303}&=+1 \cr \epsilon_{304}&=-1 \cr \epsilon_{305}&=-1
\cr \epsilon_{306}&=+1 \cr \epsilon_{307}&=+1 \cr \epsilon_{308}&=-1
\cr \hfill \cr } $

&& $\eqalign{ \hfill \cr \epsilon_{309}&=-1 \cr \epsilon_{310}&=+1
\cr \epsilon_{311}&=+1  \cr \epsilon_{312}&=-1 \cr
\epsilon_{313}&=+1 \cr \epsilon_{314}&=-1 \cr \epsilon_{315}&=-1 \cr
\epsilon_{316}&=+1 \cr \epsilon_{317}&=+1 \cr \epsilon_{318}&=-1 \cr
\epsilon_{319}&=-1 \cr \epsilon_{320}&=+1 \cr \epsilon_{321}&=+1 \cr
\epsilon_{322}&=-1 \cr \epsilon_{323}&=-1 \cr \epsilon_{324}&=+1 \cr
\epsilon_{325}&=+1 \cr \epsilon_{326}&=-1 \cr \epsilon_{327}&=-1 \cr
\epsilon_{328}&=+1 \cr \epsilon_{329}&=+1  \cr \epsilon_{330}&=-1
\cr \epsilon_{331}&=-1 \cr \epsilon_{332}&=+1 \cr \epsilon_{333}&=+1
\cr \epsilon_{334}&=-1 \cr \epsilon_{335}&=-1 \cr \epsilon_{336}&=+1
\cr \epsilon_{337}&=+1  \cr \epsilon_{338}&=-1 \cr
\epsilon_{339}&=-1 \cr \epsilon_{340}&=+1 \cr \epsilon_{341}&=+1 \cr
\epsilon_{342}&=-1 \cr \epsilon_{343}&=-1 \cr \epsilon_{344}&=+1 \cr
\epsilon_{345}&=+1  \cr \epsilon_{346}&=-1 \cr
 \hfill \cr } $

&& $\eqalign{ \hfill \cr \epsilon_{347}&=-1 \cr \epsilon_{348}&=+1
\cr \epsilon_{349}&=+1 \cr \epsilon_{350}&=-1 \cr \epsilon_{351}&=-1
\cr \epsilon_{352}&=+1 \cr \epsilon_{353}&=+1 \cr \epsilon_{354}&=-1
\cr \epsilon_{355}&=-1 \cr \epsilon_{356}&=+1 \cr \epsilon_{357}&=+1
\cr \epsilon_{358}&=-1 \cr \epsilon_{359}&=-1 \cr \epsilon_{360}&=+1
\cr \epsilon_{361}&=-1 \cr \epsilon_{362}&=+1 \cr \epsilon_{363}&=+1
\cr \epsilon_{364}&=-1 \cr \epsilon_{365}&=-1 \cr \epsilon_{366}&=+1
\cr \epsilon_{367}&=+1  \cr \epsilon_{368}&=-1 \cr
\epsilon_{369}&=-1 \cr \epsilon_{370}&=+1  \cr \epsilon_{371}&=+1
\cr \epsilon_{372}&=-1 \cr \epsilon_{373}&=-1 \cr \epsilon_{374}&=+1
\cr \epsilon_{375}&=+1 \cr \epsilon_{376}&=-1 \cr \epsilon_{377}&=-1
\cr \epsilon_{378}&=+1 \cr \epsilon_{379}&=+1  \cr
\epsilon_{380}&=-1 \cr \epsilon_{381}&=-1 \cr \epsilon_{382}&=+1 \cr
\epsilon_{383}&=+1 \cr \epsilon_{384}&=-1 \cr \hfill \cr } $ & \cr
\noalign{\hrule} }}
$$

$$ \vbox{\tabskip=0pt \offinterlineskip
\halign to425pt{\strut#& \vrule#\tabskip=0em plus1em& \hfil#&
\vrule#& \hfil#\hfil& \vrule#& \hfil#\hfil& \vrule#& \hfil#\hfil&
\vrule#& \hfil#\hfil& \vrule#& \hfil#\hfil& \vrule#& \hfil#\hfil&
\vrule#& \hfil#\hfil& \vrule#\tabskip=0pt\cr \noalign{\hrule}

&& $\eqalign{ \hfill \cr \epsilon_{385}&=-1 \cr \epsilon_{386}&=+1
\cr \epsilon_{387}&=+1 \cr \epsilon_{388}&=-1 \cr \epsilon_{389}&=-1
\cr \epsilon_{390}&=+1 \cr \epsilon_{391}&=+1 \cr \epsilon_{392}&=-1
\cr \epsilon_{393}&=-1 \cr \epsilon_{394}&=+1 \cr \epsilon_{395}&=+1
\cr \epsilon_{396}&=-1 \cr \epsilon_{397}&=-1 \cr \epsilon_{398}&=+1
\cr \epsilon_{399}&=+1 \cr \epsilon_{400}&=-1 \cr \epsilon_{401}&=-1
\cr \epsilon_{402}&=+1 \cr \epsilon_{403}&=+1 \cr \epsilon_{404}&=-1
\cr \epsilon_{405}&=-1 \cr \epsilon_{406}&=+1 \cr \epsilon_{407}&=+1
\cr \epsilon_{408}&=-1 \cr \epsilon_{409}&=+1 \cr \epsilon_{410}&=-1
\cr \epsilon_{411}&=-1 \cr \epsilon_{412}&=+1 \cr \epsilon_{413}&=+1
\cr \epsilon_{414}&=-1 \cr \epsilon_{415}&=-1 \cr \epsilon_{416}&=+1
\cr \epsilon_{417}&=+1 \cr \epsilon_{418}&=-1 \cr \epsilon_{419}&=-1
\cr \epsilon_{420}&=+1 \cr \epsilon_{421}&=+1 \cr \epsilon_{422}&=-1
\cr \hfill \cr } $

&& $\eqalign{ \hfill \cr \epsilon_{423}&=-1 \cr \epsilon_{424}&=+1
\cr \epsilon_{425}&=+1  \cr \epsilon_{426}&=-1 \cr
\epsilon_{427}&=-1 \cr \epsilon_{428}&=+1 \cr \epsilon_{429}&=+1 \cr
\epsilon_{430}&=-1 \cr \epsilon_{431}&=-1 \cr \epsilon_{432}&=+1 \cr
\epsilon_{433}&=-1 \cr \epsilon_{434}&=+1  \cr \epsilon_{435}&=+1
\cr \epsilon_{436}&=-1 \cr \epsilon_{437}&=-1 \cr \epsilon_{438}&=+1
\cr \epsilon_{439}&=+1 \cr \epsilon_{440}&=-1 \cr \epsilon_{441}&=-1
\cr \epsilon_{442}&=+1 \cr \epsilon_{443}&=+1 \cr \epsilon_{444}&=-1
\cr  \epsilon_{445}&=-1 \cr \epsilon_{446}&=+1 \cr
\epsilon_{447}&=+1  \cr \epsilon_{448}&=-1 \cr \epsilon_{449}&=-1
\cr \epsilon_{450}&=+1  \cr \epsilon_{451}&=+1 \cr
\epsilon_{452}&=-1 \cr \epsilon_{453}&=-1 \cr \epsilon_{454}&=+1 \cr
\epsilon_{455}&=+1 \cr \epsilon_{456}&=-1 \cr \epsilon_{457}&=+1 \cr
\epsilon_{458}&=-1 \cr \epsilon_{459}&=-1 \cr \epsilon_{460}&=+1 \cr
\hfill \cr } $

&& $\eqalign{ \hfill \cr \epsilon_{461}&=+1 \cr \epsilon_{462}&=-1
\cr \epsilon_{463}&=-1 \cr \epsilon_{464}&=+1 \cr \epsilon_{465}&=+1
\cr \epsilon_{466}&=-1 \cr \epsilon_{467}&=-1 \cr \epsilon_{468}&=+1
\cr \epsilon_{469}&=+1  \cr \epsilon_{470}&=-1 \cr
\epsilon_{471}&=-1 \cr \epsilon_{472}&=+1 \cr \epsilon_{473}&=+1 \cr
\epsilon_{474}&=-1 \cr \epsilon_{475}&=-1 \cr \epsilon_{476}&=+1 \cr
\epsilon_{477}&=+1  \cr \epsilon_{478}&=-1 \cr \epsilon_{479}&=-1
\cr \epsilon_{480}&=+1 \cr \epsilon_{481}&=+1 \cr \epsilon_{482}&=-1
\cr \epsilon_{483}&=-1 \cr \epsilon_{484}&=+1  \cr
\epsilon_{485}&=+1  \cr \epsilon_{486}&=-1 \cr \epsilon_{487}&=-1
\cr \epsilon_{488}&=+1 \cr \epsilon_{489}&=+1 \cr \epsilon_{490}&=-1
\cr \epsilon_{491}&=-1 \cr \epsilon_{492}&=+1 \cr \epsilon_{493}&=+1
\cr \epsilon_{494}&=-1 \cr \epsilon_{495}&=-1 \cr \epsilon_{496}&=+1
\cr \epsilon_{497}&=+1 \cr \epsilon_{498}&=-1 \cr \hfill \cr } $

&& $\eqalign{ \hfill \cr \epsilon_{499}&=-1 \cr \epsilon_{500}&=+1
\cr \epsilon_{501}&=+1 \cr \epsilon_{502}&=-1 \cr \epsilon_{503}&=-1
\cr \epsilon_{504}&=+1 \cr \epsilon_{505}&=-1 \cr \epsilon_{506}&=+1
\cr \epsilon_{507}&=+1 \cr \epsilon_{508}&=-1 \cr \epsilon_{509}&=-1
\cr \epsilon_{510}&=+1 \cr \epsilon_{511}&=+1 \cr \epsilon_{512}&=-1
\cr \epsilon_{513}&=-1 \cr \epsilon_{514}&=+1 \cr \epsilon_{515}&=+1
\cr \epsilon_{516}&=-1 \cr \epsilon_{517}&=-1 \cr \epsilon_{518}&=+1
\cr \epsilon_{519}&=+1  \cr \epsilon_{520}&=-1 \cr
\epsilon_{521}&=-1 \cr \epsilon_{522}&=+1 \cr \epsilon_{523}&=+1 \cr
\epsilon_{524}&=-1 \cr \epsilon_{525}&=-1 \cr \epsilon_{526}&=+1 \cr
\epsilon_{527}&=+1  \cr \epsilon_{528}&=-1 \cr \epsilon_{529}&=-1
\cr \epsilon_{530}&=+1 \cr \epsilon_{531}&=+1 \cr \epsilon_{532}&=-1
\cr \epsilon_{533}&=-1 \cr \epsilon_{534}&=+1 \cr \epsilon_{535}&=+1
\cr \epsilon_{536}&=-1 \cr \hfill \cr } $

&& $\eqalign{ \hfill \cr \epsilon_{537}&=-1 \cr \epsilon_{538}&=+1
\cr \epsilon_{539}&=+1  \cr \epsilon_{540}&=-1 \cr
\epsilon_{541}&=-1 \cr \epsilon_{542}&=+1 \cr \epsilon_{543}&=+1 \cr
\epsilon_{544}&=-1 \cr \epsilon_{545}&=-1 \cr \epsilon_{546}&=+1 \cr
\epsilon_{547}&=+1 \cr \epsilon_{548}&=-1 \cr \epsilon_{549}&=-1 \cr
\epsilon_{550}&=+1 \cr \epsilon_{551}&=+1 \cr \epsilon_{552}&=-1 \cr
\epsilon_{553}&=+1 \cr \epsilon_{554}&=-1 \cr \epsilon_{555}&=-1 \cr
\epsilon_{556}&=+1  \cr \epsilon_{557}&=+1 \cr \epsilon_{558}&=-1
\cr \epsilon_{559}&=-1 \cr \epsilon_{560}&=+1 \cr \epsilon_{561}&=+1
\cr \epsilon_{562}&=-1 \cr \epsilon_{563}&=-1 \cr \epsilon_{564}&=+1
\cr \epsilon_{565}&=+1 \cr \epsilon_{566}&=-1 \cr \epsilon_{567}&=-1
\cr \epsilon_{568}&=+1 \cr \epsilon_{569}&=+1 \cr \epsilon_{570}&=-1
\cr \epsilon_{571}&=-1 \cr \epsilon_{572}&=+1 \cr \epsilon_{573}&=+1
\cr \epsilon_{574}&=-1 \cr \hfill \cr } $

&& $\eqalign{ \hfill \cr \epsilon_{575}&=-1 \cr \epsilon_{576}&=+1
\cr \epsilon_{577}&=+1  \cr \epsilon_{578}&=-1 \cr
\epsilon_{579}&=-1 \cr \epsilon_{580}&=+1  \cr \epsilon_{581}&=+1
\cr \epsilon_{582}&=-1 \cr \epsilon_{583}&=-1 \cr \epsilon_{584}&=+1
\cr \epsilon_{585}&=+1  \cr \epsilon_{586}&=-1 \cr
\epsilon_{587}&=-1 \cr \epsilon_{588}&=+1  \cr \epsilon_{589}&=+1
\cr \epsilon_{590}&=-1 \cr \epsilon_{591}&=-1 \cr \epsilon_{592}&=+1
\cr \epsilon_{593}&=+1  \cr \epsilon_{594}&=-1 \cr
\epsilon_{595}&=-1 \cr \epsilon_{596}&=+1 \cr \epsilon_{597}&=+1 \cr
\epsilon_{598}&=-1 \cr \epsilon_{599}&=-1 \cr \epsilon_{600}&=+1 \cr
\epsilon_{601}&=-1 \cr \epsilon_{602}&=+1 \cr \epsilon_{603}&=+1 \cr
\epsilon_{604}&=-1 \cr \epsilon_{605}&=-1 \cr \epsilon_{606}&=+1 \cr
\epsilon_{607}&=+1  \cr \epsilon_{608}&=-1 \cr \epsilon_{609}&=-1
\cr \epsilon_{610}&=+1 \cr \epsilon_{611}&=+1 \cr \epsilon_{612}&=-1
\cr \hfill \cr } $

&& $\eqalign{ \hfill \cr \epsilon_{613}&=-1 \cr \epsilon_{614}&=+1
\cr \epsilon_{615}&=+1 \cr \epsilon_{616}&=-1 \cr \epsilon_{617}&=-1
\cr \epsilon_{618}&=+1 \cr \epsilon_{619}&=+1 \cr \epsilon_{620}&=-1
\cr \epsilon_{621}&=-1 \cr \epsilon_{622}&=+1 \cr \epsilon_{623}&=+1
\cr \epsilon_{624}&=-1 \cr \epsilon_{625}&=-1 \cr \epsilon_{626}&=+1
\cr \epsilon_{627}&=+1 \cr \epsilon_{628}&=-1 \cr \epsilon_{629}&=-1
\cr \epsilon_{630}&=+1  \cr \epsilon_{631}&=+1 \cr
\epsilon_{632}&=-1 \cr \epsilon_{633}&=-1 \cr \epsilon_{634}&=+1 \cr
\epsilon_{635}&=+1  \cr \epsilon_{636}&=-1 \cr \epsilon_{637}&=-1
\cr \epsilon_{638}&=+1  \cr \epsilon_{639}&=+1 \cr
\epsilon_{640}&=-1 \cr \epsilon_{641}&=-1 \cr \epsilon_{642}&=+1 \cr
\epsilon_{643}&=+1 \cr \epsilon_{644}&=-1 \cr \epsilon_{645}&=-1 \cr
\epsilon_{646}&=+1 \cr \epsilon_{647}&=+1 \cr \epsilon_{648}&=-1 \cr
\epsilon_{649}&=+1 \cr \epsilon_{650}&=-1 \cr \hfill \cr } $

&& $\eqalign{ \hfill \cr \epsilon_{651}&=-1 \cr \epsilon_{652}&=+1
\cr \epsilon_{653}&=+1 \cr \epsilon_{654}&=-1 \cr \epsilon_{655}&=-1
\cr \epsilon_{656}&=+1 \cr \epsilon_{657}&=+1  \cr
\epsilon_{658}&=-1 \cr \epsilon_{659}&=-1 \cr \epsilon_{660}&=+1 \cr
\epsilon_{661}&=+1 \cr \epsilon_{662}&=-1 \cr \epsilon_{663}&=-1 \cr
\epsilon_{664}&=+1 \cr \epsilon_{665}&=+1  \cr \epsilon_{666}&=-1
\cr \epsilon_{667}&=-1 \cr \epsilon_{668}&=+1 \cr \epsilon_{669}&=+1
\cr \ \epsilon_{670}&=-1 \cr \epsilon_{671}&=-1 \cr
\epsilon_{672}&=+1  \cr \epsilon_{673}&=+1 \cr \epsilon_{674}&=-1
\cr \epsilon_{675}&=-1 \cr \epsilon_{676}&=+1 \cr \epsilon_{677}&=+1
\cr \epsilon_{678}&=-1 \cr \epsilon_{679}&=-1 \cr \epsilon_{680}&=+1
\cr \epsilon_{681}&=+1  \cr \epsilon_{682}&=-1 \cr
\epsilon_{683}&=-1 \cr \epsilon_{684}&=+1 \cr \epsilon_{685}&=+1
\cr \epsilon_{686}&=-1 \cr \epsilon_{687}&=-1 \cr \epsilon_{688}&=+1
\cr  \hfill \cr } $ & \cr \noalign{\hrule} }}
$$

$$ \vbox{\tabskip=0pt \offinterlineskip
\halign to425pt{\strut#& \vrule#\tabskip=0em plus1em& \hfil#&
\vrule#& \hfil#\hfil& \vrule#& \hfil#\hfil& \vrule#& \hfil#\hfil&
\vrule#& \hfil#\hfil& \vrule#& \hfil#\hfil& \vrule#& \hfil#\hfil&
\vrule#& \hfil#\hfil& \vrule#\tabskip=0pt\cr \noalign{\hrule}

&& $\eqalign{ \hfill \cr \epsilon_{689}&=+1 \cr \epsilon_{690}&=-1
\cr \epsilon_{691}&=-1 \cr \epsilon_{692}&=+1 \cr \epsilon_{693}&=+1
\cr \epsilon_{694}&=-1 \cr \epsilon_{695}&=-1 \cr \epsilon_{696}&=+1
\cr \epsilon_{697}&=-1 \cr \epsilon_{698}&=+1 \cr \epsilon_{699}&=+1
\cr \epsilon_{700}&=-1 \cr \epsilon_{701}&=-1 \cr \epsilon_{702}&=+1
\cr \epsilon_{703}&=+1 \cr \epsilon_{704}&=-1 \cr \epsilon_{705}&=-1
\cr \epsilon_{706}&=+1  \cr \epsilon_{707}&=+1  \cr
\epsilon_{708}&=-1 \cr \epsilon_{709}&=-1 \cr \epsilon_{710}&=+1 \cr
\epsilon_{711}&=+1 \cr \epsilon_{712}&=-1 \cr \epsilon_{713}&=-1 \cr
\epsilon_{714}&=+1 \cr \epsilon_{715}&=+1  \cr \epsilon_{716}&=-1
\cr \epsilon_{717}&=-1 \cr \epsilon_{718}&=+1 \cr \epsilon_{719}&=+1
\cr \epsilon_{720}&=-1 \cr \epsilon_{721}&=+1  \cr
\epsilon_{722}&=-1 \cr \epsilon_{723}&=-1 \cr \epsilon_{724}&=+1 \cr
\epsilon_{725}&=+1  \cr \epsilon_{726}&=-1 \cr \hfill \cr } $

&& $\eqalign{ \hfill \cr \epsilon_{727}&=-1 \cr \epsilon_{728}&=+1
\cr \epsilon_{729}&=+1 \cr \epsilon_{730}&=-1 \cr \epsilon_{731}&=-1
\cr \epsilon_{732}&=+1 \cr \epsilon_{733}&=+1 \cr \epsilon_{734}&=-1
\cr \epsilon_{735}&=-1 \cr \epsilon_{736}&=+1 \cr \epsilon_{737}&=+1
\cr \epsilon_{738}&=-1 \cr \epsilon_{739}&=-1 \cr \epsilon_{740}&=+1
\cr \epsilon_{741}&=+1  \cr \epsilon_{742}&=-1 \cr
\epsilon_{743}&=-1 \cr \epsilon_{744}&=+1 \cr \epsilon_{745}&=-1 \cr
\epsilon_{746}&=+1  \cr \epsilon_{747}&=+1 \cr \epsilon_{748}&=-1
\cr \epsilon_{749}&=-1 \cr \epsilon_{750}&=+1 \cr \epsilon_{751}&=+1
\cr \epsilon_{752}&=-1 \cr \epsilon_{753}&=-1 \cr \epsilon_{754}&=+1
\cr \epsilon_{755}&=+1 \cr \epsilon_{756}&=-1 \cr \epsilon_{757}&=-1
\cr \epsilon_{758}&=+1 \cr \epsilon_{759}&=+1 \cr \epsilon_{760}&=-1
\cr \epsilon_{761}&=-1 \cr \epsilon_{762}&=+1 \cr \epsilon_{763}&=+1
\cr \epsilon_{764}&=-1 \cr \hfill \cr } $

&& $\eqalign{ \hfill \cr \epsilon_{765}&=-1 \cr \epsilon_{766}&=+1
\cr \epsilon_{767}&=+1  \cr \epsilon_{768}&=-1 \cr
\epsilon_{769}&=-1 \cr \epsilon_{770}&=+1 \cr \epsilon_{771}&=+1 \cr
\epsilon_{772}&=-1 \cr \epsilon_{773}&=-1 \cr \epsilon_{774}&=+1 \cr
\epsilon_{775}&=+1 \cr \epsilon_{776}&=-1 \cr \epsilon_{777}&=-1 \cr
\epsilon_{778}&=+1  \cr \epsilon_{779}&=+1 \cr \epsilon_{780}&=-1
\cr \epsilon_{781}&=-1 \cr \epsilon_{782}&=+1 \cr \epsilon_{783}&=+1
\cr \epsilon_{784}&=-1 \cr \epsilon_{785}&=-1 \cr \epsilon_{786}&=+1
\cr \epsilon_{787}&=+1 \cr \epsilon_{788}&=-1 \cr \epsilon_{789}&=-1
\cr \epsilon_{790}&=+1 \cr \epsilon_{791}&=+1 \cr \epsilon_{792}&=-1
\cr \epsilon_{793}&=+1 \cr \epsilon_{794}&=-1 \cr \epsilon_{795}&=-1
\cr \epsilon_{796}&=+1 \cr \epsilon_{797}&=+1 \cr \epsilon_{798}&=-1
\cr \epsilon_{799}&=-1 \cr \epsilon_{800}&=+1 \cr \epsilon_{801}&=+1
\cr \epsilon_{802}&=-1 \cr \hfill \cr } $

&& $\eqalign{ \hfill \cr \epsilon_{803}&=-1 \cr \epsilon_{804}&=+1
\cr \epsilon_{805}&=+1  \cr \epsilon_{806}&=-1 \cr
\epsilon_{807}&=-1 \cr \epsilon_{808}&=+1  \cr \epsilon_{809}&=+1
\cr \epsilon_{810}&=-1 \cr \epsilon_{811}&=-1 \cr \epsilon_{812}&=+1
\cr \epsilon_{813}&=+1  \cr \epsilon_{814}&=-1 \cr
\epsilon_{815}&=-1 \cr \epsilon_{816}&=+1 \cr \epsilon_{817}&=+1
\cr \epsilon_{818}&=-1 \cr \epsilon_{819}&=-1 \cr \epsilon_{820}&=+1
\cr \epsilon_{821}&=+1 \cr \epsilon_{822}&=-1 \cr \epsilon_{823}&=-1
\cr \epsilon_{824}&=+1 \cr \epsilon_{825}&=+1 \cr \epsilon_{826}&=-1
\cr \epsilon_{827}&=-1 \cr \epsilon_{828}&=+1 \cr \epsilon_{829}&=+1
\cr \epsilon_{830}&=-1 \cr \epsilon_{831}&=-1 \cr \epsilon_{832}&=+1
\cr \epsilon_{833}&=+1  \cr \epsilon_{834}&=-1 \cr
\epsilon_{835}&=-1 \cr \epsilon_{836}&=+1  \cr \epsilon_{837}&=+1
\cr \epsilon_{838}&=-1 \cr \epsilon_{839}&=-1 \cr \epsilon_{840}&=+1
\cr \hfill \cr } $

&& $\eqalign{ \hfill \cr \epsilon_{841}&=-1 \cr \epsilon_{842}&=+1
\cr \epsilon_{843}&=+1 \cr \epsilon_{844}&=-1 \cr \epsilon_{845}&=-1
\cr \epsilon_{846}&=+1 \cr \epsilon_{847}&=+1  \cr
\epsilon_{848}&=-1 \cr \epsilon_{849}&=-1 \cr \epsilon_{850}&=+1 \cr
\epsilon_{851}&=+1 \cr \epsilon_{852}&=-1 \cr \epsilon_{853}&=-1 \cr
\epsilon_{854}&=+1  \cr \epsilon_{855}&=+1  \cr \epsilon_{856}&=-1
\cr \epsilon_{857}&=-1 \cr \epsilon_{858}&=+1 \cr \epsilon_{859}&=+1
\cr \epsilon_{860}&=-1 \cr \epsilon_{861}&=-1 \cr \epsilon_{862}&=+1
\cr \epsilon_{863}&=+1  \cr \epsilon_{864}&=-1 \cr
\epsilon_{865}&=-1 \cr \epsilon_{866}&=+1 \cr \epsilon_{867}&=+1 \cr
\epsilon_{868}&=-1 \cr \epsilon_{869}&=-1 \cr \epsilon_{870}&=+1 \cr
\epsilon_{871}&=+1 \cr \epsilon_{872}&=-1 \cr \epsilon_{873}&=-1 \cr
\epsilon_{874}&=+1 \cr \epsilon_{875}&=+1  \cr \epsilon_{876}&=-1
\cr \epsilon_{877}&=-1 \cr \epsilon_{878}&=+1 \cr \hfill \cr } $

&& $\eqalign{ \hfill \cr \epsilon_{879}&=+1 \cr \epsilon_{880}&=-1
\cr \epsilon_{881}&=-1 \cr \epsilon_{882}&=+1 \cr \epsilon_{883}&=+1
\cr \epsilon_{884}&=-1 \cr \epsilon_{885}&=-1 \cr \epsilon_{886}&=+1
\cr \epsilon_{887}&=+1 \cr \epsilon_{888}&=-1 \cr \epsilon_{889}&=+1
\cr \epsilon_{890}&=-1 \cr \epsilon_{891}&=-1 \cr \epsilon_{892}&=+1
\cr \epsilon_{893}&=+1  \cr \epsilon_{894}&=-1 \cr
\epsilon_{895}&=-1 \cr \epsilon_{896}&=+1  \cr \epsilon_{897}&=+1
\cr \epsilon_{898}&=-1 \cr \epsilon_{899}&=-1 \cr \epsilon_{900}&=+1
\cr \epsilon_{901}&=+1  \cr \epsilon_{902}&=-1 \cr
\epsilon_{903}&=-1 \cr \epsilon_{904}&=+1  \cr \epsilon_{905}&=+1
\cr \epsilon_{906}&=-1 \cr \epsilon_{907}&=-1 \cr \epsilon_{908}&=+1
\cr \epsilon_{909}&=+1  \cr \epsilon_{910}&=-1 \cr
\epsilon_{911}&=-1 \cr \epsilon_{912}&=+1  \cr \epsilon_{913}&=-1
\cr \epsilon_{914}&=+1  \cr \epsilon_{915}&=+1  \cr
\epsilon_{916}&=-1 \cr \hfill \cr } $

&& $\eqalign{ \hfill \cr \epsilon_{917}&=-1 \cr \epsilon_{918}&=+1
\cr \epsilon_{919}&=+1 \cr \epsilon_{920}&=-1 \cr \epsilon_{921}&=-1
\cr \epsilon_{922}&=+1 \cr \epsilon_{923}&=+1 \cr \epsilon_{924}&=-1
\cr \epsilon_{925}&=-1 \cr \epsilon_{926}&=+1 \cr \epsilon_{927}&=+1
\cr \epsilon_{928}&=-1 \cr \epsilon_{929}&=-1 \cr \epsilon_{930}&=+1
\cr \epsilon_{931}&=+1  \cr \epsilon_{932}&=-1 \cr
\epsilon_{933}&=-1 \cr \epsilon_{934}&=+1 \cr \epsilon_{935}&=+1 \cr
\epsilon_{936}&=-1 \cr \epsilon_{937}&=+1 \cr \epsilon_{938}&=-1 \cr
\epsilon_{939}&=-1 \cr \epsilon_{940}&=+1 \cr \epsilon_{941}&=+1
\cr \epsilon_{942}&=-1 \cr \epsilon_{943}&=-1 \cr \epsilon_{944}&=+1
\cr \epsilon_{945}&=+1 \cr \epsilon_{946}&=-1 \cr \epsilon_{947}&=-1
\cr \epsilon_{948}&=+1 \cr \epsilon_{949}&=+1  \cr
\epsilon_{950}&=-1 \cr \epsilon_{951}&=-1 \cr \epsilon_{952}&=+1
\cr \epsilon_{953}&=+1 \cr \epsilon_{954}&=-1 \cr  \hfill \cr } $

&& $\eqalign{ \hfill \cr \epsilon_{955}&=-1 \cr \epsilon_{956}&=+1
\cr \epsilon_{957}&=+1  \cr \epsilon_{958}&=-1 \cr
\epsilon_{959}&=-1 \cr \epsilon_{960}&=+1  \cr \epsilon_{961}&=+1
\cr \epsilon_{962}&=-1 \cr \epsilon_{963}&=-1 \cr \epsilon_{964}&=+1
\cr \epsilon_{965}&=+1  \cr \epsilon_{966}&=-1 \cr
\epsilon_{967}&=-1 \cr \epsilon_{968}&=+1  \cr \epsilon_{969}&=+1
\cr \epsilon_{970}&=-1 \cr \epsilon_{971}&=-1 \cr \epsilon_{972}&=+1
\cr \epsilon_{973}&=+1  \cr \epsilon_{974}&=-1 \cr
\epsilon_{975}&=-1 \cr \epsilon_{976}&=+1  \cr \epsilon_{977}&=+1
\cr \epsilon_{978}&=-1 \cr \epsilon_{979}&=-1 \cr \epsilon_{980}&=+1
\cr \epsilon_{981}&=+1  \cr \epsilon_{982}&=-1 \cr
\epsilon_{983}&=-1 \cr \epsilon_{984}&=+1  \cr \epsilon_{985}&=-1
\cr \epsilon_{986}&=+1  \cr \epsilon_{987}&=+1  \cr
\epsilon_{988}&=-1 \cr \epsilon_{989}&=-1 \cr \epsilon_{990}&=+1 \cr
\epsilon_{991}&=+1  \cr \epsilon_{992}&=-1 \cr \hfill \cr } $ & \cr
\noalign{\hrule} }}
$$

$$ \vbox{\tabskip=0pt \offinterlineskip
\halign to415pt{\strut#& \vrule#\tabskip=0em plus1em& \hfil#&
\vrule#& \hfil#\hfil& \vrule#& \hfil#\hfil& \vrule#& \hfil#\hfil&
\vrule#& \hfil#\hfil& \vrule#& \hfil#\hfil& \vrule#& \hfil#\hfil&
\vrule#\tabskip=0pt\cr \noalign{\hrule}

&& $\eqalign{ \hfill \cr \epsilon_{993}&=-1 \cr \epsilon_{994}&=+1
\cr \epsilon_{995}&=+1 \cr \epsilon_{996}&=-1 \cr \epsilon_{997}&=-1
\cr \epsilon_{998}&=+1 \cr \epsilon_{999}&=+1 \cr
\epsilon_{1000}&=-1 \cr \epsilon_{1001}&=-1 \cr \epsilon_{1002}&=+1
\cr \epsilon_{1003}&=+1 \cr \epsilon_{1004}&=-1 \cr
\epsilon_{1005}&=-1 \cr \epsilon_{1006}&=+1 \cr \epsilon_{1007}&=+1
\cr \epsilon_{1008}&=-1 \cr \epsilon_{1009}&=-1 \cr
\epsilon_{1010}&=+1 \cr \epsilon_{1011}&=+1 \cr \epsilon_{1012}&=-1
\cr \epsilon_{1013}&=-1 \cr \epsilon_{1014}&=+1 \cr
\epsilon_{1015}&=+1 \cr \hfill \cr } $

&& $\eqalign{ \hfill \cr \epsilon_{1016}&=-1 \cr \epsilon_{1017}&=-1
\cr \epsilon_{1018}&=+1 \cr \epsilon_{1019}&=+1 \cr
\epsilon_{1020}&=-1 \cr \epsilon_{1021}&=-1 \cr \epsilon_{1022}&=+1
\cr \epsilon_{1023}&=+1 \cr \epsilon_{1024}&=-1 \cr
\epsilon_{1025}&=-1 \cr \epsilon_{1026}&=+1 \cr \epsilon_{1027}&=+1
\cr \epsilon_{1028}&=-1 \cr \epsilon_{1029}&=-1 \cr
\epsilon_{1030}&=+1 \cr \epsilon_{1031}&=+1 \cr \epsilon_{1032}&=-1
\cr \epsilon_{1033}&=+1 \cr \epsilon_{1034}&=-1 \cr
\epsilon_{1035}&=-1 \cr \epsilon_{1036}&=+1 \cr \epsilon_{1037}&=+1
\cr \epsilon_{1038}&=-1 \cr \hfill \cr } $

&& $\eqalign{ \hfill \cr \epsilon_{1039}&=-1 \cr \epsilon_{1040}&=+1
\cr \epsilon_{1041}&=+1 \cr \epsilon_{1042}&=-1 \cr
\epsilon_{1043}&=-1 \cr \epsilon_{1044}&=+1 \cr \epsilon_{1045}&=+1
\cr \epsilon_{1046}&=-1 \cr \epsilon_{1047}&=-1 \cr
\epsilon_{1048}&=+1 \cr \epsilon_{1049}&=+1 \cr \epsilon_{1050}&=-1
\cr \epsilon_{1051}&=-1 \cr \epsilon_{1052}&=+1 \cr
\epsilon_{1053}&=+1 \cr \epsilon_{1054}&=-1 \cr \epsilon_{1055}&=-1
\cr \epsilon_{1056}&=+1 \cr \epsilon_{1057}&=+1 \cr
\epsilon_{1058}&=-1 \cr \epsilon_{1059}&=-1 \cr \epsilon_{1060}&=+1
\cr \epsilon_{1061}&=+1 \cr
 \hfill \cr } $

&& $\eqalign{ \hfill \cr \epsilon_{1062}&=-1 \cr \epsilon_{1063}&=-1
\cr \epsilon_{1064}&=+1 \cr \epsilon_{1065}&=+1 \cr
\epsilon_{1066}&=-1 \cr \epsilon_{1067}&=-1 \cr \epsilon_{1068}&=+1
\cr \epsilon_{1069}&=+1 \cr \epsilon_{1070}&=-1 \cr
\epsilon_{1071}&=-1 \cr \epsilon_{1072}&=+1 \cr \epsilon_{1073}&=+1
\cr \epsilon_{1074}&=-1 \cr \epsilon_{1075}&=-1 \cr
\epsilon_{1076}&=+1 \cr \epsilon_{1077}&=+1 \cr \epsilon_{1078}&=-1
\cr \epsilon_{1079}&=-1 \cr \epsilon_{1080}&=+1 \cr
\epsilon_{1081}&=-1 \cr \epsilon_{1082}&=+1 \cr \epsilon_{1083}&=+1
\cr \epsilon_{1084}&=-1 \cr \hfill \cr } $

&& $\eqalign{ \hfill \cr \epsilon_{1085}&=-1 \cr \epsilon_{1086}&=+1
\cr \epsilon_{1087}&=+1 \cr \epsilon_{1088}&=-1 \cr
\epsilon_{1089}&=-1 \cr \epsilon_{1090}&=+1 \cr \epsilon_{1091}&=+1
\cr \epsilon_{1092}&=-1 \cr \epsilon_{1093}&=-1 \cr
\epsilon_{1094}&=+1 \cr \epsilon_{1095}&=+1 \cr \epsilon_{1096}&=-1
\cr \epsilon_{1097}&=-1 \cr \epsilon_{1098}&=+1 \cr
\epsilon_{1099}&=+1 \cr \epsilon_{1100}&=-1 \cr \epsilon_{1101}&=-1
\cr \epsilon_{1102}&=+1 \cr \epsilon_{1103}&=+1 \cr
\epsilon_{1104}&=-1 \cr \epsilon_{1105}&=-1 \cr \epsilon_{1106}&=+1
\cr \epsilon_{1107}&=+1 \cr \hfill \cr } $

&& $\eqalign{ \hfill \cr \epsilon_{1108}&=-1 \cr \epsilon_{1109}&=-1
\cr \epsilon_{1110}&=+1 \cr \epsilon_{1111}&=+1 \cr
\epsilon_{1112}&=-1 \cr \epsilon_{1113}&=-1 \cr \epsilon_{1114}&=+1
\cr \epsilon_{1115}&=+1 \cr \epsilon_{1116}&=-1 \cr
\epsilon_{1117}&=-1 \cr \epsilon_{1118}&=+1 \cr \epsilon_{1119}&=+1
\cr \epsilon_{1120}&=-1 \cr \epsilon_{1121}&=-1 \cr
\epsilon_{1122}&=+1 \cr \epsilon_{1123}&=+1 \cr \epsilon_{1124}&=-1
\cr \epsilon_{1125}&=-1 \cr \epsilon_{1126}&=+1 \cr
\epsilon_{1127}&=+1 \cr \epsilon_{1128}&=-1 \cr \epsilon_{1129}&=+1
\cr \epsilon_{1130}&=-1 \cr \hfill \cr } $

&& $\eqalign{ \hfill \cr  \epsilon_{1131}&=-1 \cr
\epsilon_{1132}&=+1 \cr \epsilon_{1133}&=+1 \cr \epsilon_{1134}&=-1
\cr \epsilon_{1135}&=-1 \cr \epsilon_{1136}&=+1 \cr
\epsilon_{1137}&=+1 \cr \epsilon_{1138}&=-1 \cr \epsilon_{1139}&=-1
\cr \epsilon_{1140}&=+1 \cr \epsilon_{1141}&=+1 \cr
\epsilon_{1142}&=-1 \cr \epsilon_{1143}&=-1 \cr \epsilon_{1144}&=+1
\cr \epsilon_{1145}&=+1 \cr \epsilon_{1146}&=-1 \cr
\epsilon_{1147}&=-1 \cr \epsilon_{1148}&=+1 \cr \epsilon_{1149}&=+1
\cr \epsilon_{1150}&=-1 \cr \epsilon_{1151}&=-1 \cr
\epsilon_{1152}&=+1 \cr \hfill \cr \hfill \cr } $ & \cr
\noalign{\hrule} }}
$$
\centerline{Table-III}

\noindent The signatures, given in this Table-III, can be easily
found to be true by comparing two expressions (II.1) and (II.3) for
the case in hand.

Let us now choose the most general specialization as in (I.5) :
$$ e^{\alpha_1} = u_1 \ \ , \ \ e^{\alpha_2} = u_2  \ \ ,
\ \ e^{\alpha_3} =u_3 \ \ , \ \ e^{\alpha_4} = u_4  \ \ . $$
\noindent As explained in sec.I, corresponding character polynomial
for an irreducible $F_4$ representation originated from a dominant
weight \ $\Lambda^+ = s_1  \lambda_1 \ + \ s_2  \lambda_2 + \ s_3
\lambda_3 + \ s_4 \lambda_4 $ will be defined in the following form:
$$ R(u_1,u_2,u_3,u_4,s_1,s_2,s_3,s_4) \equiv { P(u_1,u_2,u_3,u_4,s_1,s_2,s_3,s_4) \over
P(u_1,u_2,u_3,u_4,0,0,0,0) } \eqno(III.4) $$  \

\noindent In this most general specialization, we find inconvenient
to give $ P(u_1,u_2,u_3,u_4,s_1,s_2,s_3,s_4)$ here. It is however
given { \bf [7]} in our website.

To exemplify our result, the following moderated specialization will
however be the more suitable one :
$$ e^{\alpha_1} = x \ \ , \ \ e^{\alpha_2} = x  \ \ ,
\ \ e^{\alpha_3} = y  \ \ , \ \ e^{\alpha_4} = y  \ \ \eqno(III.5)
$$
\noindent for which (III.4) turns out to be following form :

$$ P(x,y,s_1,s_2,s_3,s_4)=P^+(x,y,s_1,s_2,s_3,s_4)+
P^-(x,y,s_1,s_2,s_3,s_4) \eqno(III.6) $$

\noindent where
$$ P^-(x,y,s_1,s_2,s_3,s_4) =P^+(1/x,1/y,s_1,s_2,s_3,s_4) \eqno(III.7) $$
\noindent and

$$\eqalign{&P^+(x,y,s_1,s_2,s_3,s_4)=x^{-2 s_1-s_2-s_3-s_4} y^{-13-4
s_1-6 s_2-5 s_3-3 s_4} ( \cr &-x^{4 s_1+s_4} y^{4 s_1+s_4}
(-1+y^{2+s_3+s_4}) (-1+y^{9+2 s_1+4 s_2+2 s_3+s_4}) (-1+y^{11+2
s_1+4 s_2+3 s_3+2 s_4}) \cr &+x^{3 s_1+2 s_4} y^{8+4 s_1+2 s_2+3
s_3+4 s_4} (-1+y^{5+2 s_1+2 s_2+s_3}) (-1+y^{4+2 s_2+s_3+s_4})
(-1+y^{9+2 s_1+4 s_2+2 s_3+s_4}) \cr &-x^{1+4 s_1+2 s_2} y^{1+4
s_1+2 s_2} (-1+y^{3+2 s_2+s_3}) (-1+y^{7+2 s_1+2 s_2+2 s_3+s_4})
(-1+y^{10+2 s_1+4 s_2+3 s_3+s_4}) \cr &-x^{5+4 s_1+4 s_2+2 s_3}
y^{8+4 s_1+6 s_2+3 s_3} (-1+y^{1+s_3}) (-1+y^{6+2 s_1+2
s_2+s_3+s_4}) (-1+y^{7+2 s_1+2 s_2+2 s_3+s_4}) \cr &+x^{4+3 s_1+4
s_2+2 s_3} y^{8+4 s_1+6 s_2+3 s_3} (-1+y^{1+s_3}) (-1+y^{4+2
s_2+s_3+s_4}) (-1+y^{5+2 s_2+2 s_3+s_4}) \cr &+x^{3+4 s_1+2 s_2+2
s_3} y^{4+4 s_1+2 s_2+3 s_3} (-1+y^{3+2 s_2+s_3}) (-1+y^{6+2 s_1+2
s_2+s_3+s_4}) (-1+y^{9+2 s_1+4 s_2+2 s_3+s_4}) \cr &+x^{2+4 s_1+2
s_2+s_4} y^{2+4 s_1+2 s_2+s_4} (-1+y^{4+2 s_2+s_3+s_4}) (-1+y^{7+2
s_1+2 s_2+2 s_3+s_4}) \times \cr &(-1+y^{11+2 s_1+4 s_2+3 s_3+2
s_4})-x^{2+3 s_1+2 s_3+2 s_4} y^{9+4 s_1+2 s_2+4 s_3+4 s_4}
(-1+y^{5+2 s_1+2 s_2+s_3}) \times \cr &(-1+y^{5+2 s_2+2 s_3+s_4})
(-1+y^{10+2 s_1+4 s_2+3 s_3+s_4}) +x^{1+4 s_1+s_3+s_4} y^{2+4 s_1+2
s_3+s_4} (-1+y^{1+s_4}) \times \cr &(-1+y^{10+2 s_1+4 s_2+3
s_3+s_4}) (-1+y^{11+2 s_1+4 s_2+3 s_3+2 s_4})-x^{1+3 s_1+s_2+2 s_3}
y^{4+4 s_1+2 s_2+3 s_3} \times \cr &(-1+y^{3+2 s_2+s_3})
(-1+y^{2+s_3+s_4}) (-1+y^{5+2 s_2+2 s_3+s_4}) -x^{1+3 s_1+s_2+2 s_4}
y^{12+4 s_1+6 s_2+3 s_3+4 s_4} \times \cr &(-1+y^{5+2 s_1+2
s_2+s_3}) (-1+y^{2+s_3+s_4}) (-1+y^{7+2 s_1+2 s_2+2 s_3+s_4})+x^{1+3
s_1+2 s_3+s_4} y^{6+4 s_1+2 s_2+4 s_3+s_4} \times \cr &(-1+y^{6+2
s_1+2 s_2+s_3+s_4}) (-1+y^{5+2 s_2+2 s_3+s_4}) (-1+y^{11+2 s_1+4
s_2+3 s_3+2 s_4}) }
$$

$$
\eqalign{&-x^{1+2 s_2+2 s_3+2 s_4} y^{7+4 s_2+4 s_3+4 s_4}
(-1+y^{1+s_3}) (-1+y^{9+2 s_1+4 s_2+2 s_3+s_4}) (-1+y^{10+2 s_1+4
s_2+3 s_3+s_4}) \cr &+x^{23+7 s_1+10 s_2+7 s_3+4 s_4} y^{41+10
s_1+18 s_2+12 s_3+6 s_4} (-1+y^{1+s_3}) (-1+y^{1+s_4})
(-1+y^{2+s_3+s_4}) \cr &-x^{22+7 s_1+9 s_2+7 s_3+4 s_4} y^{37+10
s_1+14 s_2+12 s_3+6 s_4} (-1+y^{3+2 s_2+s_3}) (-1+y^{1+s_4})
(-1+y^{4+2 s_2+s_3+s_4}) \cr & -x^{22+6 s_1+10 s_2+7 s_3+4 s_4}
y^{41+10 s_1+18 s_2+12 s_3+6 s_4} (-1+y^{1+s_3}) (-1+y^{1+s_4})
(-1+y^{2+s_3+s_4}) \cr & +x^{21+7 s_1+9 s_2+6 s_3+4 s_4} y^{34+10
s_1+14 s_2+9 s_3+6 s_4} (-1+y^{3+2 s_2+s_3}) (-1+y^{2+s_3+s_4})
(-1+y^{5+2 s_2+2 s_3+s_4}) \cr &-x^{20+7 s_1+9 s_2+6 s_3+3 s_4}
y^{32+10 s_1+14 s_2+9 s_3+4 s_4} (-1+y^{1+s_3}) (-1+y^{4+2
s_2+s_3+s_4}) (-1+y^{5+2 s_2+2 s_3+s_4}) \cr &-x^{20+7 s_1+8 s_2+6
s_3+4 s_4} y^{32+10 s_1+12 s_2+9 s_3+6 s_4} (-1+y^{1+s_3})
(-1+y^{4+2 s_2+s_3+s_4}) (-1+y^{5+2 s_2+2 s_3+s_4}) \cr &+x^{20+6
s_1+8 s_2+7 s_3+4 s_4} y^{37+10 s_1+14 s_2+12 s_3+6 s_4} (-1+y^{3+2
s_2+s_3}) (-1+y^{1+s_4}) (-1+y^{4+2 s_2+s_3+s_4}) \cr &+x^{20+5
s_1+9 s_2+7 s_3+4 s_4} y^{33+6 s_1+14 s_2+12 s_3+6 s_4} (-1+y^{5+2
s_1+2 s_2+s_3}) (-1+y^{1+s_4}) \times \cr &(-1+y^{6+2 s_1+2
s_2+s_3+s_4}) +x^{19+7 s_1+8 s_2+6 s_3+3 s_4} y^{30+10 s_1+12 s_2+9
s_3+4 s_4} (-1+y^{3+2 s_2+s_3}) \times \cr &(-1+y^{2+s_3+s_4})
(-1+y^{5+2 s_2+2 s_3+s_4}) -x^{19+5 s_1+9 s_2+6 s_3+4 s_4} y^{30+6
s_1+14 s_2+9 s_3+6 s_4} \times \cr &(-1+y^{5+2 s_1+2 s_2+s_3})
(-1+y^{2+s_3+s_4}) (-1+y^{7+2 s_1+2 s_2+2 s_3+s_4}) -x^{19+5 s_1+8
s_2+7 s_3+4 s_4} \times \cr &y^{33+6 s_1+14 s_2+12 s_3+6 s_4}
(-1+y^{5+2 s_1+2 s_2+s_3}) (-1+y^{1+s_4}) (-1+y^{6+2 s_1+2
s_2+s_3+s_4}) \cr &-x^{18+7 s_1+8 s_2+5 s_3+3 s_4} y^{29+10 s_1+12
s_2+8 s_3+4 s_4} (-1+y^{3+2 s_2+s_3}) (-1+y^{1+s_4}) (-1+y^{4+2
s_2+s_3+s_4}) \cr &-x^{18+6 s_1+8 s_2+5 s_3+4 s_4} y^{34+10 s_1+14
s_2+9 s_3+6 s_4} (-1+y^{3+2 s_2+s_3}) (-1+y^{2+s_3+s_4}) (-1+y^{5+2
s_2+2 s_3+s_4}) \cr &+x^{18+5 s_1+9 s_2+6 s_3+3 s_4} y^{28+6 s_1+14
s_2+9 s_3+4 s_4} (-1+y^{1+s_3}) (-1+y^{6+2 s_1+2 s_2+s_3+s_4})
\times \cr &(-1+y^{7+2 s_1+2 s_2+2 s_3+s_4}) +x^{17+7 s_1+7 s_2+5
s_3+3 s_4} y^{29+10 s_1+12 s_2+8 s_3+4 s_4} (-1+y^{1+s_3})
(-1+y^{1+s_4}) \times \cr &(-1+y^{2+s_3+s_4}) +x^{17+5 s_1+8 s_2+5
s_3+4 s_4} y^{30+6 s_1+14 s_2+9 s_3+6 s_4} (-1+y^{5+2 s_1+2
s_2+s_3}) (-1+y^{2+s_3+s_4}) \times \cr &(-1+y^{7+2 s_1+2 s_2+2
s_3+s_4}) +x^{17+4 s_1+8 s_2+6 s_3+4 s_4} y^{26+4 s_1+12 s_2+9 s_3+6
s_4} (-1+y^{1+s_3}) \times \cr &(-1+y^{6+2 s_1+2 s_2+s_3+s_4})
(-1+y^{7+2 s_1+2 s_2+2 s_3+s_4}) +x^{16+6 s_1+8 s_2+5 s_3+2 s_4}
y^{32+10 s_1+14 s_2+9 s_3+4 s_4} \times \cr &(-1+y^{1+s_3})
(-1+y^{4+2 s_2+s_3+s_4}) (-1+y^{5+2 s_2+2 s_3+s_4})+x^{16+6 s_1+6
s_2+5 s_3+4 s_4} \times \cr &y^{32+10 s_1+12 s_2+9 s_3+6 s_4}
(-1+y^{1+s_3}) (-1+y^{4+2 s_2+s_3+s_4}) (-1+y^{5+2 s_2+2 s_3+s_4})
\cr &+x^{16+5 s_1+6 s_2+6 s_3+4 s_4} y^{24+6 s_1+8 s_2+9 s_3+6 s_4}
(-1+y^{5+2 s_1+2 s_2+s_3}) (-1+y^{4+2 s_2+s_3+s_4}) \times \cr
&(-1+y^{9+2 s_1+4 s_2+2 s_3+s_4}) -x^{16+4 s_1+8 s_2+6 s_3+3 s_4}
y^{24+4 s_1+12 s_2+9 s_3+4 s_4} (-1+y^{5+2 s_1+2 s_2+s_3}) \times
\cr &(-1+y^{2+s_3+s_4}) (-1+y^{7+2 s_1+2 s_2+2 s_3+s_4})-x^{15+5
s_1+8 s_2+5 s_3+2 s_4} y^{28+6 s_1+14 s_2+9 s_3+4 s_4}
(-1+y^{1+s_3}) \times \cr &(-1+y^{6+2 s_1+2 s_2+s_3+s_4}) (-1+y^{7+2
s_1+2 s_2+2 s_3+s_4}) -x^{15+5 s_1+6 s_2+6 s_3+3 s_4} y^{22+6 s_1+8
s_2+9 s_3+4 s_4} \times \cr &(-1+y^{3+2 s_2+s_3}) (-1+y^{6+2 s_1+2
s_2+s_3+s_4}) (-1+y^{9+2 s_1+4 s_2+2 s_3+s_4}) +x^{15+4 s_1+8 s_2+5
s_3+3 s_4} \times \cr &y^{23+4 s_1+12 s_2+8 s_3+4 s_4} (-1+y^{5+2
s_1+2 s_2+s_3}) (-1+y^{1+s_4}) (-1+y^{6+2 s_1+2 s_2+s_3+s_4})\cr
&-x^{15+4 s_1+6 s_2+6 s_3+4 s_4} y^{22+4 s_1+8 s_2+9 s_3+6 s_4}
(-1+y^{3+2 s_2+s_3}) (-1+y^{6+2 s_1+2 s_2+s_3+s_4}) \times \cr
&(-1+y^{9+2 s_1+4 s_2+2 s_3+s_4})-x^{14+6 s_1+6 s_2+5 s_3+2 s_4}
y^{30+10 s_1+12 s_2+9 s_3+4 s_4} (-1+y^{3+2 s_2+s_3}) \times \cr &
(-1+y^{2+s_3+s_4}) (-1+y^{5+2 s_2+2 s_3+s_4})   }
$$

$$
\eqalign{&-x^{14+5 s_1+6 s_2+4 s_3+4 s_4} y^{21+6 s_1+8 s_2+6 s_3+6
s_4} (-1+y^{5+2 s_1+2 s_2+s_3}) (-1+y^{5+2 s_2+2 s_3+s_4}) \times
\cr &(-1+y^{10+2 s_1+4 s_2+3 s_3+s_4}) -x^{14+5 s_1+5 s_2+5 s_3+4
s_4} y^{24+6 s_1+8 s_2+9 s_3+6 s_4} (-1+y^{5+2 s_1+2 s_2+s_3})
\times \cr &(-1+y^{4+2 s_2+s_3+s_4}) (-1+y^{9+2 s_1+4 s_2+2
s_3+s_4}) +x^{14+4 s_1+6 s_2+6 s_3+3 s_4} y^{20+4 s_1+8 s_2+9 s_3+4
s_4} \times \cr &(-1+y^{5+2 s_1+2 s_2+s_3}) (-1+y^{4+2 s_2+s_3+s_4})
(-1+y^{9+2 s_1+4 s_2+2 s_3+s_4}) +x^{13+5 s_1+5 s_2+4 s_3+4 s_4}
\times \cr &y^{21+6 s_1+8 s_2+6 s_3+6 s_4} (-1+y^{5+2 s_1+2
s_2+s_3}) (-1+y^{5+2 s_2+2 s_3+s_4}) (-1+y^{10+2 s_1+4 s_2+3
s_3+s_4}) \cr &+x^{13+4 s_1+6 s_2+4 s_3+4 s_4} y^{19+4 s_1+8 s_2+6
s_3+6 s_4} (-1+y^{3+2 s_2+s_3}) (-1+y^{7+2 s_1+2 s_2+2 s_3+s_4})
\times \cr &(-1+y^{10+2 s_1+4 s_2+3 s_3+s_4}) -x^{13+3 s_1+7 s_2+5
s_3+3 s_4} y^{23+4 s_1+12 s_2+8 s_3+4 s_4} (-1+y^{1+s_3})
(-1+y^{1+s_4}) \times \cr &(-1+y^{2+s_3+s_4}) -x^{13+3 s_1+6 s_2+5
s_3+4 s_4} y^{26+4 s_1+12 s_2+9 s_3+6 s_4} (-1+y^{1+s_3}) (-1+y^{6+2
s_1+2 s_2+s_3+s_4}) \times \cr &(-1+y^{7+2 s_1+2 s_2+2 s_3+s_4})
+x^{12+6 s_1+6 s_2+3 s_3+2 s_4} y^{29+10 s_1+12 s_2+8 s_3+4 s_4}
(-1+y^{3+2 s_2+s_3}) \times \cr &(-1+y^{1+s_4}) (-1+y^{4+2
s_2+s_3+s_4}) +x^{12+5 s_1+6 s_2+3 s_3+3 s_4} y^{17+6 s_1+8 s_2+4
s_3+4 s_4} (-1+y^{3+2 s_2+s_3}) \times \cr &(-1+y^{7+2 s_1+2 s_2+2
s_3+s_4}) (-1+y^{10+2 s_1+4 s_2+3 s_3+s_4}) +x^{12+5 s_1+5 s_2+5
s_3+2 s_4} y^{22+6 s_1+8 s_2+9 s_3+4 s_4} \times \cr &(-1+y^{3+2
s_2+s_3}) (-1+y^{6+2 s_1+2 s_2+s_3+s_4}) (-1+y^{9+2 s_1+4 s_2+2
s_3+s_4}) +x^{12+3 s_1+5 s_2+5 s_3+4 s_4} \times \cr &y^{22+4 s_1+8
s_2+9 s_3+6 s_4} (-1+y^{3+2 s_2+s_3}) (-1+y^{6+2 s_1+2 s_2+s_3+s_4})
(-1+y^{9+2 s_1+4 s_2+2 s_3+s_4}) \cr &+x^{11+5 s_1+6 s_2+4 s_3+s_4}
y^{16+6 s_1+8 s_2+6 s_3+s_4} (-1+y^{6+2 s_1+2 s_2+s_3+s_4})
(-1+y^{5+2 s_2+2 s_3+s_4}) \times \cr &(-1+y^{11+2 s_1+4 s_2+3 s_3+2
s_4}) -x^{11+4 s_1+6 s_2+3 s_3+3 s_4} y^{15+4 s_1+8 s_2+4 s_3+4 s_4}
(-1+y^{5+2 s_1+2 s_2+s_3}) \times \cr &(-1+y^{5+2 s_2+2 s_3+s_4})
(-1+y^{10+2 s_1+4 s_2+3 s_3+s_4}) -x^{11+4 s_1+4 s_2+5 s_3+3 s_4}
y^{17+4 s_1+6 s_2+8 s_3+4 s_4} \times \cr &(-1+y^{5+2 s_1+2
s_2+s_3}) (-1+y^{1+s_4}) (-1+y^{6+2 s_1+2 s_2+s_3+s_4}) -x^{11+4
s_1+4 s_2+4 s_3+4 s_4} \times \cr &y^{17+4 s_1+6 s_2+6 s_3+6 s_4}
(-1+y^{1+s_3}) (-1+y^{9+2 s_1+4 s_2+2 s_3+s_4}) (-1+y^{10+2 s_1+4
s_2+3 s_3+s_4}) \cr &+x^{11+3 s_1+6 s_2+5 s_3+2 s_4} y^{24+4 s_1+12
s_2+9 s_3+4 s_4} (-1+y^{5+2 s_1+2 s_2+s_3}) (-1+y^{2+s_3+s_4})
\times \cr &(-1+y^{7+2 s_1+2 s_2+2 s_3+s_4}) -x^{11+3 s_1+5 s_2+4
s_3+4 s_4} y^{19+4 s_1+8 s_2+6 s_3+6 s_4} (-1+y^{3+2 s_2+s_3})
\times \cr &(-1+y^{7+2 s_1+2 s_2+2 s_3+s_4}) (-1+y^{10+2 s_1+4 s_2+3
s_3+s_4}) -x^{10+6 s_1+4 s_2+3 s_3+2 s_4} y^{29+10 s_1+12 s_2+8
s_3+4 s_4} \times \cr &(-1+y^{1+s_3}) (-1+y^{1+s_4})
(-1+y^{2+s_3+s_4}) -x^{10+5 s_1+6 s_2+3 s_3+s_4} y^{14+6 s_1+8 s_2+4
s_3+s_4} \times \cr &(-1+y^{4+2 s_2+s_3+s_4}) (-1+y^{7+2 s_1+2 s_2+2
s_3+s_4}) (-1+y^{11+2 s_1+4 s_2+3 s_3+2 s_4}) -x^{10+5 s_1+5 s_2+4
s_3+s_4} \times \cr &y^{16+6 s_1+8 s_2+6 s_3+s_4} (-1+y^{6+2 s_1+2
s_2+s_3+s_4}) (-1+y^{5+2 s_2+2 s_3+s_4}) (-1+y^{11+2 s_1+4 s_2+3
s_3+2 s_4}) \cr &-x^{10+4 s_1+6 s_2+4 s_3+s_4} y^{14+4 s_1+8 s_2+6
s_3+s_4} (-1+y^{4+2 s_2+s_3+s_4}) (-1+y^{7+2 s_1+2 s_2+2 s_3+s_4})
\times \cr &(-1+y^{11+2 s_1+4 s_2+3 s_3+2 s_4}) -x^{10+3 s_1+5 s_2+5
s_3+2 s_4} y^{20+4 s_1+8 s_2+9 s_3+4 s_4} (-1+y^{5+2 s_1+2 s_2+s_3})
\times \cr &(-1+y^{4+2 s_2+s_3+s_4}) (-1+y^{9+2 s_1+4 s_2+2
s_3+s_4}) +x^{10+3 s_1+4 s_2+5 s_3+3 s_4} y^{17+4 s_1+6 s_2+8 s_3+4
s_4} \times \cr &(-1+y^{3+2 s_2+s_3}) (-1+y^{1+s_4}) (-1+y^{4+2
s_2+s_3+s_4})+x^{10+3 s_1+4 s_2+4 s_3+4 s_4} y^{17+4 s_1+6 s_2+6
s_3+6 s_4} \times \cr &(-1+y^{1+s_3}) (-1+y^{9+2 s_1+4 s_2+2
s_3+s_4}) (-1+y^{10+2 s_1+4 s_2+3 s_3+s_4})-x^{9+5 s_1+5 s_2+2 s_3+2
s_4} \times \cr &y^{17+6 s_1+8 s_2+4 s_3+4 s_4} (-1+y^{3+2 s_2+s_3})
(-1+y^{7+2 s_1+2 s_2+2 s_3+s_4}) (-1+y^{10+2 s_1+4 s_2+3 s_3+s_4})
\cr &-x^{9+5 s_1+3 s_2+3 s_3+3 s_4} y^{13+6 s_1+4 s_2+4 s_3+4 s_4}
(-1+y^{1+s_3}) (-1+y^{9+2 s_1+4 s_2+2 s_3+s_4}) \times \cr
&(-1+y^{10+2 s_1+4 s_2+3 s_3+s_4}) }
$$

$$
\eqalign{&+x^{9+4 s_1+6 s_2+3 s_3+s_4} y^{12+4 s_1+8 s_2+4 s_3+s_4}
(-1+y^{6+2 s_1+2 s_2+s_3+s_4}) \times \cr &(-1+y^{5+2 s_2+2
s_3+s_4}) (-1+y^{11+2 s_1+4 s_2+3 s_3+2 s_4}) -x^{9+3 s_1+6 s_2+3
s_3+2 s_4} y^{23+4 s_1+12 s_2+8 s_3+4 s_4} \times \cr &(-1+y^{5+2
s_1+2 s_2+s_3}) (-1+y^{1+s_4}) (-1+y^{6+2 s_1+2 s_2+s_3+s_4})
+x^{8+5 s_1+5 s_2+2 s_3+s_4} y^{14+6 s_1+8 s_2+4 s_3+s_4} \times \cr
&(-1+y^{4+2 s_2+s_3+s_4}) (-1+y^{7+2 s_1+2 s_2+2 s_3+s_4})
(-1+y^{11+2 s_1+4 s_2+3 s_3+2 s_4}) +x^{8+4 s_1+4 s_2+4 s_3+s_4}
\times \cr &y^{12+4 s_1+6 s_2+6 s_3+s_4} (-1+y^{2+s_3+s_4})
(-1+y^{9+2 s_1+4 s_2+2 s_3+s_4}) (-1+y^{11+2 s_1+4 s_2+3 s_3+2 s_4})
\cr &+x^{8+4 s_1+4 s_2+2 s_3+3 s_4} y^{12+4 s_1+6 s_2+3 s_3+4 s_4}
(-1+y^{5+2 s_1+2 s_2+s_3}) (-1+y^{2+s_3+s_4}) \times \cr &(-1+y^{7+2
s_1+2 s_2+2 s_3+s_4}) +x^{8+3 s_1+5 s_2+4 s_3+s_4} y^{14+4 s_1+8
s_2+6 s_3+s_4} (-1+y^{4+2 s_2+s_3+s_4}) \times \cr &(-1+y^{7+2 s_1+2
s_2+2 s_3+s_4}) (-1+y^{11+2 s_1+4 s_2+3 s_3+2 s_4}) +x^{7+5 s_1+3
s_2+3 s_3+s_4} y^{10+6 s_1+4 s_2+4 s_3+s_4} \times \cr
&(-1+y^{2+s_3+s_4}) (-1+y^{9+2 s_1+4 s_2+2 s_3+s_4}) (-1+y^{11+2
s_1+4 s_2+3 s_3+2 s_4}) +x^{7+4 s_1+2 s_2+3 s_3+3 s_4} \times \cr
&y^{9+4 s_1+2 s_2+4 s_3+4 s_4} (-1+y^{5+2 s_1+2 s_2+s_3}) (-1+y^{5+2
s_2+2 s_3+s_4}) (-1+y^{10+2 s_1+4 s_2+3 s_3+s_4}) \cr &+x^{7+3 s_1+5
s_2+2 s_3+2 s_4} y^{15+4 s_1+8 s_2+4 s_3+4 s_4} (-1+y^{5+2 s_1+2
s_2+s_3}) (-1+y^{5+2 s_2+2 s_3+s_4}) \times \cr &(-1+y^{10+2 s_1+4
s_2+3 s_3+s_4}) -x^{7+3 s_1+4 s_2+4 s_3+s_4} y^{12+4 s_1+6 s_2+6
s_3+s_4} (-1+y^{2+s_3+s_4}) \times \cr &(-1+y^{9+2 s_1+4 s_2+2
s_3+s_4}) (-1+y^{11+2 s_1+4 s_2+3 s_3+2 s_4}) -x^{7+3 s_1+4 s_2+2
s_3+3 s_4} y^{12+4 s_1+6 s_2+3 s_3+4 s_4} \times \cr &(-1+y^{3+2
s_2+s_3}) (-1+y^{2+s_3+s_4}) (-1+y^{5+2 s_2+2 s_3+s_4}) +x^{6+5
s_1+2 s_2+2 s_3+2 s_4} y^{13+6 s_1+4 s_2+4 s_3+4 s_4} \times \cr
&(-1+y^{1+s_3}) (-1+y^{9+2 s_1+4 s_2+2 s_3+s_4}) (-1+y^{10+2 s_1+4
s_2+3 s_3+s_4}) -x^{6+4 s_1+2 s_2+2 s_3+3 s_4} \times \cr &y^{8+4
s_1+2 s_2+3 s_3+4 s_4} (-1+y^{5+2 s_1+2 s_2+s_3}) (-1+y^{4+2
s_2+s_3+s_4}) (-1+y^{9+2 s_1+4 s_2+2 s_3+s_4}) \cr &-x^{6+3 s_1+5
s_2+2 s_3+s_4} y^{12+4 s_1+8 s_2+4 s_3+s_4} (-1+y^{6+2 s_1+2
s_2+s_3+s_4}) (-1+y^{5+2 s_2+2 s_3+s_4}) \times \cr &(-1+y^{11+2
s_1+4 s_2+3 s_3+2 s_4})-x^{5+5 s_1+3 s_2+s_3+s_4} y^{8+6 s_1+4 s_2+2
s_3+s_4} (-1+y^{1+s_4}) \times \cr &(-1+y^{10+2 s_1+4 s_2+3
s_3+s_4}) (-1+y^{11+2 s_1+4 s_2+3 s_3+2 s_4}) -x^{5+5 s_1+2 s_2+2
s_3+s_4} y^{10+6 s_1+4 s_2+4 s_3+s_4} \times \cr &(-1+y^{2+s_3+s_4})
(-1+y^{9+2 s_1+4 s_2+2 s_3+s_4}) (-1+y^{11+2 s_1+4 s_2+3 s_3+2 s_4})
-x^{5+4 s_1+4 s_2+s_3+s_4} \times \cr &y^{8+4 s_1+6 s_2+2 s_3+s_4}
(-1+y^{1+s_4}) (-1+y^{10+2 s_1+4 s_2+3 s_3+s_4}) (-1+y^{11+2 s_1+4
s_2+3 s_3+2 s_4}) \cr &-x^{5+4 s_1+2 s_2+3 s_3+s_4} y^{6+4 s_1+2
s_2+4 s_3+s_4} (-1+y^{6+2 s_1+2 s_2+s_3+s_4}) (-1+y^{5+2 s_2+2
s_3+s_4}) \times \cr &(-1+y^{11+2 s_1+4 s_2+3 s_3+2 s_4})
+x^{5+s_1+4 s_2+3 s_3+2 s_4} y^{23+4 s_1+12 s_2+8 s_3+4 s_4}
(-1+y^{1+s_3}) (-1+y^{1+s_4}) \times \cr &(-1+y^{2+s_3+s_4})
+x^{5+s_1+3 s_2+3 s_3+3 s_4} y^{7+4 s_2+4 s_3+4 s_4} (-1+y^{1+s_3})
(-1+y^{9+2 s_1+4 s_2+2 s_3+s_4}) \times \cr &(-1+y^{10+2 s_1+4 s_2+3
s_3+s_4}) +x^{4+5 s_1+2 s_2+s_3+s_4} y^{8+6 s_1+4 s_2+2 s_3+s_4}
(-1+y^{1+s_4}) \times \cr &(-1+y^{10+2 s_1+4 s_2+3 s_3+s_4})
(-1+y^{11+2 s_1+4 s_2+3 s_3+2 s_4}) +x^{4+3 s_1+4 s_2+s_3+s_4}
y^{8+4 s_1+6 s_2+2 s_3+s_4} \times \cr &(-1+y^{1+s_4}) (-1+y^{10+2
s_1+4 s_2+3 s_3+s_4}) (-1+y^{11+2 s_1+4 s_2+3 s_3+2 s_4})+x^{4+3
s_1+s_2+3 s_3+2 s_4} \times \cr &y^{17+4 s_1+6 s_2+8 s_3+4 s_4}
(-1+y^{5+2 s_1+2 s_2+s_3}) (-1+y^{1+s_4}) (-1+y^{6+2 s_1+2
s_2+s_3+s_4}) \cr &+x^{4+3 s_1+s_2+2 s_3+3 s_4} y^{8+4 s_1+2 s_2+3
s_3+4 s_4} (-1+y^{1+s_3}) (-1+y^{4+2 s_2+s_3+s_4}) (-1+y^{5+2 s_2+2
s_3+s_4}) \cr &-x^{4+s_1+2 s_2+3 s_3+3 s_4} y^{5+2 s_2+4 s_3+4 s_4}
(-1+y^{3+2 s_2+s_3}) (-1+y^{7+2 s_1+2 s_2+2 s_3+s_4}) \times \cr
&(-1+y^{10+2 s_1+4 s_2+3 s_3+s_4})-x^{3+s_1+3 s_2+3 s_3+s_4} y^{4+4
s_2+4 s_3+s_4} (-1+y^{2+s_3+s_4}) \times \cr &(-1+y^{9+2 s_1+4 s_2+2
s_3+s_4}) (-1+y^{11+2 s_1+4 s_2+3 s_3+2 s_4})  }
$$

$$
\eqalign{&+x^{3+s_1+2 s_2+2 s_3+3 s_4} y^{4+2 s_2+3 s_3+4 s_4}
\times \cr &(-1+y^{3+2 s_2+s_3}) (-1+y^{6+2 s_1+2 s_2+s_3+s_4})
(-1+y^{9+2 s_1+4 s_2+2 s_3+s_4}) +x^{2+s_1+2 s_2+3 s_3+s_4} \times
\cr &y^{2+2 s_2+4 s_3+s_4} (-1+y^{4+2 s_2+s_3+s_4}) (-1+y^{7+2 s_1+2
s_2+2 s_3+s_4}) (-1+y^{11+2 s_1+4 s_2+3 s_3+2 s_4}) \cr
&-x^{2+s_1+s_2+3 s_3+2 s_4} y^{17+4 s_1+6 s_2+8 s_3+4 s_4}
(-1+y^{3+2 s_2+s_3}) (-1+y^{1+s_4}) (-1+y^{4+2 s_2+s_3+s_4}) \cr
&-x^{2+s_1+s_2+2 s_3+3 s_4} y^{4+2 s_2+3 s_3+4 s_4} (-1+y^{1+s_3})
(-1+y^{6+2 s_1+2 s_2+s_3+s_4}) (-1+y^{7+2 s_1+2 s_2+2 s_3+s_4}) \cr
&+x^{1+s_1+3 s_2+s_3+s_4} y^{2+4 s_2+2 s_3+s_4} (-1+y^{1+s_4})
(-1+y^{10+2 s_1+4 s_2+3 s_3+s_4}) \times \cr &(-1+y^{11+2 s_1+4
s_2+3 s_3+2 s_4}) ) }  \eqno(III.8)
$$

In view of (I.8), it is seen that (III.6) gives us
$$ \eqalign{&P(x,y,0,0,0,0)={1 \over x^{23} \ y^{32}} (1 + x) (1 + y) (-1 + x)^3 (-1 + y)^3
(-1 + x y)^4 (1 + x y)^2 (-1 + x^2 y) \times \cr &(-1 + x y^2)^4 (1
+ x y^2) (1 + x y + x^2 y^2) (-1 + x^3 y^2) (-1 + x y^3) (-1 + x^2
y^3)^2 (1 + x^2 y^3) \times \cr &(-1 + x y^4) (1 + x y^2 + x^2 y^4)
(-1 + x^3 y^4)^2 (-1 + x^3 y^5) (-1 + x^5 y^6)} \eqno(III.9) $$

Following examples will be helpful here to illustrate our work:
$$  \eqalign{
&Ch(\lambda_1)=R(x,y,1,0,0,0)={1 \over x^5 y^6} (1 + x^2 y^2) (1 + x
+ x^2 + x^2 y + x^2 y^2 + x^2 y^3 + x^3 y^3 + x^2 y^4 \cr & \ \ \ \
\ \ \ \ \ \ \ \ + 2 x^3 y^4 + x^4 y^4 + x^5 y^4 + x^3 y^5 + x^5 y^5
+ x^3 y^6 + x^4 y^6 + 2 x^5 y^6 + x^6 y^6 + x^5 y^7 \cr & \ \ \ \ \
\ \ \ \ \ \ \ +x^6 y^7 + x^6 y^8 + x^6 y^9 + x^6 y^{10} + x^7 y^{10}
+ x^8 y^{10}) \cr
&Ch(\lambda_3+\lambda_4)=R(x,y,0,0,1,1)={1 \over
x^{9} y^{14}} (1 +y)^2 (1 + x y)^2 (1 + x^2 y) (1 + y^2) \times \cr
& \ \ \ \ \ \ \ \ \ \ \ \ \ \ \ \ \ (1 + x y^2) (1 + x^2 y^2) (1 - x
y + x^2 y^2) (1 + x y^3) (1 + x^2 y^3) (1 + x^3 y^4) (1+x^3 y^5)}
$$

A useful application of (III.4) is in the calculation of tensor
coupling coefficients {\bf [8]}. We note, for instance, that the
coefficients in decomposition
$$ \eqalign{V(\lambda_1) &\otimes V(\lambda_3 + \lambda_4)=
V(\lambda_1 + \lambda_3 + \lambda_4) \oplus V(\lambda_3 + 2 \
\lambda_4) \oplus V(2 \ \lambda_3) \oplus V(\lambda_2 + \lambda_4)
\cr & \ \ \ \ \oplus V(\lambda_1 +2 \ \lambda_4) \oplus V(\lambda_1
+ \lambda_3) \oplus 2 \ V(\lambda_3 + \lambda_4) \oplus V(3 \
\lambda_4) \oplus V(\lambda_2) \oplus V(\lambda_1 + \lambda_4) \cr &
\ \ \ \ \oplus V(2 \ \lambda_4) \oplus V(\lambda_3)}
$$
\noindent can be calculated easily by
$$ \eqalign{R(x,y,1,0,0,0) &\otimes R(x,y,0,0,1,1)=
R(x,y,1,0,1,1) \oplus R(x,y,0,0,1,2) \oplus R(x,y,0,0,2,0) \cr
&\oplus R(x,y,0,1,0,1) \oplus R(x,y,1,0,0,2) \oplus  R(x,y,1,0,1,0)
\oplus 2 \ R(x,y,0,0,1,1) \cr &\oplus R(x,y,0,0,0,3) \oplus
R(x,y,0,1,0,0) \oplus R(x,y,1,0,0,1) \oplus R(x,y,0,0,0,2) \cr
&\oplus R(x,y,0,0,1,0)}
$$
\noindent where the terms above are already known by (III.4).

\vskip 3mm
\noindent {\bf{CONCLUSION}}
\vskip 3mm

As is promised, the case for $F_4$ Lie algebra is studied here
explicitly. In a subsequent work, we will study the case for $E_6$
Lie algebra. This first three cases complete the exceptional chain
of finite Lie algebras. Beyond these, although applicable, the
method seem to be unpractical. We have shown however that there is
another method {\bf [9]} which works equally well for $E_7$ and
$E_8$.

\bigskip
\noindent{\bf {REFERENCES}}
\vskip5mm

\item{[1]} M.Gungormez and H.R.Karadayi, J. Geometry and Physics 57 (2007) 2533-2538

\item{[2]} H.Weyl, The Classical Groups, N.J. Princeton Univ. Press (1946)

\item{[3]} H.R.Karadayi and M.Gungormez, J.Phys.A:Math.Gen. 32 (1999) 1701-1707

\item{[4]} J.E.Humphreys, Introduction to Lie Algebras and Representation Theory, \hfill\break
N.Y., Springer-Verlag (1972)

\item{[5]} V.G.Kac, Infinite Dimensional Lie Algebras, N.Y., Cambridge Univ. Press (1990)

\item{[6]} Weyl dimension formula cited in p.139 of ref [4]

\item{[7]} http://atlas.cc.itu.edu.tr/$\sim$ gungorm

\item{[8]} Steinberg formula cited in p.140 of ref [4]

\item{[9]} H.R.Karadayi and M.Gungormez, Summing over the Weyl Groups of $E_7$ and $E_8$, \hfill\break
math-ph/9812014

\end